\documentclass[iop]{emulateapj}
\usepackage{xcolor}
\usepackage{multirow}

\newcommand{\hii}{H\,{\sc ii} }
\newcommand{\pv}{$p-v$ }
\slugcomment{}
\shorttitle{Influence of Wolf--Rayet stars on surrounding star-forming molecular clouds}
\shortauthors{T. Baug et al.}
\begin{document}

\title{Influence of Wolf--Rayet stars on surrounding star-forming molecular clouds}
\author{T. Baug\altaffilmark{1}}
\altaffiltext{1}{Kavli Institute for Astronomy and Astrophysics, Peking University, Beijing 100871, People's Republic of China}
\author{Richard de Grijs\altaffilmark{2,3,4}}
\altaffiltext{2}{Department of Physics and Astronomy, Macquarie University, Balaclava Road, Sydney, NSW 2109, Australia}
\altaffiltext{3}{Research Centre for Astronomy, Astrophysics and Astrophotonics, Macquarie University, Balaclava Road, Sydney, NSW 2109, Australia}
\altaffiltext{4}{International Space Science Institute--Beijing, 1 Nanertiao, Zhongguancun, Hai Dian District, Beijing 100190, People's Republic of China}
\author{L.~K. Dewangan\altaffilmark{5}}
\altaffiltext{5}{Physical Research Laboratory, Navrangpura, Ahmedabad 380 009, India}
\author{Gregory J. Herczeg\altaffilmark{1}}
%
\author{D.~K. Ojha\altaffilmark{6}}
\altaffiltext{6}{Department of Astronomy and Astrophysics, Tata Institute of Fundamental Research, Homi Bhabha Road, Mumbai 400 005, India}
\author{Ke Wang\altaffilmark{1}}
%
\author{Licai Deng\altaffilmark{7}}
\altaffiltext{7}{Key Laboratory for Optical Astronomy, National Astronomical Observatories, Chinese Academy of Sciences, 20A Datun Road, Chaoyang District, Beijing 100012, People's Republic of China}
\author{B.~C. Bhatt\altaffilmark{8}}
\altaffiltext{8}{Indian Institute of Astrophysics, II Block, Koramangala, Bangalore 560 034, India}

\email{tapas.polo@gmail.com; tapas@pku.edu.cn}

\begin{abstract}
We investigate the influence of Wolf--Rayet (W--R) stars on their surrounding star-forming molecular clouds. We study five regions containing W--R
 stars in the inner Galactic plane ($l\sim$[14$^\circ$--52$^\circ$]), using multi-wavelength data from near-infrared to radio wavelengths.
 Analysis of $^{13}$CO line data reveals that these W--R stars have developed gas-deficient cavities in addition to molecular shells with
 expansion velocities of a few km s$^{-1}$. The pressure owing to stellar winds primarily drives these  expanding shells and sweeps up
 the surrounding matter to distances of a few pc. The column densities of shells are enhanced by a minimum of 14\% for one region to
 a maximum of 88\% for another region with respect to the column densities within their central cavities. No active star formation --- including
 molecular condensations, protostars, or ionized gas --- is found inside the cavities, whereas such features are observed around the molecular
 shells. Although the expansion of ionized gas is considered an effective mechanism to trigger star formation, the dynamical ages of
 the \hii regions in our sample are generally not sufficiently long to do so efficiently. Overall, our results hint at the possible
 importance of negative W--R wind-driven feedback on the gas-deficient cavities, where star formation is quenched as a consequence.
 In addition, the presence of active star formation around the molecular shells indicates that W--R stars may also assist in
 accumulating molecular gas, and that they could initiate star formation around those shells.
\end{abstract}


\section{Introduction}
\label{sec:intro}
Wolf--Rayet (W--R) stars are post-main-sequence stars that have descended from their early O-type progenitors ($m_\ast \gtrsim 20
 M_\odot$). W--R stars are characterized by enormous mass-loss rates, $\sim10^{-5} M_\odot$ yr$^{-1}$, with a wind velocity of
 1000--5000 km s$^{-1}$ \citep{crowther07}. Given such strong energetics, W--R stars can significantly affect their natal
 environments despite their short life span \citep[$\sim$5 Myr;][]{lamers99}. In fact, the energetics associated with less massive
 OB stars ($m_\ast \gtrsim 8 M_\odot$ but $m_\ast <20 M_\odot$), such as stellar winds and ultraviolet (UV) radiation, are observationally
 found to influence and shape their surrounding interstellar medium \citep[ISM;][]{motte18}. Their strong energetic impact may either
 trigger the formation of the next generation of stars or disperse the surrounding molecular gas into the ISM, and thus halt further
 star formation \citep{deharveng10}. Two primary theories are thought to explain the triggered star formation mechanism. One is the
 `collect and collapse' scenario, where molecular gas is collected between the ionization and shock fronts developed by the massive
 star and eventually collapses to form stars once a critical density is reached \citep{elmegreen77, elmegreen98}. The alternative theory
 is `radiation-driven implosion', where UV photons from massive stars initiate the formation of the next generation of stars in the
 surrounding, pre-existing cores \citep{reipurth83,bertoldi89,lefloch94}. Observational evidence of next generation star formation
 triggered by massive stars is found in several Galactic star-forming regions \citep[see e.g.,][and references therein]{dewangan16,
 zavagno10a, zavagno10b, pomares09}. For example, evidence of the `collect and collapse' process is seen around several \hii regions
 where second-generation stars were found in the periphery of the ionized gas \citep{zavagno07,deharveng08}. The `radiation-driven
 implosion' process is also notably observed in several bright-rimmed clouds, as evidenced by e.g., aligned and sequential star
 formation with respect to the energetic source \citep[see e.g.][ and references therein]{ogura07, urquhart07, morgan10, panwar14}.

The influence of W--R stars on their parent clouds might be significantly different from what is typically seen around less
 massive OB stars or \hii regions. The effect of a W--R star on its parent molecular cloud could either be positive or negative in the
 context of next generation star formation triggers. Owing to its energetic stellar wind, a W--R star may create a surrounding cavity
 and develop wind-blown expanding shells of parsecs to tens of parsecs scales with typical expansion velocities of a few km s$^{-1}$
 \citep{marston96}. Triggering mechanisms may occur at the boundary of the expanding molecular shell \citep{whitworth94}. However,
 based on hydrodynamic simulations of turbulent giant molecular clouds, \citet{dale13} reported that the efficiency of wind-driven
 triggering to initiate star formation is low compared with that of the ionized gas. The presence of a W--R star instead may destroy
 favorable conditions for further star formation by driving away the surrounding molecular gas into the ISM \citep{dale13,sokal16}.
 \citet{sokal16} studied several massive star clusters associated with W--R stars and suggested that the presence of a W--R star may
 accelerate cluster emergence. On the other hand, a positive impact of W--R stars on their parent clouds, leading to further star
 formation, was also found in a few Galactic regions \citep[see][and references therein]{liu12, dewangan16}.

Observational studies of star-forming regions associated with W--R stars are limited mainly because of their rarity and also since a
 large fraction of W--R stars are still unknown in the Milky Way \citep{crowther07}. Recent systematic surveys aimed at identifying
 Galactic W--R stars \citep[e.g.][]{shara12, kanarek15} have increased their number by a factor of two with respect to the previous
 decade. This study aims to explore a few potential star-forming regions that could be associated with W--R stars and, subsequently, to
 examine whether the presence of W--R stars has any impact on the surrounding molecular gas. We perform an analysis of multi-wavelength
 data from near-infrared (NIR) to radio wavelengths to investigate the star formation activity in the clouds surrounding W--R stars.
 Note that massive star-forming regions are often associated with \hii regions, which may also have a large impact on triggered star
 formation \citep{deharveng10}.

Since the advent of the {\sl Spitzer} Space Telescope, thousands of parsec-scale mid-infrared (MIR) Galactic bubbles have been identified
 in {\sl Spitzer}-IRAC 8 $\mu m$ images \citep{churchwell06, churchwell07, simpson12}. \citet{deharveng10} reported active star formation
 in many of those MIR Galactic bubbles. Thus, MIR bubbles could be suitable regions for our study, provided they are associated
 with W--R stars. This study is organized as follows. The details of the selected regions, their associated MIR Galactic bubbles, and
 spatially overlapping W--R stars are presented in Section~\ref{sec:regions} (see also Table~\ref{table1}). Section~\ref{sec:data}
 describes the details of the multi-wavelength data sets used in the analysis. In Section~\ref{sec:association}, we explore the possibility
 that these W--R stars may be associated with individual molecular clouds. Then, in Section~\ref{sec:kin} we perform an analysis to examine
 the dynamics of molecular gas using molecular-line data and ongoing star formation activity, if any, toward these regions. A
 detailed discussion of the possible influence of W--R stars on their surrounding molecular clouds is offered in Section~\ref{sec:SF}.
 Finally, we provide our conclusions in Section~\ref{sec:conclusions}.

\section{Selected regions}
\label{sec:regions}
We cross-matched the available catalogs of W--R stars \citep{vander01, vander06, hadfield07, shara12, kanarek15} with the
 {\sl Spitzer}--Galactic Legacy Infrared Mid-Plane Survey Extraordinaire \citep[GLIMPSE;][]{benjamin03} MIR Galactic bubble catalogs
 \citep{churchwell06, churchwell07, simpson12}, employing a search radius of twice the effective bubble radius. This particular radius
 was chosen for our cross-matching, assuming that a W--R star located outside the bubble periphery but at a separation of twice the
 radius of the MIR bubble might have a strong impact on the gas around the bubble --- even though it might not be the primary driving
 source for the bubble to form. The primary aim of this study is not to search for the effect of W--R stars on MIR bubbles but
 to examine their influence on surrounding star-forming parental clouds. The role of MIR bubbles here is only to help us in identifying
 regions exhibiting active star formation. It is important to note that \citet{deharveng10} found that at least 86\% of the bubbles
 are associated with ionized gas. Thus, most of these bubbles can be considered active massive star-forming regions.
 
 We found spatial matches for about 30 regions. We further limited the identified regions to only those for
 which molecular-line ($^{13}$CO) data are available with a good velocity resolution (i.e. data from the Galactic Ring Survey, GRS;
 $l\sim$[14$^\circ$--55$^\circ$]; see Section~\ref{GRS}). W--R stars with poorly constrained spectral types have also been excluded
 from this study. Finally, our combined selection criteria yield six potential regions for further study. However, one of these (MIR
 bubble N46) has already been studied extensively by \citet{dewangan16}. Thus, in this paper we explore five Galactic regions to
 examine the influence of W--R stars on their parent molecular clouds. Additional analysis of the well-studied bubble N46
 \citep{dewangan16} is performed primarily to serve as a reference. An overview of our five selected regions and the bubble N46
 (region G27 in this paper) is presented below (see Table~\ref{table1} for a summary).

\subsection{G15.010--0.570}
\begin{figure}
\epsscale{1.2}
\plotone{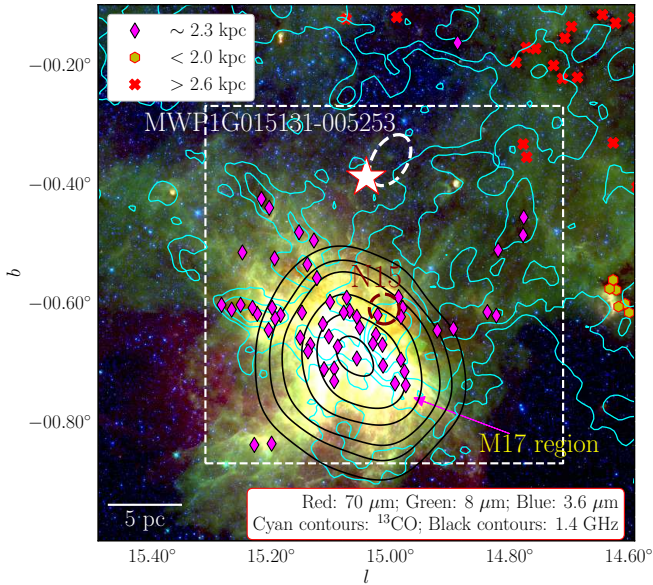}
\caption{\scriptsize Three-color composite image (red: 70 $\mu$m; green: 8 $\mu$m; blue: 3.6 $\mu$m) around the G15 region.
 The position of the W--R star, 2MASS J18192219--1603123, is marked by a white star. Two MIR bubbles, N15 and MWP1G015131--005253,
 are marked by brown and white dashed ellipses, respectively. The 1.4 GHz VGPS contours are shown in black at levels of 100, 150,
 200, 300, 400, and 500 mJy beam$^{-1}$. Cyan contours represent the integrated  $^{13}$CO gas of the host molecular cloud toward the region. A scale bar
 corresponding to a projected separation of 5 pc is also shown in the bottom left-hand corner. The dust clumps identified by
 \citet{urquhart18} are marked by diamonds, hexagons, and crosses, depending on their distances (see legend). The area shown
 in the white dashed box is the region selected for further analysis.}
\label{fig1}
\end{figure}

The region G15.010--0.570 (hereafter G15) is located at Galactic coordinates $l$=15$^\circ$.010, $b$=$-$0$^\circ$.570. Figure~\ref{fig1}
 shows a three-color composite image of a large area around the region (red: {\sl Herschel} 70 $\mu$m; green: {\sl Spitzer}-IRAC 8 $\mu$m;
 blue: {\sl Spitzer}-IRAC 3.6 $\mu$m). The Jansky Very Large Array (VLA) Galactic Plane survey (VGPS)\footnote{http://www.ras.ucalgary.ca/VGPS/}
 radio continuum contours at 1.4 GHz are also superimposed. Contours of the integrated $^{13}$CO map of the host molecular cloud
 (the identification procedure of host molecular cloud is discussed in Section~\ref{sec:association}; see also Table~\ref{table1}) are also overlaid.
 The region hosts two MIR bubbles, N15 and MWP1G015131--005253, and a W--R star, 2MASS J18192219--1603123, located at the edge of the
 bubble MWP1G015131--005253. This W--R star was first identified and classified as a WN7o star by \citet{hadfield07}. The W--R star has
 a {\sl Gaia} parallax of 0.42$\pm$0.14 milli-arcsec \citep{gaia18}, which corresponds to a distance of 2.3$^{+1.3}_{-0.6}$ kpc
 \citep[estimated using a Bayesian approach by][]{bailer18}.

The famous Omega Nebula \citep[Messier 17 or M17; distance $\sim$2.0 kpc;][]{wu14} is spatially located adjacent to the W--R star. M17
 is regarded one of the most active massive Galactic star-forming regions, with a star-formation rate $\ge$ $0.004~M_\odot$ yr$^{-1}
 $\citep[see][and references therein]{povich16}. \citet{chibueze16} identified hundreds of H$_2$O maser sources in the M17
 region spanning a local standard of rest velocity ($v_\mathrm{LSR}$) range of 14.0--22.0 km s$^{-1}$. They also reported a distance to
 the region of 2.04$^{+0.16}_{-0.17}$ kpc, based on trigonometric parallaxes of the H$_2$O masers. This similarity in distances
 and the spatial proximity of the W--R star to M17 imply that they could be part of the same molecular cloud. The G15 region also hosts
 a large-scale gaseous filament known as F18 at a distance of 2.0 kpc \citep[see][]{wang16}.
 
Recently, \citet{urquhart18} reported distances for thousands of 870 $\mu$m dust clumps toward the inner Galactic plane. They resolved
 the distance ambiguities of these clumps using measurements from H{\sc i} line observations, maser parallaxes, and spectroscopic
 observations. Several dust clumps were identified toward this region, primarily located at distances of 2.0--2.6 kpc in the
 $v_\mathrm{LSR}$ range of 16--30 km s$^{-1}$. Finally, considering the distances and the distribution of the dust clumps, we  
 selected a 36$'\times$36$'$ area centered at $l$=15$^\circ$.010, $b$=$-$0$^\circ$.570 for further analysis (see Figure~\ref{fig1}).

\subsection{G24.750+0.100}
\begin{figure}
\epsscale{1.2}
\plotone{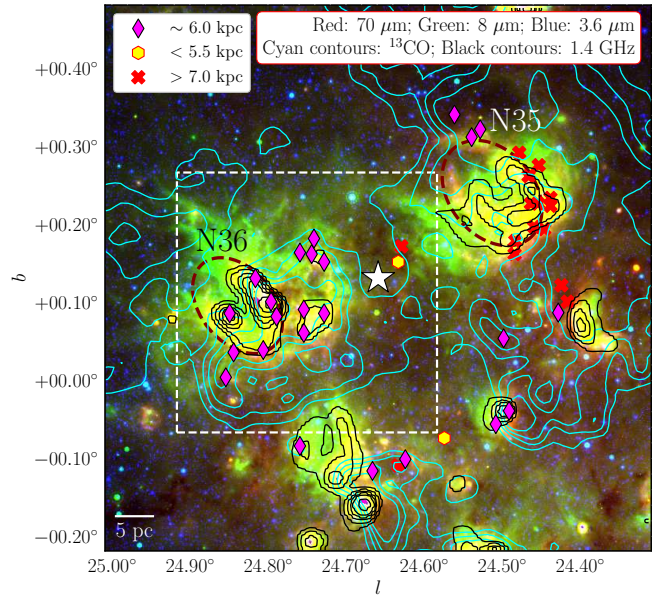}
\caption{\scriptsize Three-color composite image (red: 70 $\mu$m; green: 8 $\mu$m; blue: 3.6 $\mu$m) around the G24 region. The
 position of the W--R star, 1477--55L, is marked by a white star. Two MIR bubbles, N35 and N36, are marked by brown dashed ellipses.
 The 1.4 GHz VGPS contours are shown in black at levels of 40, 50, 65, 80, 100, 150, and 200 mJy beam$^{-1}$. In addition, cyan
 contours show the distribution of the $^{13}$CO gas of the host cloud. The remaining symbols are similar to those in Figure~\ref{fig1}.}
\label{fig2}
\end{figure}

A three-color composite image (red: {\sl Herschel} 70 $\mu$m; green: {\sl Spitzer}-IRAC 8 $\mu$m; blue: {\sl Spitzer}-IRAC 3.6 $\mu$m)
 of a large area around the G24.750+0.100 region (hereafter G24) is shown in Figure~\ref{fig2}. The distribution of ionized gas in the
 region is shown by the VGPS 1.4 GHz radio continuum contours. The distribution of $^{13}$CO gas of the host molecular
 cloud is also shown. The region includes two MIR bubbles, N35 and N36, and the W--R
 star 1477-55L. This W--R star was first identified by \citet{shara12}, and classified as a WC9 star. Although, \citet{shara12} reported
 a foreground extinction to the W--R star of $A_{K_{\mathrm s}} \sim 3.03$ mag, but no distance estimate was given by these authors. The
 region is also associated with a large gaseous filament, F27, located at a distance of 5.6 kpc \citep{wang16}. The presence of two
 molecular clumps, U24.50--0.04 and C24.48+0.21, toward the region are found at $v_\mathrm{LSR}$ of 110.3 and 117.5 km s$^{-1}$, respectively.
 The corresponding distances are 9.1 and 8.6 kpc \citep{anderson12}. 
 
The ionized gas associated with these clumps is characterized by $v_\mathrm{LSR}$ of 108.1 and 115.7 km s$^{-1}$, respectively. The dust
 clumps toward this region \citep{urquhart18} are located in two different velocity ranges. Clumps at a distance of $\sim$6 kpc have
 $v_\mathrm{LSR}$ in the range 98--114 km s$^{-1}$, whereas clumps located further than 7 kpc typically have $v_\mathrm{LSR}$ greater than
 115 km s$^{-1}$. Accordingly, Figure~\ref{fig2} shows that the bubbles N35 and N36 may be associated with two different distances. The
 distance estimate to the W--R star, 1477--55L, of 5.7$\pm$0.5 kpc (see Section~\ref{spec-phot-dist}; see also Table~\ref{table1}) implies
 that it is likely associated with the cloud hosting the MIR bubble N36. 
 
Recently, \citet{dewangan18} performed a detailed study of this region, primarily to reveal the formation of the N36 bubble. They found
 evidence of a collision between two nearby molecular clouds toward N36. Even so the primary goal and the area of this study are
 significantly different from those of \citet{dewangan18}. Based on the distribution as well as the distances to the dust clumps, we selected
 a 20$'\times$20$'$ area centered at $l$=24$^\circ$.750, $b$=+0$^\circ$.100 for further analysis.
\subsection{G34.260+0.169}
\begin{figure}
\epsscale{1.2}
\plotone{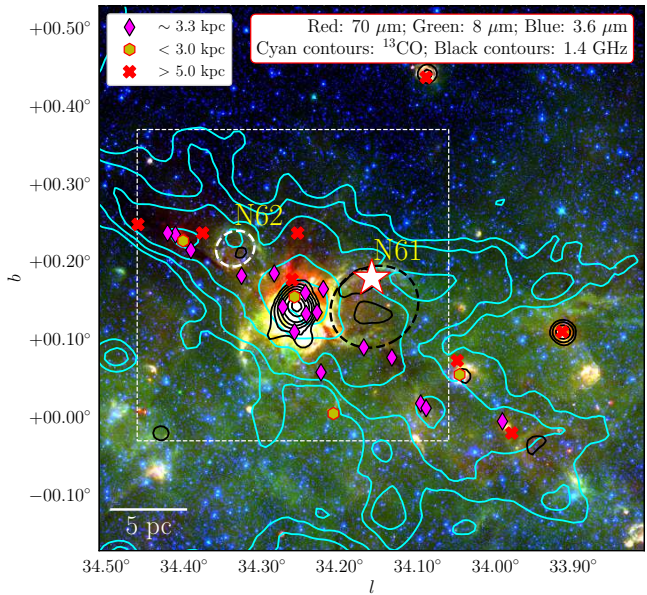}
\caption{\scriptsize Three-color composite image of the region around G34. The position of the W--R star, 1553--15DF, is marked by
 a white star. The region hosts two MIR bubbles, N61 and N62, which are marked by dashed black and white ellipses, respectively.
 The 1.4 GHz VGPS contours (in black) are shown at levels of 28, 42, 70, 120, 200, 400, and 600 mJy beam$^{-1}$, while cyan contours
 show the distribution of $^{13}$CO gas of the host cloud. The remaining symbols are similar to those in Figure~\ref{fig1}.}
\label{fig3}
\end{figure}

A three-color composite image of the G34.260+0.169 region (hereafter G34) is shown in Figure~\ref{fig3}. This region hosts a WC8-type W--R
 star, 1553-15DF \citep{kanarek15}. \citet{wang16} reported the presence of a gaseous filament, F36, at a distance of 1.6 kpc projected
 toward this region. The G34 region also harbors an infrared dark cloud (IRDC), G34.43+0.24, which was explored to look for high-mass
 star-forming cores by \citet{xu16} and \citet{sakai18}. \citet{xu16} found that the IRDC is located at a $v_\mathrm{LSR}$ range of
 53--63 km s$^{-1}$ and suggested that it could be divided into three parts based on the prevailing evolutionary stages. The IRDC has
 multiple distance measurements, ranging from 1.56 kpc to 3.9 kpc \citep{faundez04, rathborne06, simon06, kurayama11, foster12}. However,
 the distance of 1.56 kpc, measured using a H$_2$O maser parallax by \citet{kurayama11}, is controversial since there was only a single
 background source for reference \citep{foster12}. 

The region also hosts two MIR bubbles, N61 and N62, and at least two molecular clumps
 \citep[U34.26+0.15 and  U34.40+0.23;][]{anderson12} at a $v_\mathrm{LSR}$ of about 57 km s$^{-1}$. Clump U34.26+0.15 is located at
 a distance of 3.4$\pm$0.5 kpc \citep{anderson12}. \citet{xu16} suggested that the bubble N61 may be associated with the
 IRDC G34.43+0.24. However, the other bubble, N62, is located at the far kinematic distance ($d_\mathrm{Far}$) of 10.55 kpc, and has no
 physical association with the IRDC \citep{devine18}. Dust clumps near the W--R star are projected onto a filamentary structure (see
 Figure~\ref{fig3}), and these clumps have distances of $\sim$3.3 kpc ($v_\mathrm{LSR}$ $\in 53$--59 km s$^{-1}$) which are similar to
 the distance to the W--R star (2.9$\pm$0.3 kpc; see Table~\ref{table1} and Section~\ref{spec-phot-dist}). Based on the distribution of
 the dust clumps, we selected a 24$'\times$24$'$ area centered at $l$=34$^\circ$.260, $b$=+0$^\circ$.169 for further analysis.
\subsection{G35.598--0.032}
\label{sec:G35}
\begin{figure}
\epsscale{1.2}
\plotone{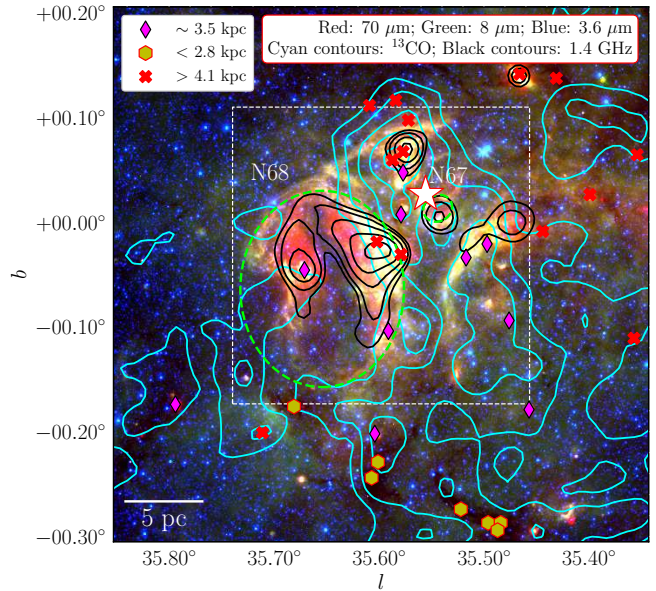}
\caption{\scriptsize Three-color composite image of the G35 region. The position of the W--R star, 1567--51L, is marked by a white
 star. This region also hosts two MIR bubbles, N67 and N68, which are marked by dashed green ellipses. The 1.4 GHz VGPS contours (in
 black) are overplotted at levels of 30, 35, 45, 60, and 80 mJy beam$^{-1}$. Distribution of the $^{13}$CO gas of the host cloud is
 also shown in cyan contours. The remaining symbols are similar to those in Figure~\ref{fig1}.}
\label{fig4}
\end{figure}

The G35.598--0.032 region (hereafter G35) hosts a WC7 star, 1567-51L \citep{shara12}. It also harbors two MIR Galactic bubbles, N67 and
 N68. The W--R star is located toward the edge of N67 (see Figure~\ref{fig4}). \citet{zhang13} studied N68 and its environment using
 multi-wavelength data sets. They concluded that the expansion of the \hii region has affected the surrounding molecular gas for
 further star formation. 

Several cold clumps are found around this region \citep{anderson12}. They are located at a $v_\mathrm{LSR}$
 range of 51--56 km s$^{-1}$. Dust clumps \citep{urquhart18} spatially close to the W--R star in this region are found at distances of
 $\sim$3.5 kpc ($v_\mathrm{LSR}$ range of 46--64 km s$^{-1}$). Considering the distance to the W--R star (3.8$\pm$0.4 kpc; see
 Table~\ref{table1}) and the associated dust clumps, we have selected a 17$'\times$17$'$ area centered at $l$=35$^\circ$.598,
 $b$=$-$0$^\circ$.032 for further analysis.
 
\subsection{G51.840+0.410}
\label{sec:G51}
\begin{figure}
\epsscale{1.2}
\plotone{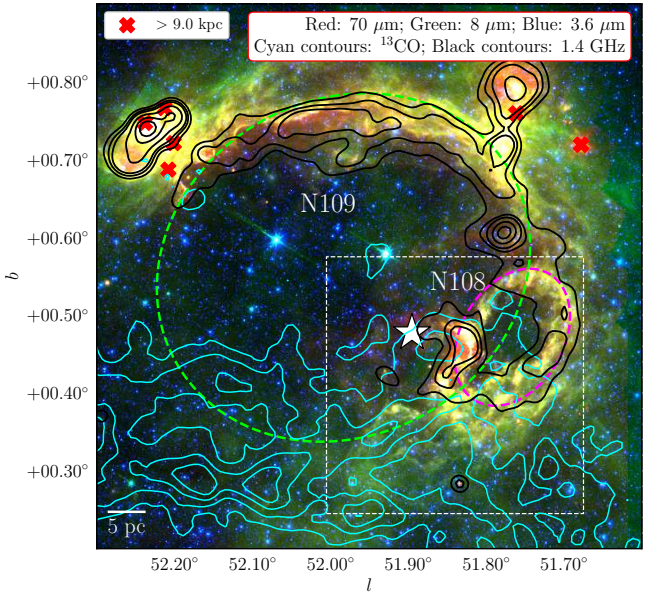}
\caption{\scriptsize Three-color composite image of the G51 region. The position of the W--R star, 1697--38F, is marked by a white
 star. This region also hosts two MIR bubbles, N108 and N109, which are marked by dashed blue and green ellipses, respectively. The
 1.4 GHz VGPS contours (in black) are marked at levels of 11, 13, 15, 21, 30, and 40 mJy beam$^{-1}$. Cyan contours correspond to
 the integrated $^{13}$CO gas of the possible host molecular cloud toward the region. The remaining symbols are similar to those in Figure~\ref{fig1}.}
\label{fig5}
\end{figure}

The G51.840+0.410 region (hereafter G51) harbors a W--R star, 1697-38F, which is classified as WC9 \citep{kanarek15} and hosts two MIR
 bubbles, N108 and N109 (see Figure~\ref{fig5}). \citet{bania12} noted star formation toward the edge of N109, and identified this
 as one of the largest \hii regions in the Milky Way. The region is reported to be part of a large molecular filamentary
 structure spanning the velocity range of $-$5.0 to 17.4 km s$^{-1}$ \citep{li13}. 

Several authors reported that the \hii regions associated with the bubbles are located at a far kinematic distance of 9.8$\pm$0.5
 kpc \citep{watson03,anderson09,bania12}. Dust clumps identified by \citet{urquhart18} are overplotted in Figure~\ref{fig5}. But
 dust clumps are only found toward the upper edge of N109, and those clumps are located at distances greater than 9 kpc \citep{urquhart18}.
 No dust clumps are identified around the W--R star. The host molecular cloud also shows no association with these particular clumps (see Figure~\ref{fig5}).
 Thus, for this region, we have selected a small 20$'\times$20$'$ area centered at
 $l$=51$^\circ$.840, $b$=+0$^\circ$.410 for further analysis, excluding the clumps located at distances $>$9 kpc.
\subsection{Reference region: G27.323--0.144}
\label{sec:G27}
The G27.323--0.144 region (hereafter G27) hosts the MIR bubble N46 (see Figure~\ref{fig6}a). The area (24$'\times$17$'$) studied by 
 \citet{dewangan16} is marked by a dashed box in Figure~\ref{fig6}a. They found that the presence of the W--R star, 1503-160L, has
 played an important role in the formation of young stellar objects (YSOs) in this region. Several signatures of the presence of the
 W--R star are adopted as references for the study of our selected regions.
%

\begin{deluxetable*}{ccccccccccccc}
\tablewidth{0pt}
\tabletypesize{\scriptsize}
\tablecaption{Details of the selected regions with parameters of the associated W--R stars and molecular clouds. \label{table1}}
\tablehead{
\colhead{Region} & \multicolumn{5}{c}{Associated W--R Star}                                                 && \multicolumn{5}{c}{Associated molecular cloud} \\ 
\cline{2-6} \cline{8-12}\\
Gll.lll+b.bbb    & \colhead{RA (J2000)}  &\colhead{Dec (J2000)} &\colhead{Spec.}& $A_K$   & Distance        && \colhead{Vel. range}   &\colhead{Trough Vel.}  &\colhead{Near KD$^a$}& \colhead{Far KD} & $v_\mathrm{exp}$\\
	         & \colhead{(hh:mm:ss.s)}&\colhead{(dd:mm:ss.s)}&\colhead{Type} & (mag)   & (kpc)      	    && \colhead{(km s$^{-1}$)}&\colhead{(km s$^{-1}$)}& $d\pm$1$\sigma$ kpc & $d\pm$1$\sigma$ kpc & km s$^{-1}$}
\startdata
G15.010--0.570    &      18:19:22.2       &    --16:03:12.4      &  WN7o         & 0.56    & 2.3$^{+1.3} _{-0.6}$$^b$ && 20--38        &         29.0          &    2.8$\pm$0.3      &13.2$\pm$0.4  & 4.0$\pm$1.8  \\
G24.750+0.100     &      18:35:47.6       &    --07:17:49.9      &  WC9          & 3.35    & 5.7$\pm$0.5     &&     95--115            &        105.5          &    6.0$\pm$0.6      & 9.2$\pm$0.4  & 5.0$\pm$3.0  \\
G34.260+0.169     &      18:53:02.6       &     +01:10:22.9      &  WC8          & 4.41    & 2.9$\pm$0.3     &&     49--63             &         53.5          &    3.3$\pm$0.4      &10.5$\pm$0.5  & --   \\
G35.598--0.032    &      18:56:07.9       &     +02:20:48.9      &  WC7          & 2.56    & 3.8$\pm$0.4     &&     44--62             &         56.4          &    3.5$\pm$0.4      &10.2$\pm$0.4  & 5.0$\pm$3.5  \\
G51.840+0.410     &      19:25:18.1       &     +17:02:15.9      &  WC9          & 2.16    & 5.9$\pm$0.5     &&     42--52             &         47.3          &    3.7$\pm$0.6      & 6.8$\pm$0.7  & 2.0$\pm$1.8  \\
G27.323--0.144$^c$&      18:41:34.1       &    --05:04:01.4      &  WN7$^d$      & 1.21$^d$& 5.2$\pm$0.5$^d$ &&     88--96$^d$         &         92.0          &    5.2$\pm$0.7      & 9.4$\pm$0.6  & 3.0$^d$
\enddata
\tablenotetext{$a$}{KD: Kinematic distance}
\tablenotetext{$b$}{Distance from {\sl Gaia} parallax \citep{bailer18}}
\tablenotetext{$c$}{Reference region; \citet{dewangan16}}
\tablenotetext{$d$}{Parameters from \citet{dewangan16}}
\end{deluxetable*}

\section{Data}
\label{sec:data}
Details of the multi-wavelength data sets used in our analysis are presented below.
\subsection{Near-infrared data}
NIR magnitudes of the selected W--R stars were obtained from the Two Micron All Sky Survey catalog \citep[2MASS catalog;][]{skrutskie06,
 cutri03}, mainly to allow us to calculate their spectro-photometric distances.
\subsection{Mid-infrared data}
The MIR images (with a spatial resolution of $\sim$2$\arcsec$) and photometric magnitudes of point-like sources were obtained from the
 {\sl Spitzer}--GLIMPSE survey archive \citep[i.e. the GLIMPSE-I Spring '07 highly reliable catalog;][]{benjamin03} . Multiband Infrared
 Photometer for {\sl Spitzer} (MIPS) Inner Galactic Plane Survey \citep[MIPSGAL;][]{carey05} 24 $\mu$m photometric magnitudes of point
 sources \citep{gutermuth15} are also used.
\subsection{Far-infrared data}
Level2\_5 processed {\sl Herschel} 70--500 $\mu$m images were used. The data were obtained from the ESA-{\sl Herschel} science
 archive (P.I. S. Molinari). The {\sl Herschel} images have beam sizes of 5$\farcs$8, 12$\arcsec$, 18$\arcsec$, 25$\arcsec$, and 
 37$\arcsec$ at 70, 160, 250, 350, and 500 $\mu$m, respectively \citep{griffin10,poglitsch10}. These multi-band images helped us to
 construct column density maps with a final spatial resolution of 37$\arcsec$.
\subsection{Molecular line data}
\label{GRS}
To perform a detailed investigation of the molecular gas associated with the selected regions, we obtained $^{13}$CO ($J$=1--0) line data
 from the GRS \citep{jackson06}. GRS line data have a velocity resolution of 0.21~km\,s$^{-1}$, with an angular resolution of 45$\arcsec$.
 The data have a main beam efficiency ($\eta_{\rm mb}$) of $\sim$0.48, with a typical rms sensitivity (1$\sigma$) of $\approx0.13$~K and
 a velocity coverage from $-$5 to 135~km~s$^{-1}$ \citep{jackson06}.

In addition, we used $^{12}$CO ($J$=1--0), $^{13}$CO ($J$=1--0), and C$^{18}$O ($J$=1--0) line data from the FOREST \citep[i.e. the four-beam
 receiver system on the Nobeyama 45 m telescope;][]{minamidani16} unbiased Galactic plane imaging survey \citep[FUGIN;][]{umemoto17}. The 45 m
 radio telescope is operated by the Nobeyama Radio Observatory. The FUGIN survey data cover the Galactic longitude ranges 10$^\circ$--50$^\circ$
 and 198$^\circ$--236$^\circ$. The data have a velocity resolution of 1.3~km\,s$^{-1}$ and an angular resolution of 21$''$, with a (1$\sigma$)
 rms sensitivity of 0.12~K.
\subsection{Radio continuum data}
To examine the distribution of the ionized gas, we retrieved the VGPS 1.4 GHz continuum maps. The VGPS maps have an angular resolution
 of 60$''$ and an rms of 11 mJy beam$^{-1}$ \citep{stil06}. In addition, to identify the detailed structures of
 the ionized gas, we also used the 1.4 GHz continuum maps from National Radio Astronomy Observatory (NRAO) Jansky Very Large Array (VLA)
 Sky Survey (NVSS) as it provides a better angular resolution with a beam size of $\sim$45$''$ and a sensitivity of
 $\sim$0.45 mJy beam$^{-1}$ \citep{condon98}.
\section{Search for molecular clouds associated with W--R stars}
\label{sec:association}
Identification of the host molecular clouds and establishing their association with W--R stars is essential before proceeding to search for
 any possible influence of the W--R stars. Thus, we first determined the distances to the selected W--R stars, and the molecular
 clouds in the velocity range of the dust clumps (see Section~\ref{sec:regions}) in order to establish their possible association. 
 We considered a molecular cloud to be the host cloud of a W--R star if the kinematic distance of that cloud is matched with
 the distance of the W–R star within an uncertainty of 1$\sigma$.
\subsection{Distances to W--R stars}
\label{spec-phot-dist}
The only W--R star in our sample that has {\sl Gaia} parallax measurements is the one in the G15 region. Thus, the distances to
 the remaining W--R stars in our sample are determined using the spectro-photometric method following \citet{shara12}. For completeness, the
 procedure is briefly outlined here. \citet{crowther06} extensively studied the W--R star population in the cluster Westerlund 1, and 
 reported the absolute magnitudes of W--R stars of different spectral types. Spectro-photometric distances to W--R stars in our sample 
 were estimated after correcting for foreground extinction following a similar procedure as adopted by \citet{dewangan16}. Estimated
 values of the extinction and distance to each W--R star are listed in Table~\ref{table1}. However, for the W--R star in the G15 region, we
 used the distance from {\sl Gaia} astrometry \citep{bailer18}.

\subsection{Identification of molecular clouds associated with W--R stars}
In this section, we identify those molecular clouds that have similar distances as the W--R stars and thus, could be physically
 associated with each other. We first look for the molecular clouds that overlap with the $v_\mathrm{LSR}$ of the ATLASGAL dust
 clumps reported by \citet{urquhart18}. Also, because of strong energetics, these W--R stars are able to develop cavities in
 their parent molecular clouds. Once such a cavity develops, its signature could also be present in the corresponding $^{13}$CO spectrum
 if the molecular gas still exists in the foreground and background of the W--R star. In this scenario, $^{13}$CO spectrum toward the
 W--R star should show a double-peaked emission profile separated by a few km s$^{-1}$. The velocity corresponding to the position of
 the W--R star should fall in the trough between the double-peaked emission features (i.e. the part of the cloud devoid of molecular gas).
 The separation between the two emission peaks depends on the age and wind velocity of the driving W--R star
 and also on the initial density of the surrounding molecular cloud.

\begin{figure}
\epsscale{1.2}
\plotone{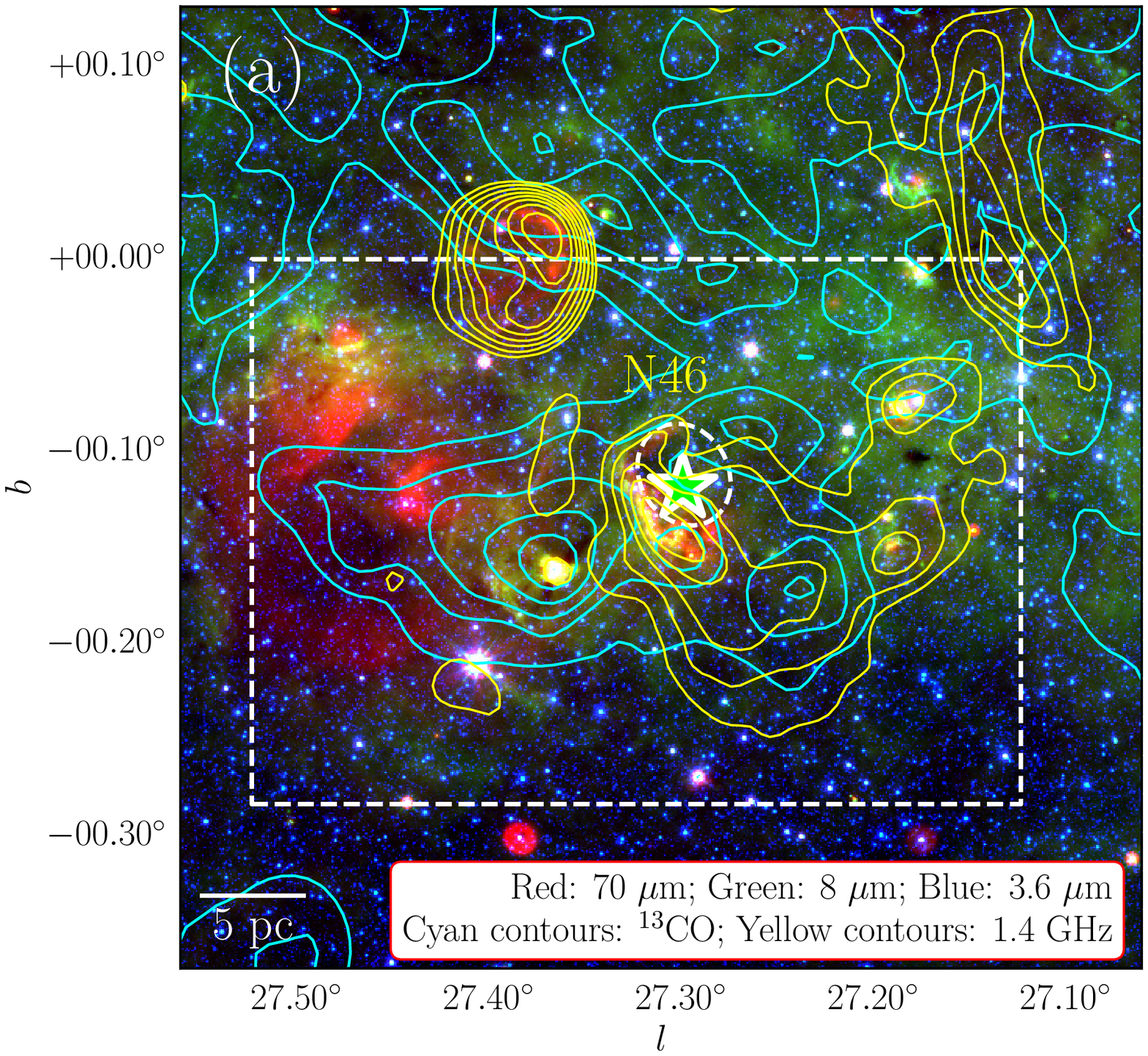}\\
\plotone{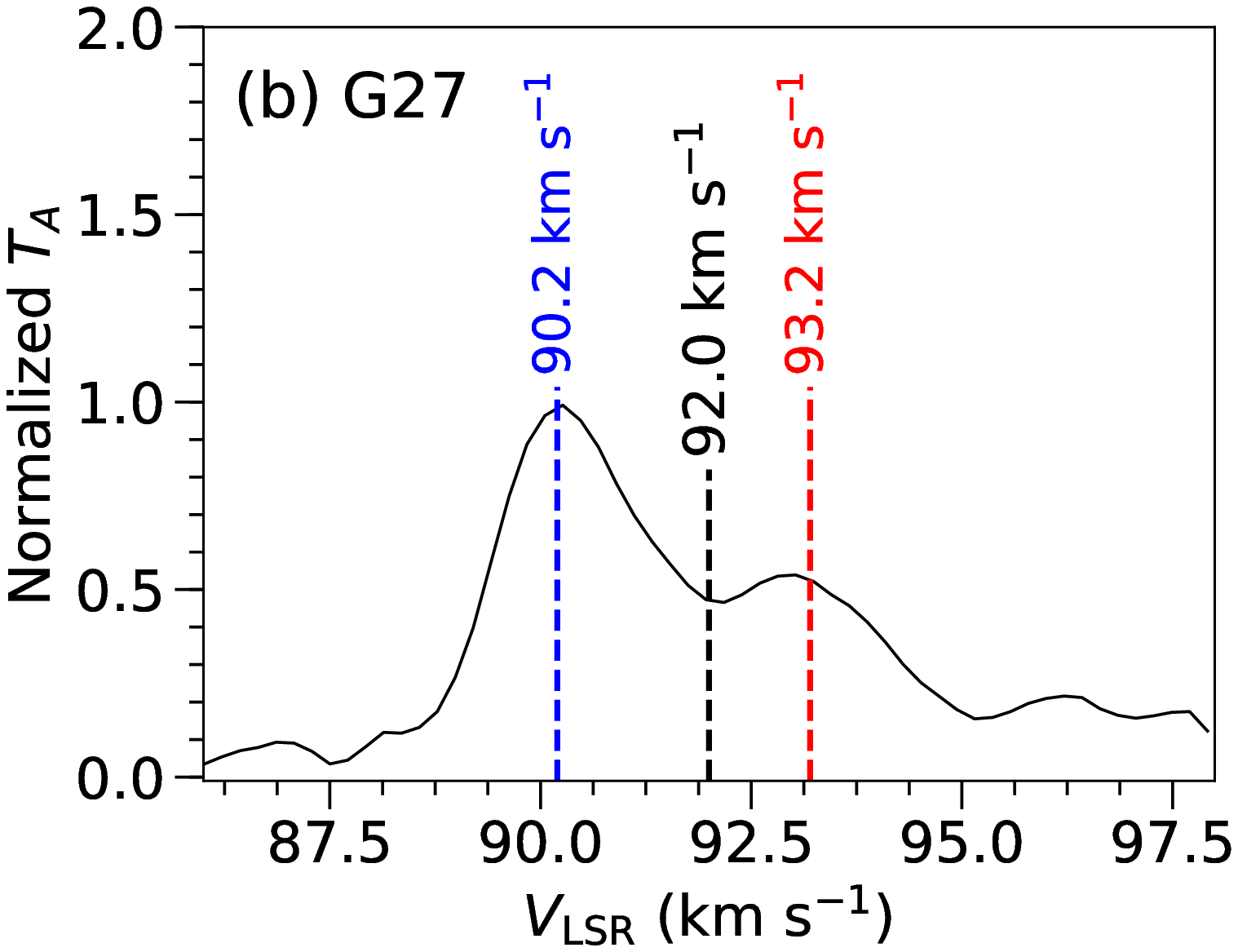}
\caption{\scriptsize Three-color composite image of the reference region, G27. The position of the W--R star, 1503--160L, is marked
 by a green star. The region hosts a MIR Galactic bubble, N46, which is marked by a dashed white ellipse. The 1.4 GHz VGPS contours
 (in yellow) are overplotted at levels of 28, 32, 37, 43, 50, 58, 70, and 90 mJy beam$^{-1}$, and cyan contours show the distribution
 of $^{13}$CO gas. A scale bar corresponding to 5 pc is shown in the bottom left-hand corner. (b) $^{13}$CO spectrum toward the W--R
 star in the reference region. The spectrum was obtained by averaging all emission within a circle of a 45$''$ diameter centered on the
 coordinates of the W--R star. Two emission peaks are clearly seen. We have marked the velocities corresponding to the emission peaks
 and the trough by blue and red, and black dashed lines, respectively.}
\label{fig6}
\end{figure}

Figure~\ref{fig6}b shows the $^{13}$CO spectrum of the G27 region. The molecular cloud hosting this particular region is 
 located at a velocity range of 88--95 km s$^{-1}$ \citep[see][]{dewangan16}. The spectrum is constructed by combining all $^{13}$CO
 emission within a 45$''$ diameter (i.e., the beam size of the GRS data)
 centered on the W--R star. Two emission peaks are clearly visible in the spectrum. 
 We estimated the kinematic distance corresponding to the velocity of the trough ($v_\mathrm{LSR}~\sim~$92.0 km s$^{-1}$;
 see Figure~\ref{fig6}b) using the `Kinematic Distance Calculation Tool'\footnote{http://www.treywenger.com/kd/index.php} of
 \citet{wenger18} which evaluates a Monte Carlo kinematic distance adopting the solar Galactocentric distance of 8.31$\pm$0.16 kpc
 \citep{reid14}. The calculation yields a near kinematic distance ($d_\mathrm{Near}$) of 5.2$\pm$0.6 kpc.
 
The spectro-photometric distance of the W--R star associated with the reference region is 5.2$\pm$0.5 kpc \citep[][see also Table~\ref{table1}]{dewangan16}. A 
 well-matched distance of a W--R star with the velocity of the trough in the $^{13}$CO spectrum implies a physical association of the W--R
 star with this particular molecular cloud and, more precisely, with the trough at a $v_\mathrm{LSR}$ of 92.0 km s$^{-1}$. The $^{13}$CO
 spectrum (Figure~\ref{fig6}b) also exhibits two emission peaks corresponding to blue-shifted (90.2 km s$^{-1}$) and red-shifted (93.2
 km s$^{-1}$) parts of the cloud, possibly dispersed by the energetics of the W--R star. Evidence of the influence of the W--R star
 was already found for this region by \citet{dewangan16}. Hence, any region with a double-peaked $^{13}$CO spectrum toward the position of
 a W--R star exhibiting a trough might be explored for any impact of W--R stars on their parent molecular clouds.

\begin{figure*} 
\epsscale{1.0}
\plottwo{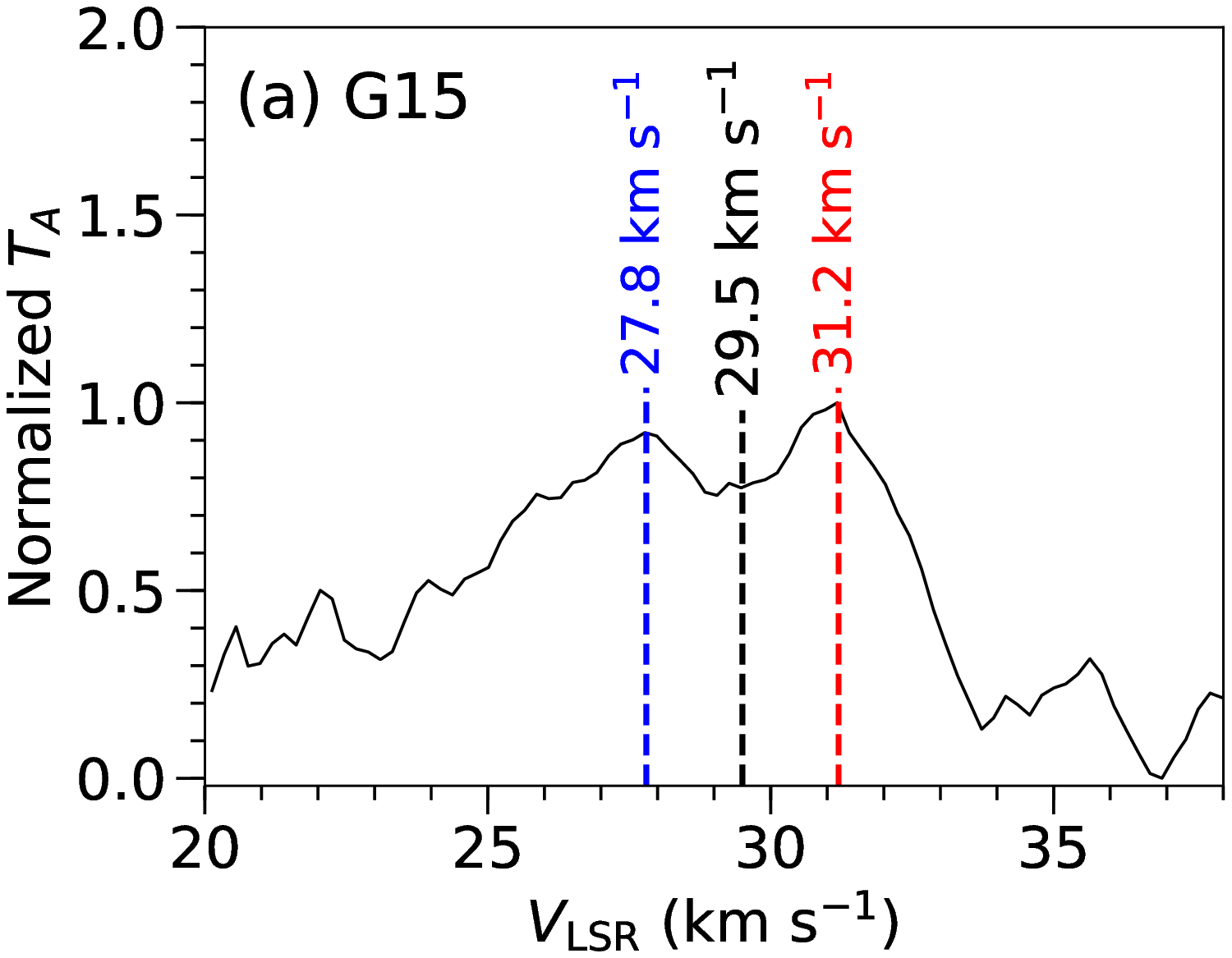}{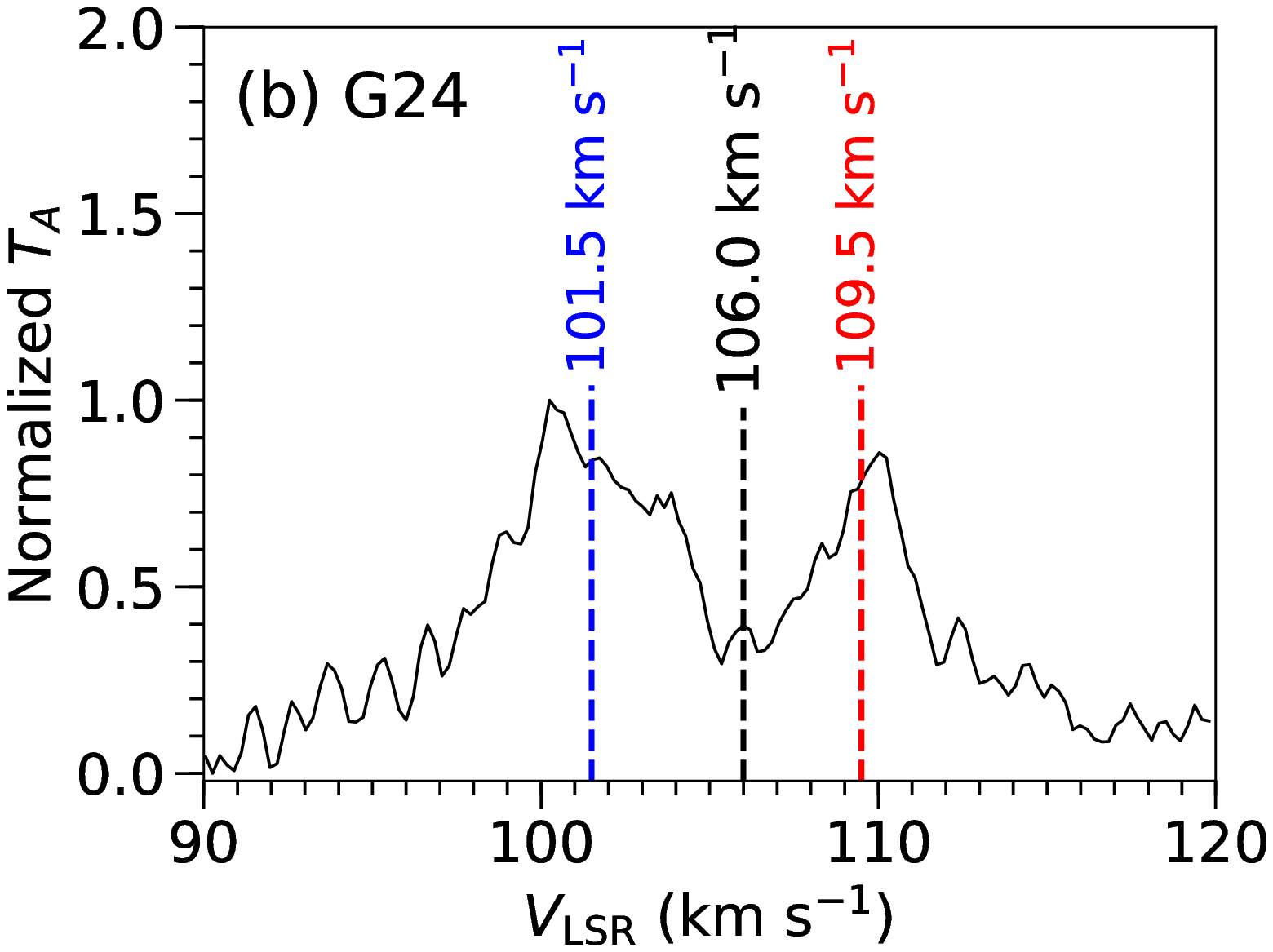}\\
\plottwo{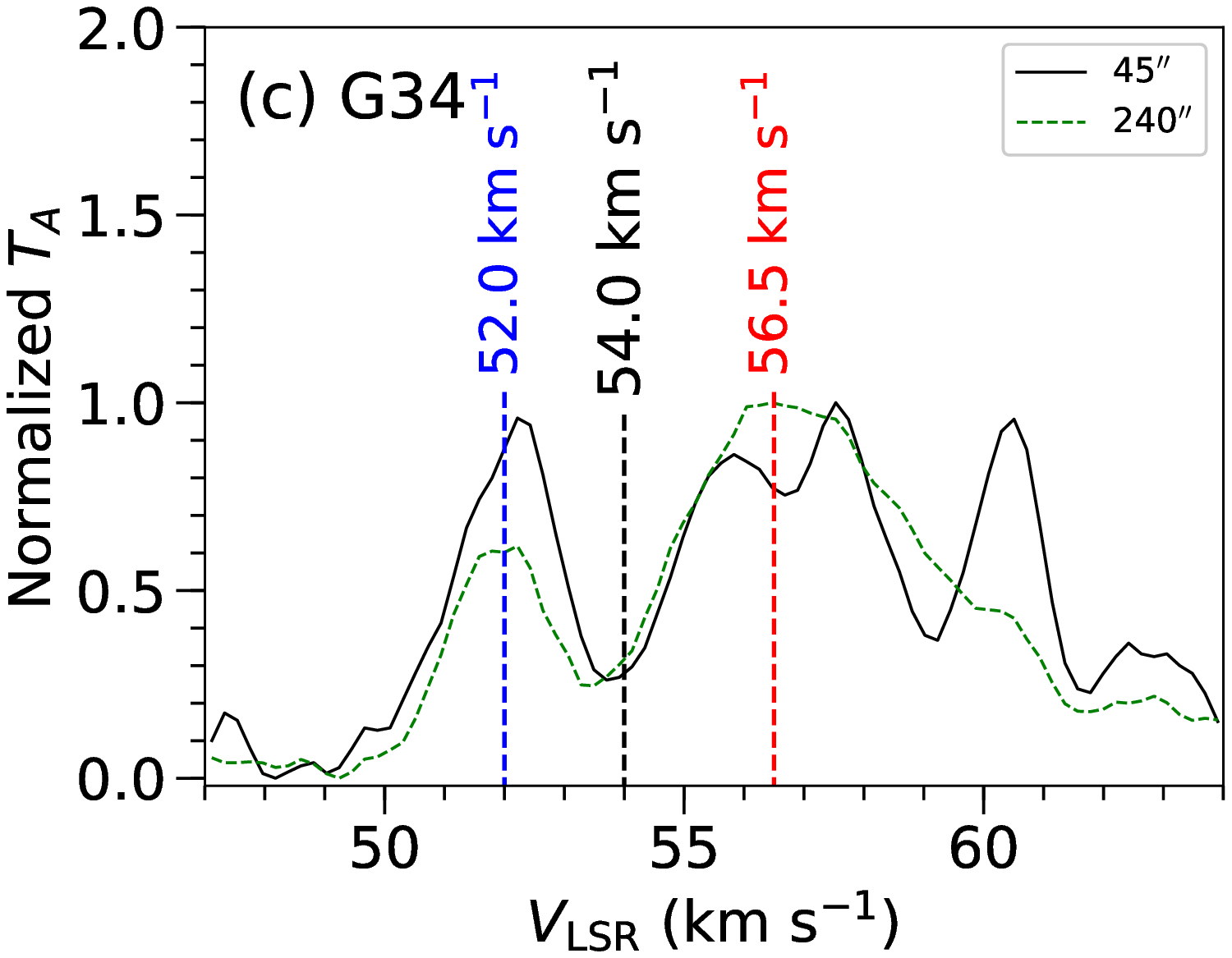}{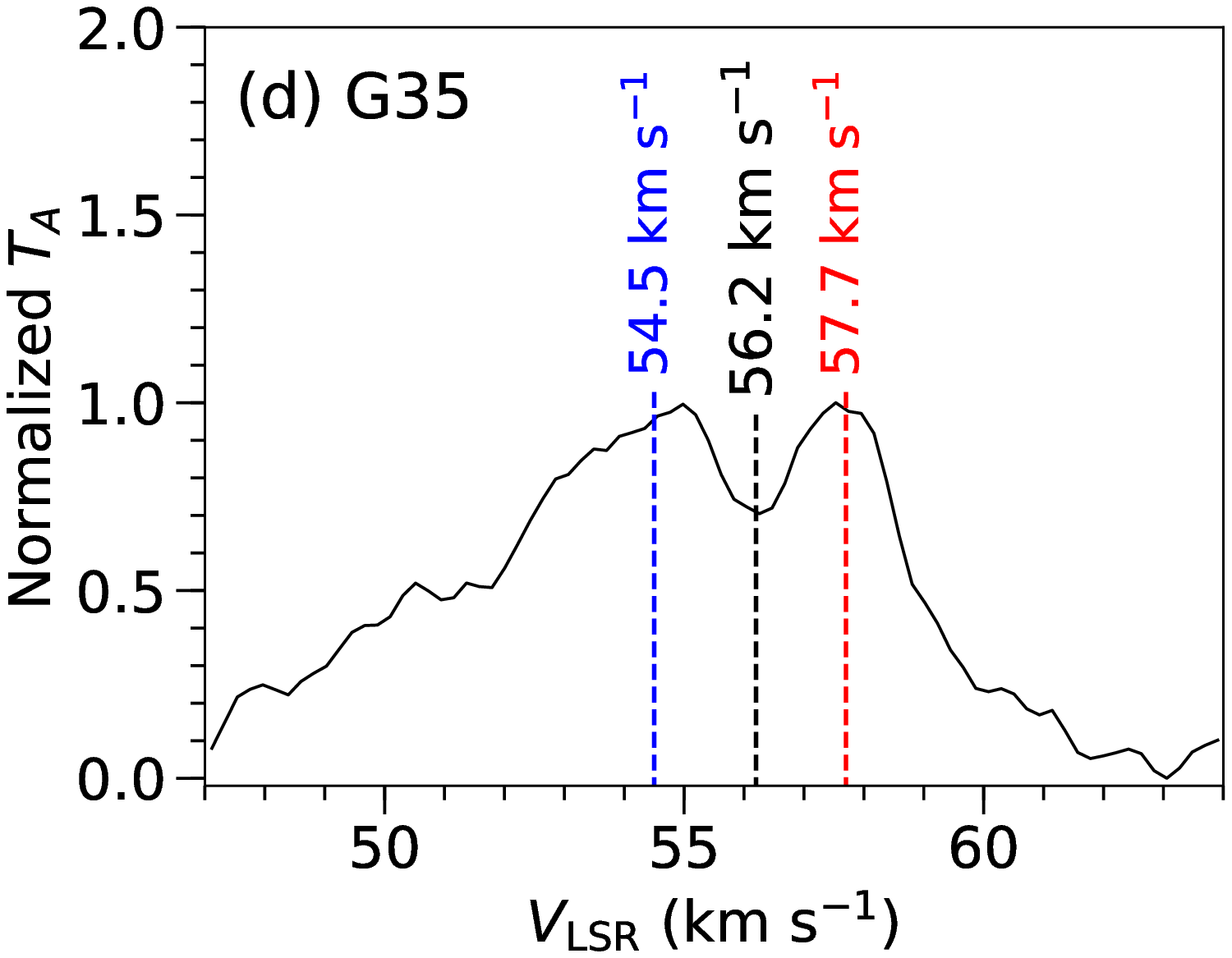}\\
\epsscale{0.5}
\plotone{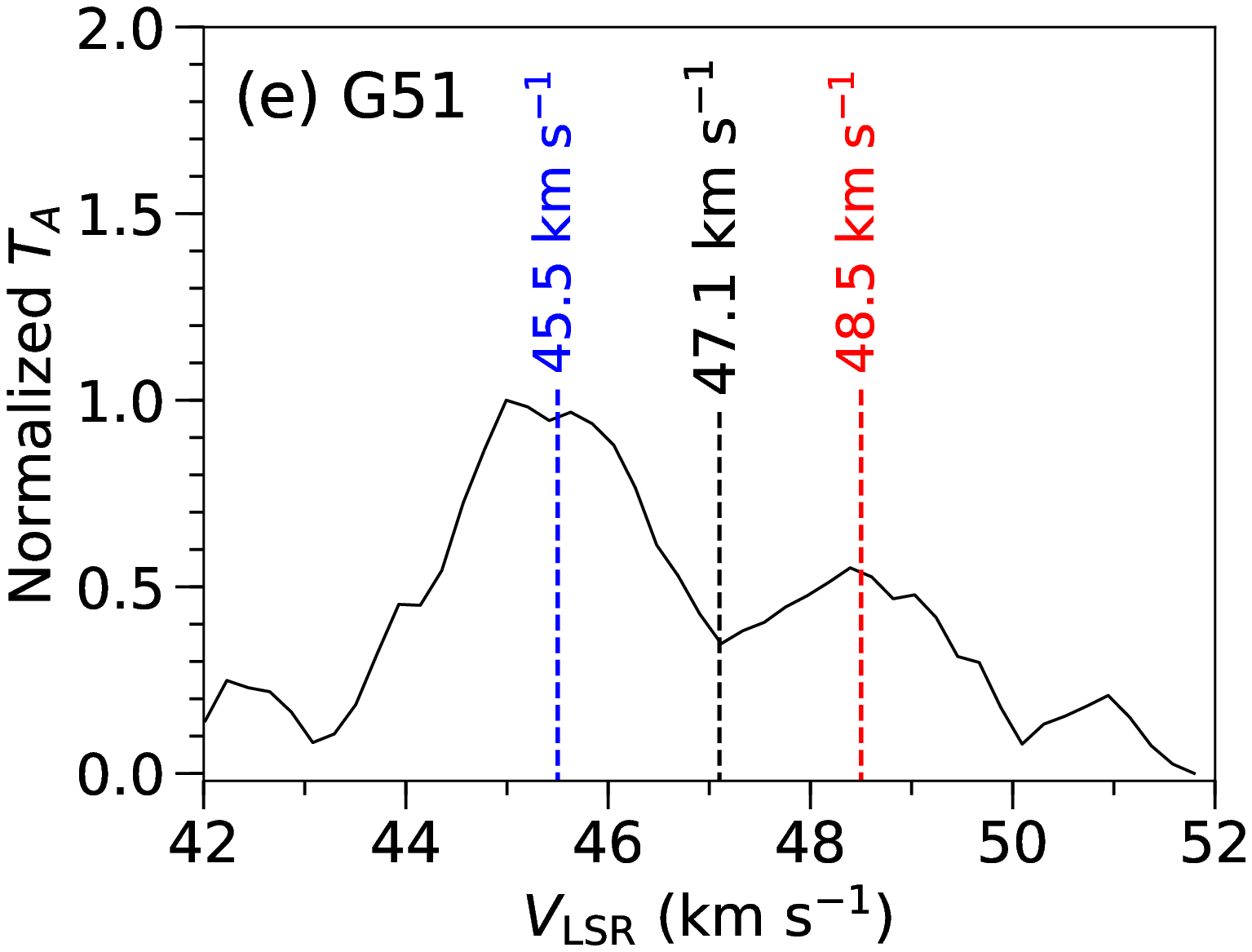}
\caption{\scriptsize $^{13}$CO spectra of our five selected regions (a) G15, (b) G24, (c) G34, (d) G35, and (e) G51. The spectra were
 constructed following the same procedure as that applied to the reference region, G27 (see the caption of Figure~\ref{fig6}).
 Velocities corresponding to the blue- and red-shifted emission peaks, and to the trough, are marked by blue and red, and black-dashed
 lines, respectively. The $^{13}$CO spectrum of the G34 region constructed for a 45$''$ diameter shows three peaks. The third peak at
 $\sim$61 km s$^{-1}$ dissolves if the spectrum is constructed for a 240$''$ diameter.}
\label{fig7}
\end{figure*}

The $^{13}$CO spectra for our selected regions are constructed by integrating all emission within a 45$''$ diameter, centered on the
 positions of the W--R stars (Figure~\ref{fig7}) for velocity ranges where dust clumps are generally identified (see Section~\ref{sec:regions}).
 All spectra show similar double-peaked spectral signatures as seen in the spectrum toward the reference region (see Figure~\ref{fig6}b)
 except for the G34 region. The spectrum for the G34 region constructed for a 45$''$ diameter shows three peaks (see Figure~\ref{fig7}c).
 However, the highest velocity peak (at $v_\mathrm{LSR}$ of $\sim$61 km s$^{-1}$) dissolves when the spectrum is produced for a 240$''$ diameter.
 This implies a small component of molecular gas along the sight line to the W--R star which is peaking at a $v_\mathrm{LSR}$ of 61 km s$^{-1}$.
 The emission peaks, troughs, and corresponding velocities are marked and labeled in Figure~\ref{fig7}.
 
The kinematic distances (i.e. $d_\mathrm{Near}$)
 of all but one trough (in the G51 region) agree well with the distances of the respective W--R stars, indicating that they are the host
 clouds of these W--R stars (see Table~\ref{table1} for the velocity ranges of the identified clouds). The scenario is different for G51
 as no dust clumps were identified near the W--R star in this region \citep{urquhart18}. Upon exploring the GRS $^{13}$CO spectrum for the
 entire observed velocity range, we identified a molecular cloud in a velocity range of 43--51 km s$^{-1}$ with the expected double-peaked
 spectral signature (Figure~\ref{fig7}e). For this region, the $d_\mathrm{Far}$ (6.8$\pm$0.7 kpc)
 corresponding to the velocity of the trough lies within the range of the spectro-photometric distance (5.9$\pm$0.5 kpc) of the W--R star.
 It is also possible for this region that the surrounding area of this W--R star is devoid of molecular gas. However, we consider
 that the W--R star toward the G51 region is possibly associated with this particular molecular cloud located in the $v_\mathrm{LSR}$
 range of 43--51 km s$^{-1}$.

Note that double peaked structures in the molecular spectra may arise for several reasons. Thus, we explored the molecular spectra
 of five lines of sight within the field for each region presented in Appendix~\ref{appendix}. Such structures are indeed present in the
 molecular spectra including in the spectrum toward the W--R star in G24, but at a different $v_\mathrm{LSR}$ range. However, the origin
 of such structures is better clarified based on a position--velocity ($p-v$) analysis. All field regions that show double-peaked spectral
 features typically have two cloud components along the line of sight. However, the \pv analysis revealed that those clouds (except for one)
 do not exhibit the signature that resembles the expansion of the molecular gas (unlike the signatures discussed in the next section). Only
 one field region (in the G24 field) shows a signature of expanding molecular gas, possibly driven by an intermediate mass YSO (for details
 see Appendix~\ref{appendix}).

\section{Kinematics of molecular gas and identification of young sources}
\label{sec:kin}
In the previous section, analysis of the molecular spectra helped us to identify the host clouds of all five W--R stars. 
 In the subsequent sections, we explore the dynamics of the host molecular clouds and also search for signatures of active
 star formation (i.e. the presence of cold clumps and YSOs) around these W--R stars.

\subsection{Dynamics of the molecular gas}
\label{sec:dynamics}
The \pv diagram of molecular line data (e.g., $^{13}$CO) is a powerful tool to probe the dynamics of the molecular gas. Expansion,
 inflows, and outflows in molecular clouds have different imprints in the \pv diagram \citep[see e.g.][]{arce11, dewangan16, baug18}. For
 example, ring-like, U-like, or inverted U-like structures in the \pv diagram are typically indicative of the presence of expanding
 molecular shells \citep[see][]{arce11,fontani12,feddersen18}. In the G27 region, \citet{dewangan16} reported the presence of an expanding
 molecular shell inferred from a similar feature in the \pv diagram. They also estimated that the molecular shell in the G27 has an expansion
 velocity ($v_\mathrm{exp}$) of 3 km s$^{-1}$.

\begin{figure*}
\epsscale{0.9}
\plottwo{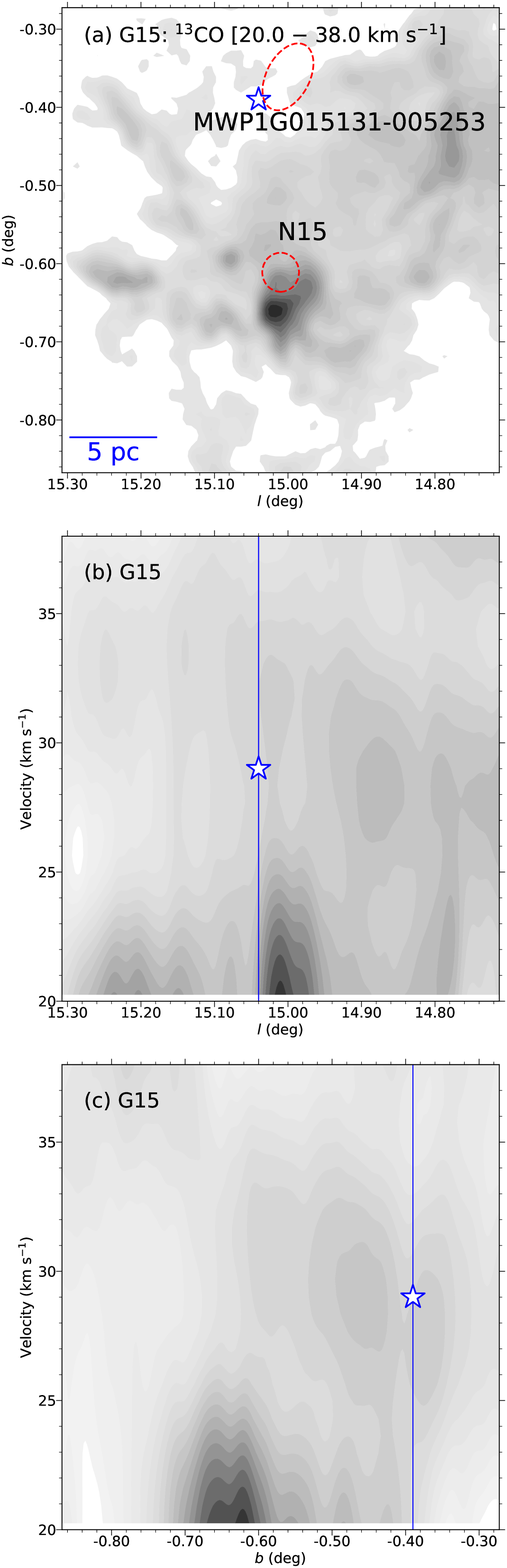}{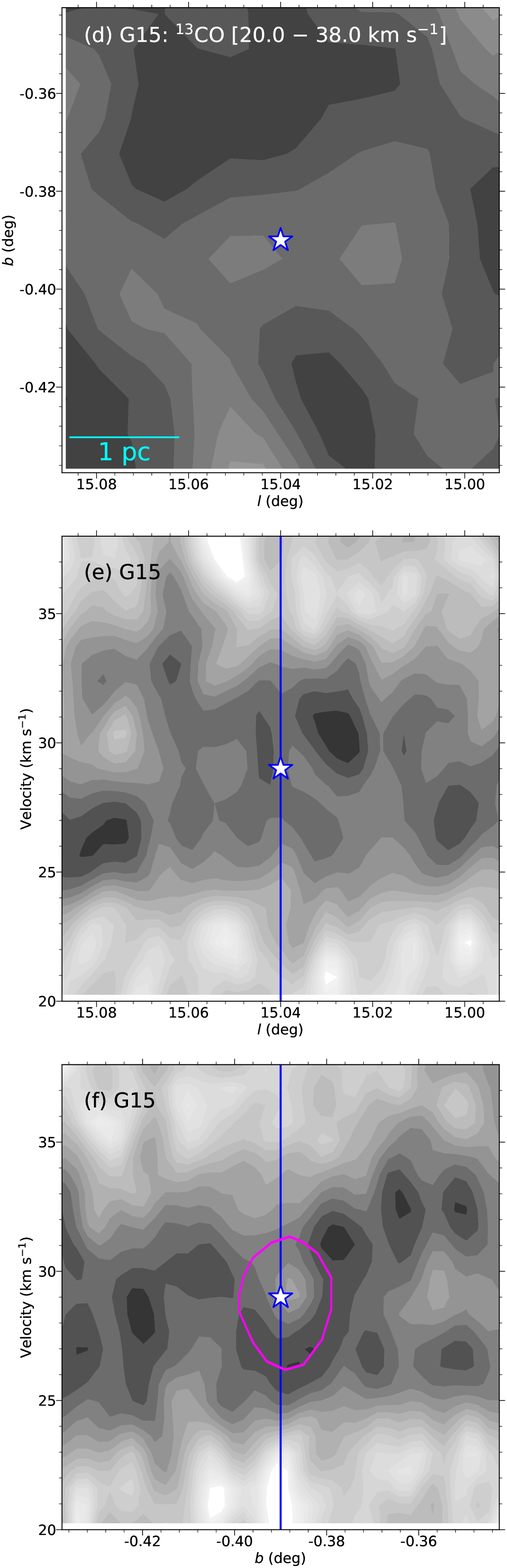}
\caption{\scriptsize (a) Integrated intensity map (solid gray contours) of the G15 region for the 36$'\times$36$'$ area around
 $l$=15$^\circ$.010, $b$=$-$0$^\circ$.570, for a velocity range of 20--38 km s$^{-1}$. (b) and (c) Corresponding \pv diagrams
 (solid gray contours) of the selected area around the G15 region for the same velocity range. No special feature is visible in the \pv
 diagrams. (d), (e), and (f) are, respectively, the integrated intensity map and \pv diagrams (solid gray contours) of the G15 region
 for the same velocity range of 20--38 km s$^{-1}$ but for a smaller area of 6$'\times$6$'$ area around $l$=15$^\circ$.040,
 $b$=$-$0$^\circ$.390. A ring-like structure can be easily discerned in panel (f). The position of the W--R star is marked by stars
 in the spatial maps and by lines in all \pv diagrams. The stars marked in the \pv diagrams indicate the probable positions of the
 W--R star if it is located in the trough marked in Figure~\ref{fig7}a.}
\label{fig8}
\end{figure*}

\begin{figure*}
\epsscale{0.9}
\plottwo{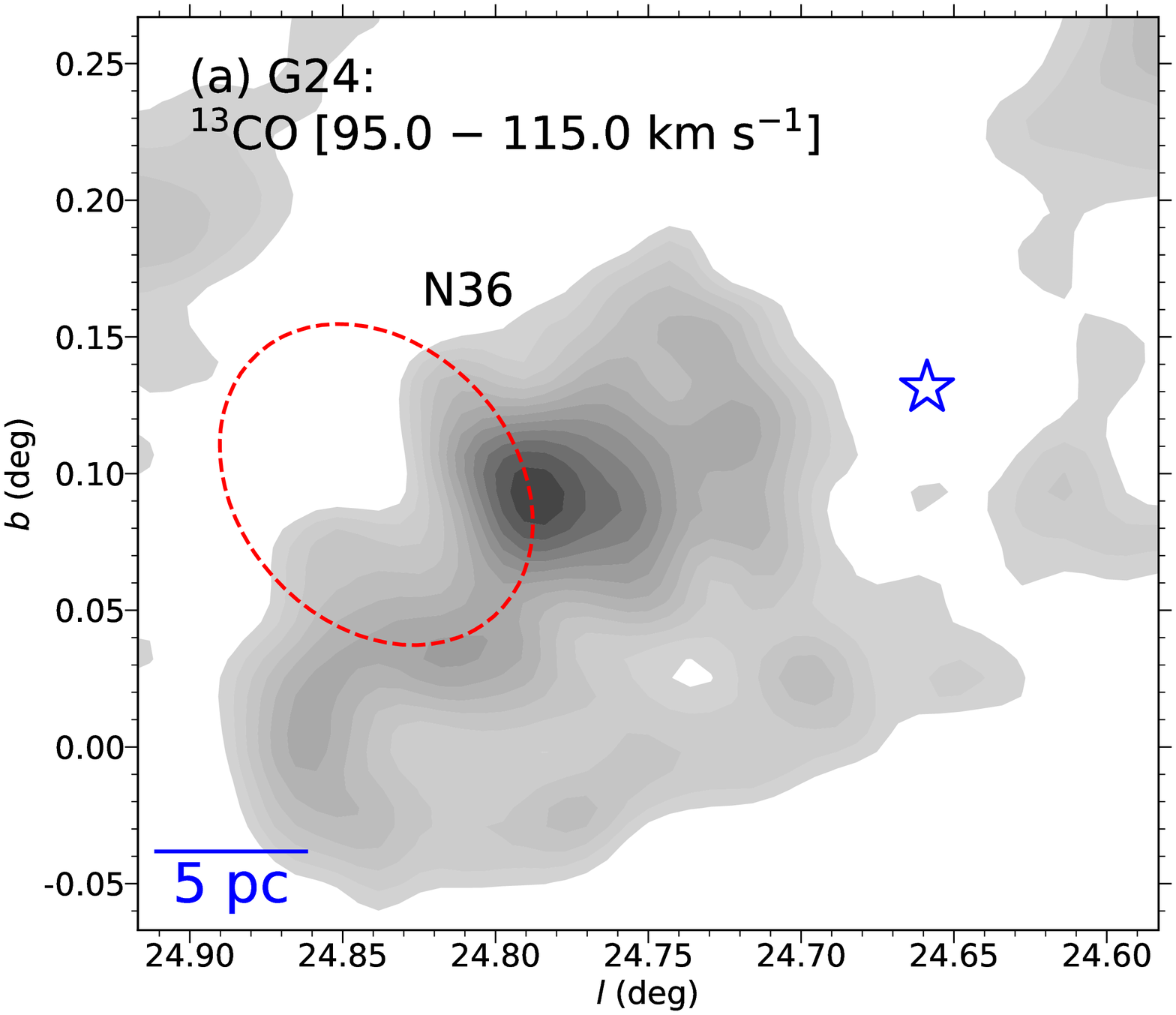}{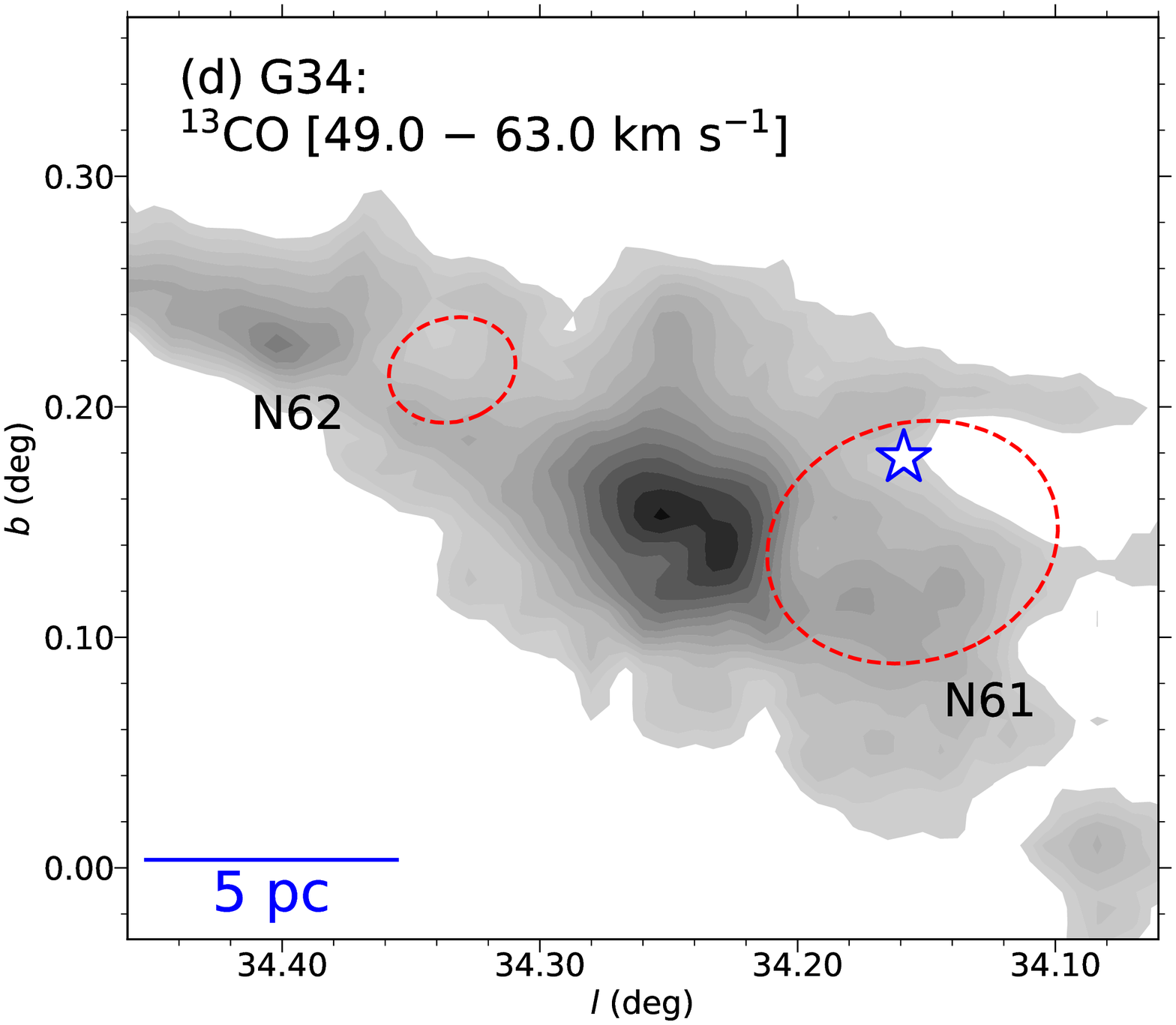}\\
\plottwo{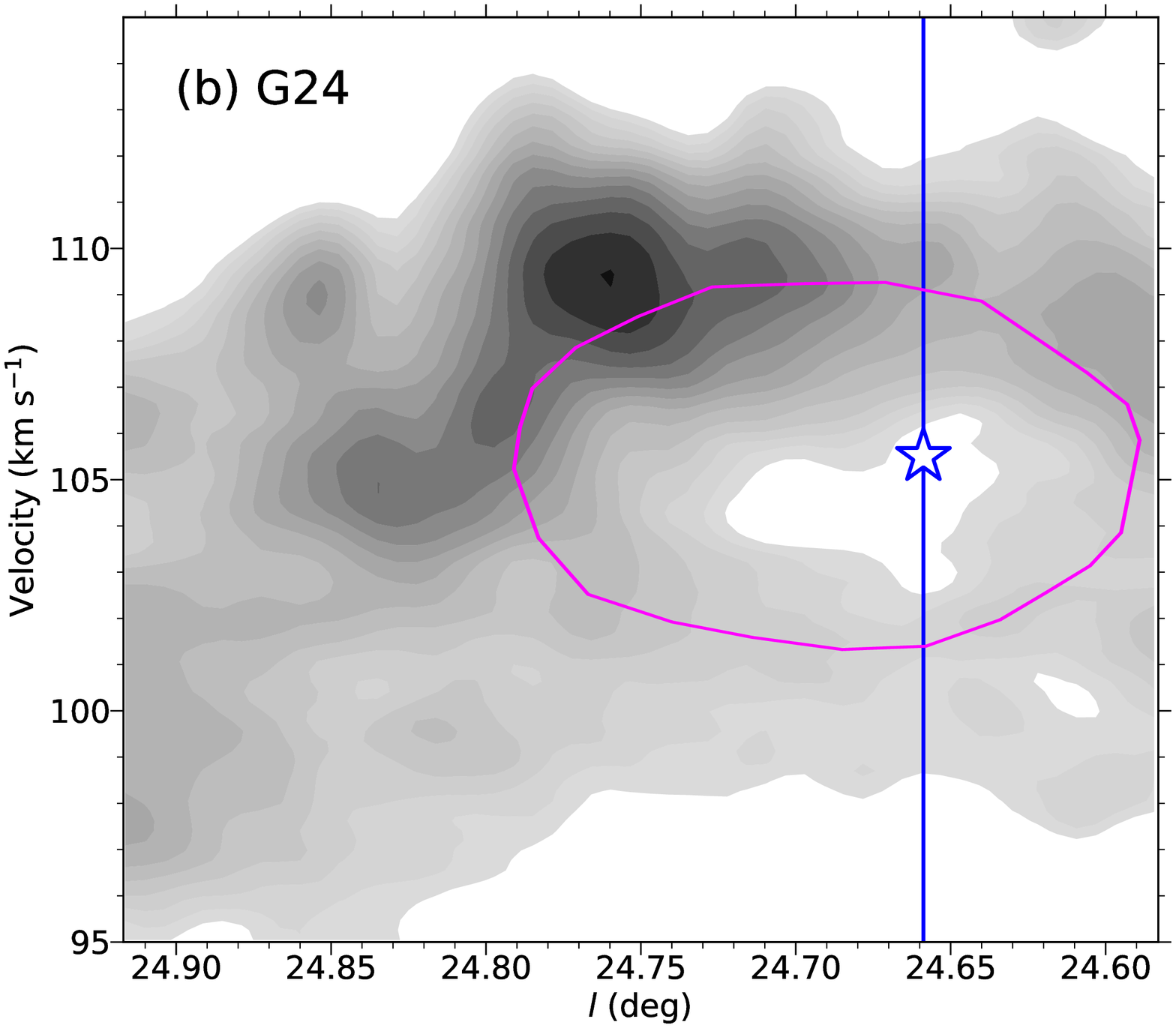}{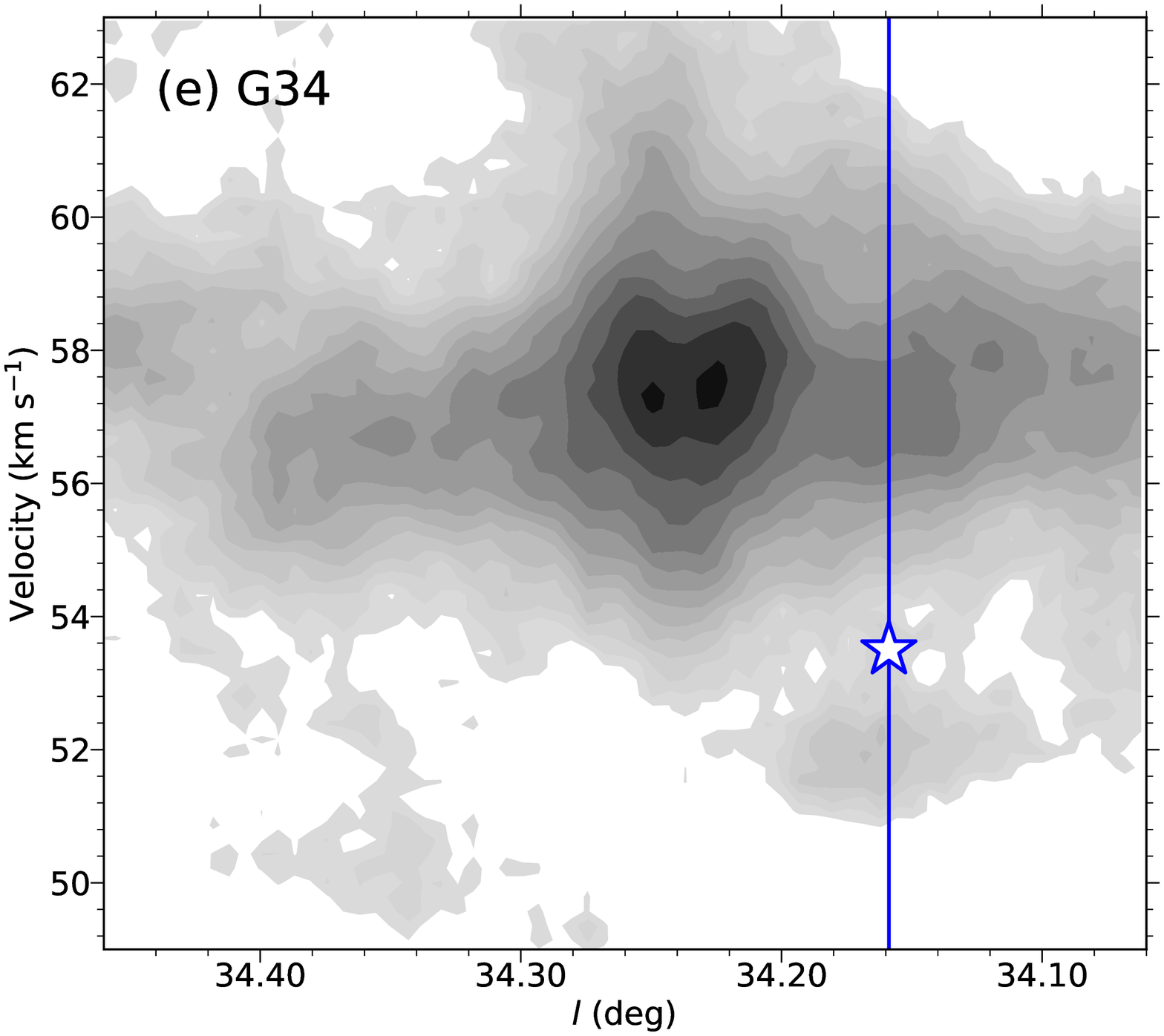}\\
\plottwo{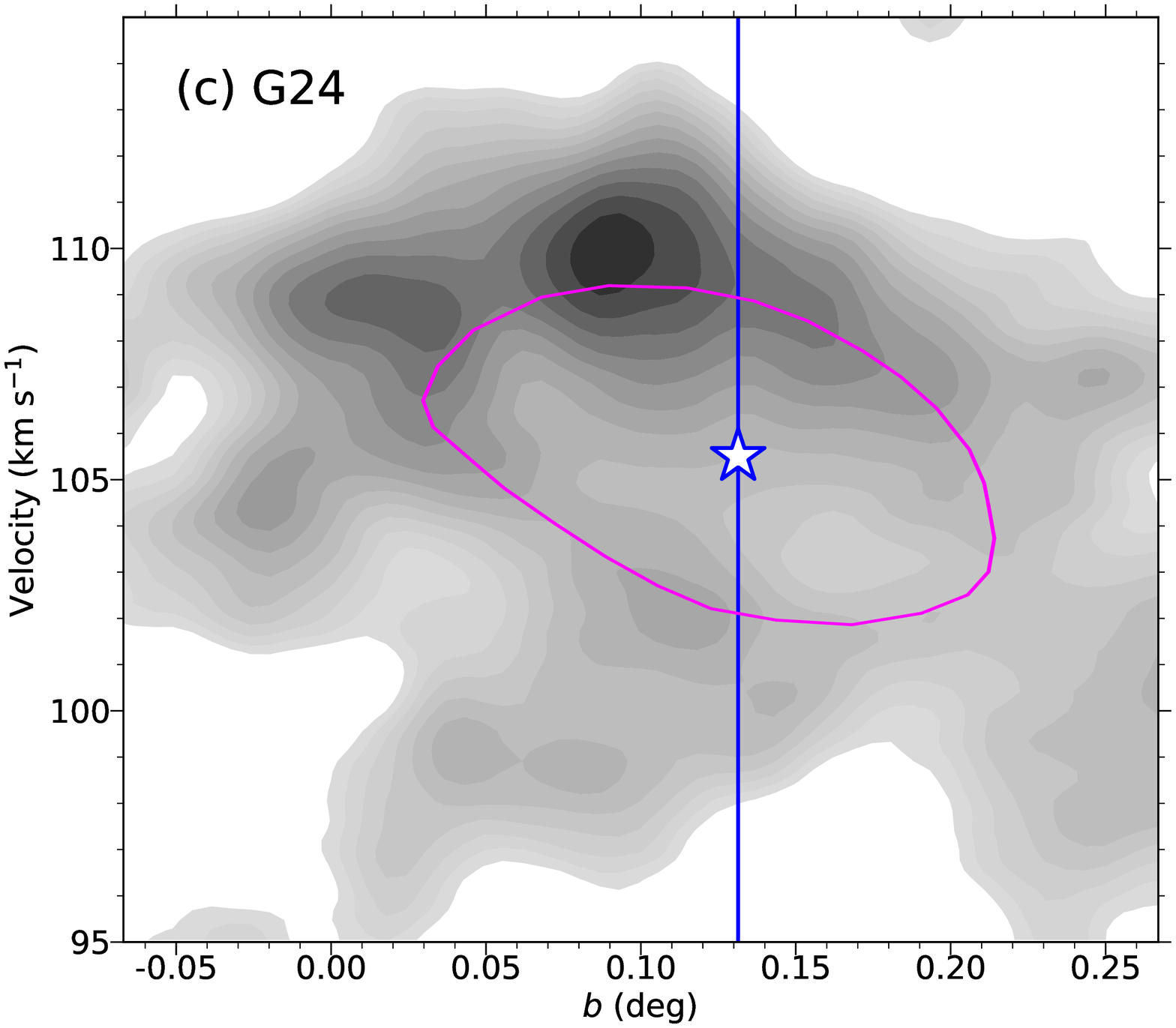}{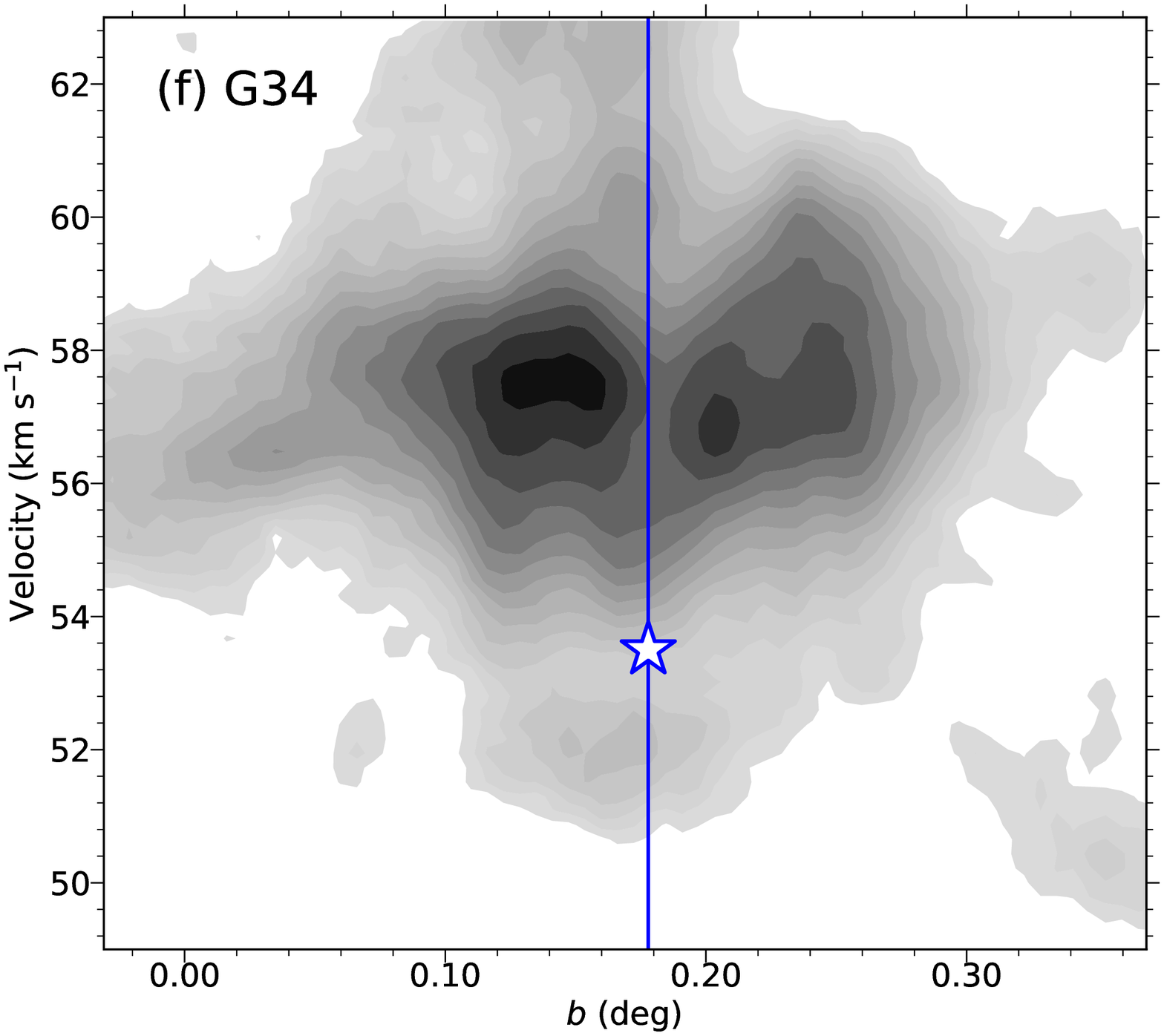}
\caption{\scriptsize (a) Integrated intensity map (solid gray contours) of the G24 region for a velocity range of 92--118 km s$^{-1}$.
 (b) and (c) \pv diagrams (solid gray contours) of the G24 region for the same velocity range. (d), (e), and (f) are, respectively, the
 integrated intensity map and \pv diagrams (solid gray contours) of the G34 region for a velocity range of 49--63 km s$^{-1}$. The
 positions of the W--R stars are marked by stars in the spatial maps and by lines in the \pv diagrams of both regions. The stars
 marked in \pv diagrams indicate the probable positions of the W--R stars if they are located in the troughs marked in
 Figures~\ref{fig7}b and c.}
\label{fig9}
\end{figure*}

\begin{figure*}
\epsscale{0.9}
\plottwo{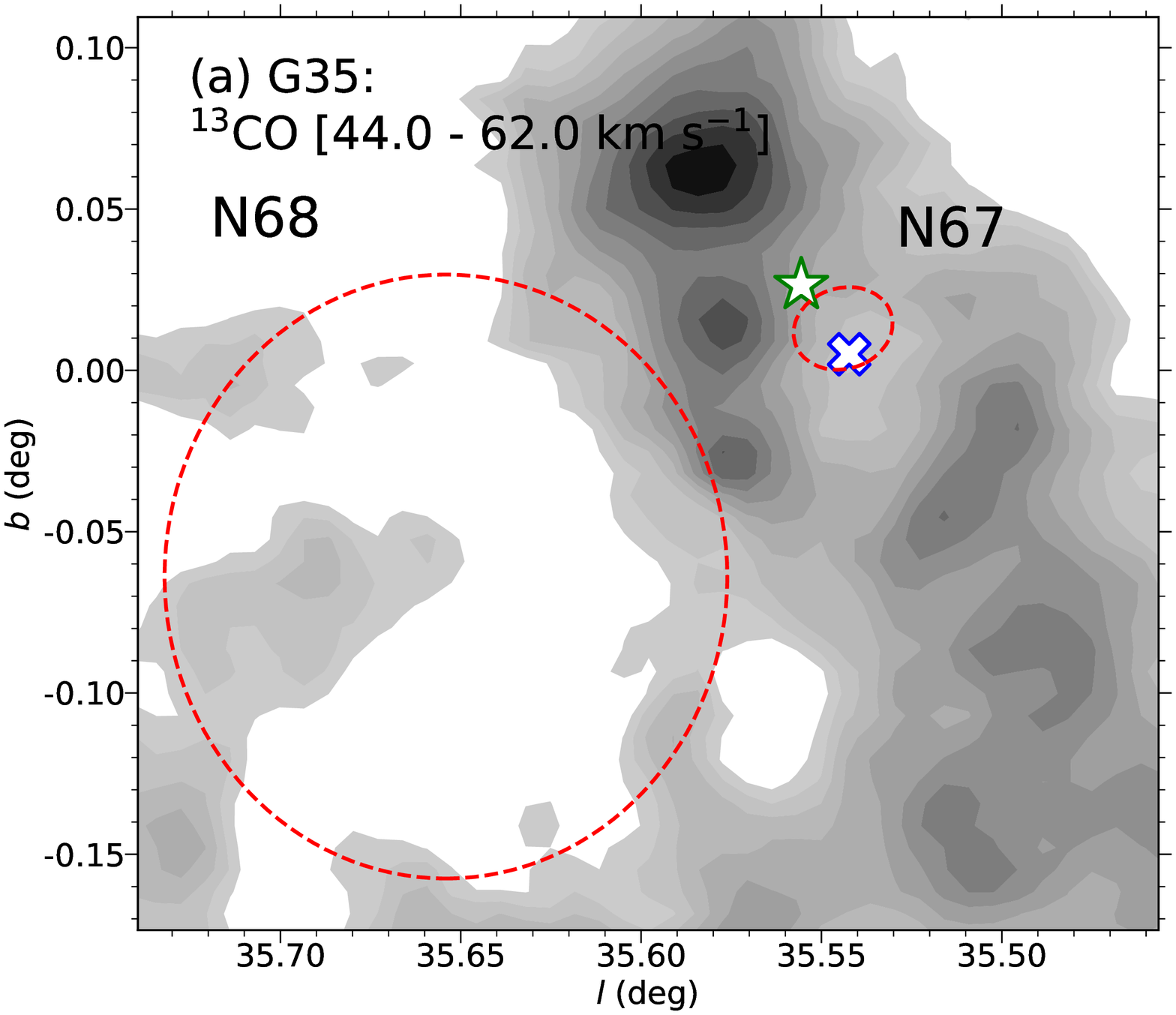}{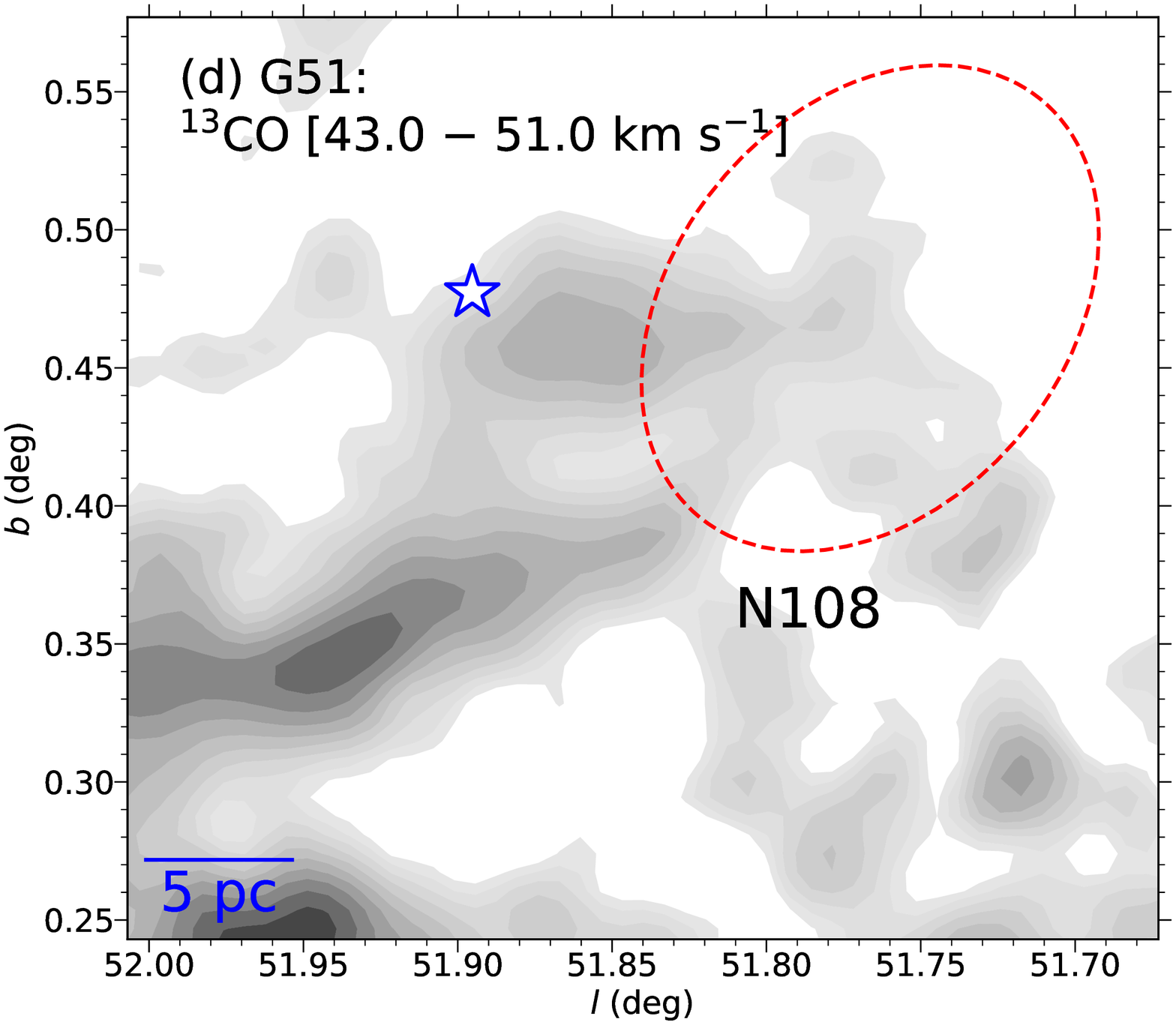}\\
\plottwo{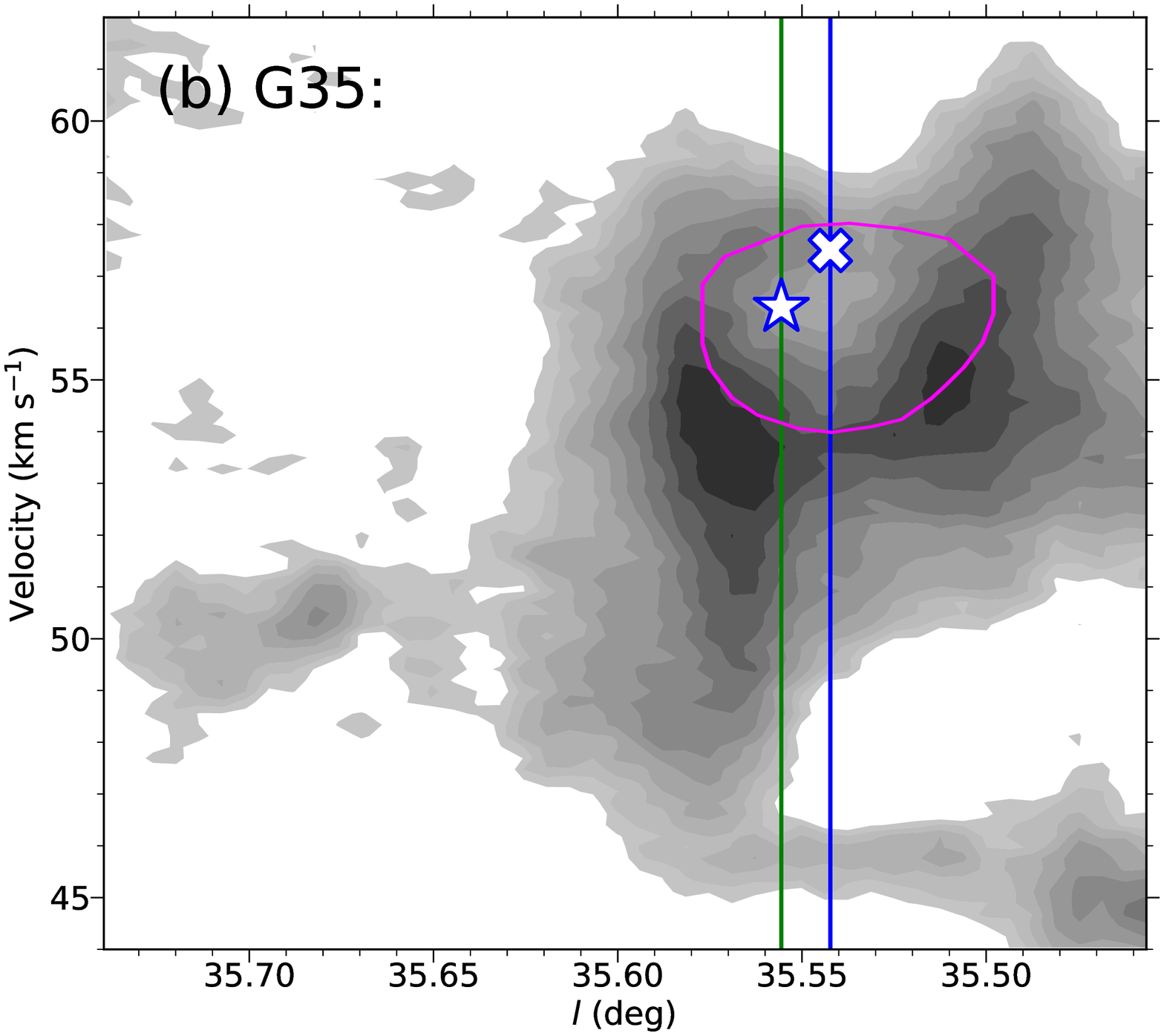}{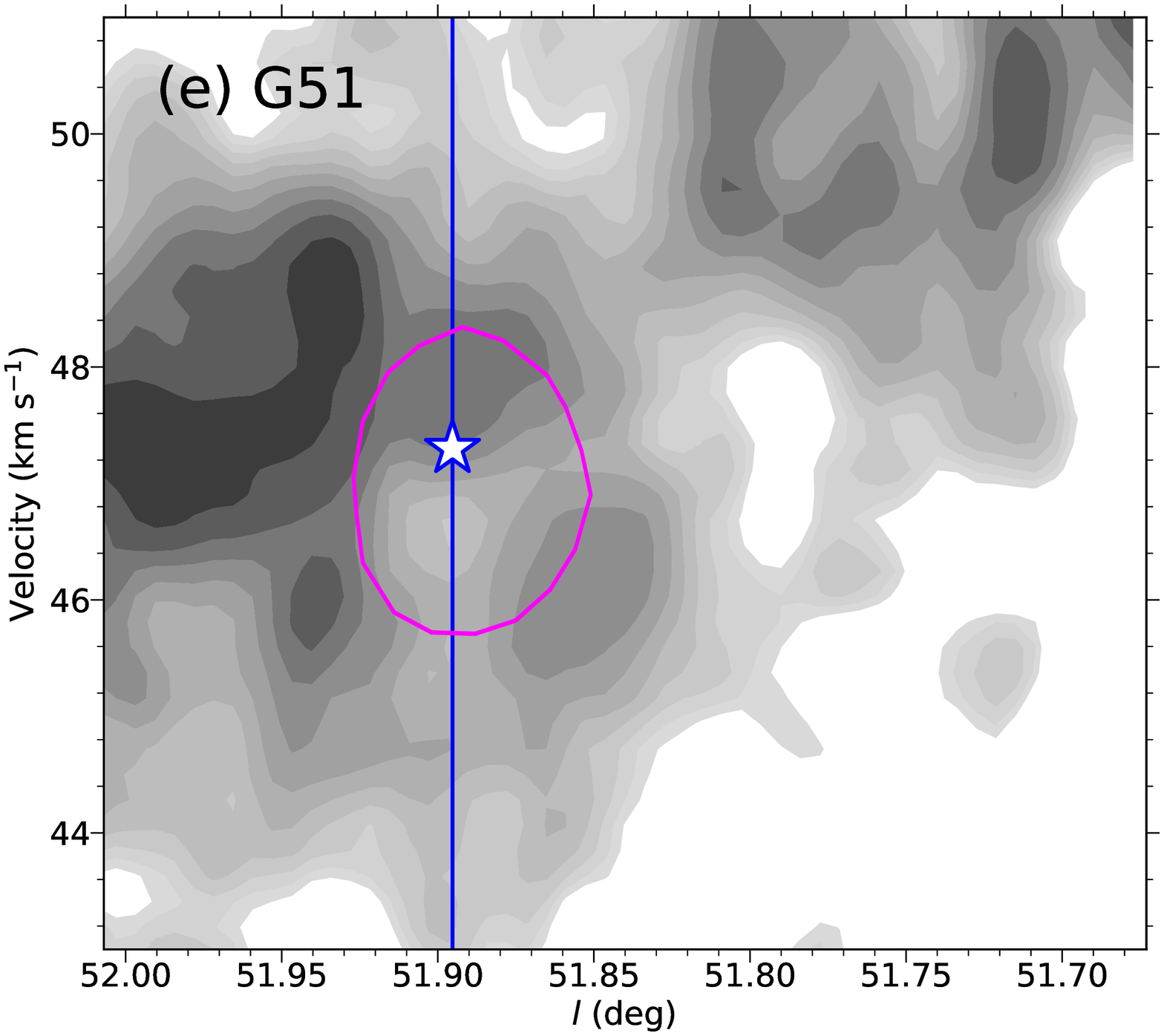}\\
\plottwo{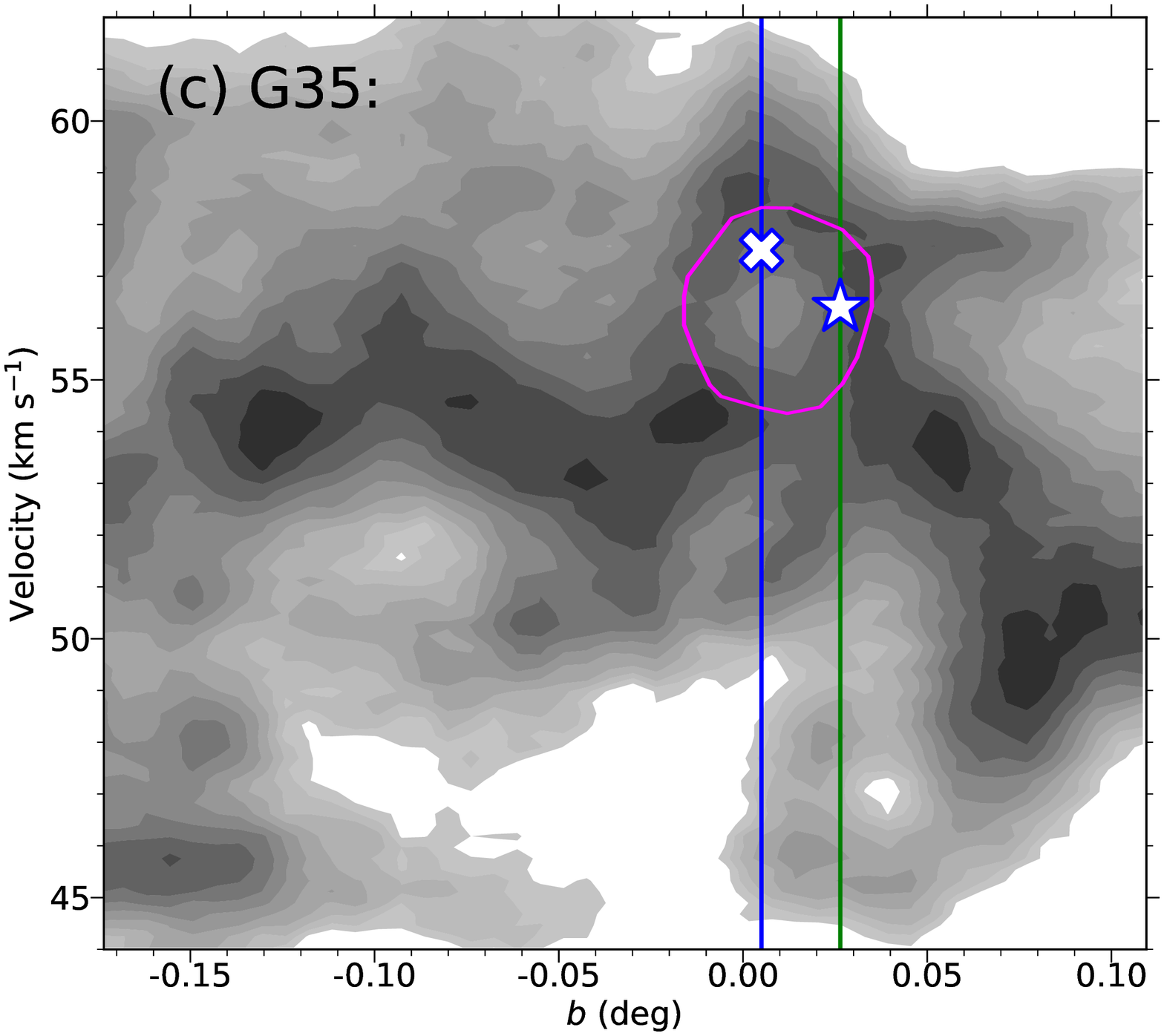}{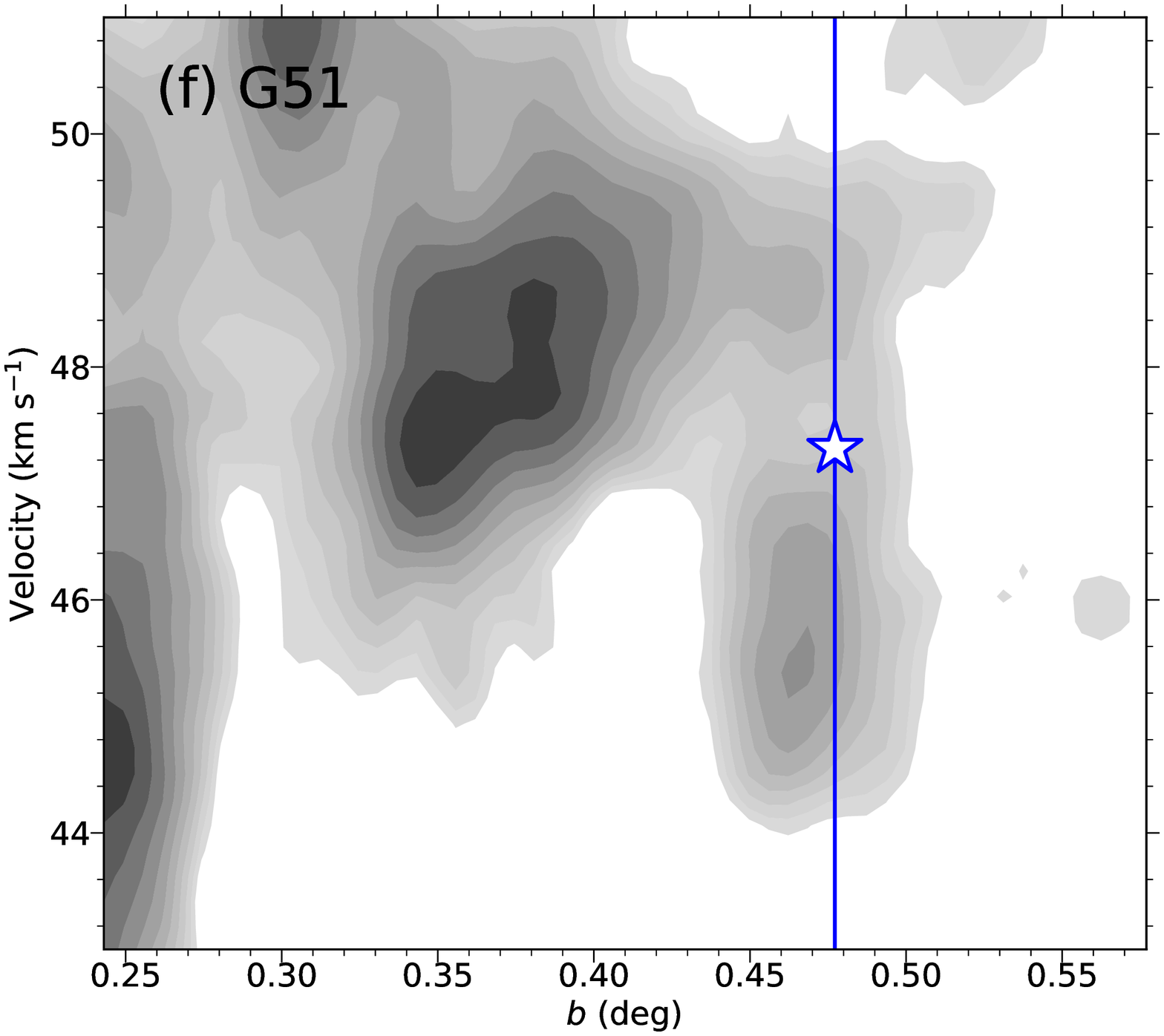}
\caption{\scriptsize (a), (b), and (c) Integrated intensity map and \pv diagrams (solid gray contours) of the G35 region for a velocity
 range of 44--62 km s$^{-1}$. (d), (e), and (f) Integrated intensity map and \pv diagrams (solid gray contours) of the G51 region for a
 velocity range of 42--52 km s$^{-1}$. The positions of the W--R stars are indicated by stars in the spatial maps and by lines in the
 \pv diagrams for both regions. The stars marked in the \pv diagrams indicate the probable positions of the W--R stars if they are
 located in the troughs marked in Figures~\ref{fig7}d and e. Toward G35, an \hii region located near the W--R star which could also be
 the driving source of the expanding shell. Possible location of the \hii region is also marked by a cross in the integrated map as
 well as in the \pv diagrams.}
\label{fig10}
\end{figure*}

The integrated intensity maps and the corresponding \pv diagrams of all five regions are shown in Figures~\ref{fig8}--\ref{fig10}. Ring-like
 structures are seen in the \pv diagrams of all but one region (see the magenta lines in Figures~\ref{fig8}--\ref{fig10}). The coordinates
 of the W--R stars (i.e. $l$ or $b$) and the velocities of the corresponding troughs are also marked in all \pv diagrams. No significant
 feature is seen in the G15 region when the \pv diagram is constructed for the full selected area (see Figure~\ref{fig8}b,c). But, a clear
 ring-like structure can be discerned when it is constructed for a 6$'\times$6$'$ area centered on the W--R star.
 This could be because the \pv analysis for the larger area includes strong $^{13}$CO emission from M17, which may drown out
 weak (e.g., ring-like) features. These ring-like structures in the \pv diagrams are typical signatures of the presence of expanding molecular
 shells, and the presence of a W--R star toward the center of this ring-like structure indicates that the expanding shell has possibly
 developed from the influence of the W--R star.

No ring-like or U-like structures are seen in the \pv diagram of the G34 region despite the physical association of the W--R star with the
 corresponding molecular cloud. A possible reason could be that the red-shifted part of the cloud is larger and more intense than its blue-shifted
 counterpart (see Figure~\ref{fig7}c). Thus, the red-shift cloud has sufficient intensity to surpass the faint features in the \pv diagram.
 Note that the ring-like structure in the \pv diagram of the G24 region is deformed. Also, in the G35 region, the W--R star is not exactly
 at the center of the ring-like structure. As can be seen in the integrated $^{13}$CO map of the G24 region (see Figure~\ref{fig9}a), the molecular
 gas is mainly concentrated near the bubble N36. Such an asymmetrical distribution of molecular gas may lead to a deformed ring-like structure in
 the \pv diagrams. The displacement of the W--R star with respect to the ring-like structures may arise either from differential expansion of the
 blue- and red-shifted parts of the molecular cloud because of their density contrast, or if the W--R star has a differential motion with respect to the
 surrounding clouds. For example, the cloud toward the N36 bubble with respect to the W--R star is about 3--4 times more intense compared to the cloud in the
 other side of the W--R star. Such contrast in the molecular cloud density cloud deform the ring-like features in the \pv diagram. 
 
For G35, an \hii region was spatially found toward the ring-like structure. However, $v_\mathrm{exp}$ of this particular ionized gas
 is 57.5$\pm$0.1 km s$^{-1}$ \citep{anderson11}, and hence, does not appear at the center of the ring-like structure in the \pv diagram
 (see Figure~\ref{fig10}a,b,c). It is, thus, possible that both W--R star and ionized gas played important role in expanding the surrounding
 gas. However, the dynamical age of this \hii region is ($\sim$0.5 Myr; calculated in Section~\ref{sec:SF}) much lower compared to the age of
 the W--R star ($\sim$5 Myr). Thus, we consider that the W--R star is the primary driving source of the expanding gas. It is also possible
 that the W--R star in the G35 region has a differential plane of the sky motion with respect to the surrounding molecular cloud which has
 made it to appear off-centered. We found that in a life-time of a W--R star ($\sim$5 Myr) a differential motion of 10 $\mu as$ yr$^{-1}$ may
 lead to the off-set seen in the \pv diagram of the G35 region. Understanding the origin of this scenario requires a detailed analysis of the
 proper motion of the W--R star as well as of the surrounding stars that are formed in the same host cloud.
 The expansion velocities of the molecular shells, inferred from the difference between the central velocity and the velocity of the outer
 edge of the ring-like structures, are also estimated for all regions (see Table~\ref{table1}). The shells identified here are expanding
 with velocities of 2--5 km s$^{-1}$.
\subsection{Identification of dust clumps}
\label{sec:herschel}
Multiband {\sl Herschel} images (160, 250, 350, and 500 $\mu$m) are useful to construct column density maps. The detailed procedure to 
 construct a column density map is not presented in this paper but can be found elsewhere \citep{mallick15, baug18}. Figures corresponding to
 the column density maps are not shown here. For clump identification, we employed the python-based {\sc astrodendro}
 package\footnote{https://dendrograms.readthedocs.io/en/stable/index.html} \citep{rosolowsky08}, which uses the dendrogram technique to identify
 hierarchical structures (or clumps). We considered only those clumps as real clumps that have effective areas of more than 9 pixels of column
 density maps and peak flux levels in excess of 5$\sigma$ compared to the surrounding background flux (where $\sigma$ is rms determined from
 dark patches in the maps). The traditional {\sc clumpfind} method \citep{williams94} was also applied to the same set of column density maps,
 and we found that the number of detected clumps and their central coordinates are generally similar. Henceforth, we will only consider those
 clumps that were identified using the dendrogram method. Statistics of the identified clumps in all five regions are presented in Table~\ref{table2}.

The mass of each clump is estimated using \citep[see also][]{mallick15}:
\begin{equation}
	M_\mathrm{clump} = \mu_\mathrm{H_2} m_\mathrm{H} Area_\mathrm{pix} \Sigma N(\mathrm{H_2}),
\end{equation}
where $Area_\mathrm{pix}$ is the area subtended by a single pixel. Identified clumps have a wide range of masses (see Table~\ref{table2}).
 The dust clumps identified toward the G24 region are generally one order of magnitude more massive compared with the other four
 regions. Note that this particular region is located at a greater distance and, hence, multiple small clumps may appear as a single clump
 because of resolution limitations. Also, multiple clouds are present toward the G24 region (as seen in the $^{13}$CO spectrum for
 the full GRS velocity range). Thus, flux from multiple clouds may increase the flux levels in the {\sl Herschel} maps and, hence, the
 estimated column density.
\subsection{Young stellar sources}
\label{sec:yso}
Young star-forming regions are typically associated with large numbers of YSOs. YSOs in all selected regions are identified using the MIR
 color--magnitude and color--color schemes. The {\sl Spitzer}-IRAC and MIPS point sources are employed to identify and classify
 YSOs using [3.6]$-$[24]/[3.6] color--magnitude, and [5.8]$-$[8.0]/[3.6]$-$[4.5] and [3.6]$-$[4.5]/[4.5]$-$[5.8] color--color schemes
 (the corresponding figures are not shown). The detailed procedure of YSO identification schemes can be found in \citet{baug16}. 
 The methods used may suffer from significant contamination in distant regions, since multiple sources may appear as a
 single source because of the limited spatial resolution of the {\sl Spitzer} images.
 In addition, AGB stars and background galaxies often appear with excess flux in the MIR bands, and thus, may mimic as YSOs.
 However, here our main interest is to statistically identify areas with active star formation rather than characterize the
 properties of individual YSOs. Thus, even for distant regions, these schemes may help us to identify areas with ongoing star formation.
 The statistics of the identified YSOs for all regions are listed in Table~\ref{table2}.

\begin{deluxetable*}{ccccccccc}
\tablewidth{0pt}
\tabletypesize{\scriptsize} 
\tablecaption{Details of the identified cold clumps, YSOs, and {\sl Gaia} sources in all five regions.\label{table2}}
\tablehead{
\colhead{Region} & Number of & \multicolumn{3}{c}{Clump Mass ($M_{\odot}$)}       & \multicolumn{3}{c}{YSOs} \\ 
		 &   Clumps  & Minimum  & Maximum & Median & Class I & Flat Spec. & Class II                 }
\startdata
G15              &    70     &  107     &  3207   &  323   &   95    &   20       &   198     \\
G24              &    27     &  701     & 16241   & 1838   &   92    &   11       &   168     \\
G34              &    29     &   42     &  3642   &  390   &  136    &   17       &   152     \\
G35              &    20     &  105     &  4095   &  445   &   57    &    6       &    55     \\
G51              &    19     &  168     &  3955   &  337   &   27    &   10       &    33     
\enddata
\end{deluxetable*}

\subsubsection{Surface density analysis of YSOs}
\label{sec:surface_density}
\begin{figure} 
\epsscale{1.2}
\plotone{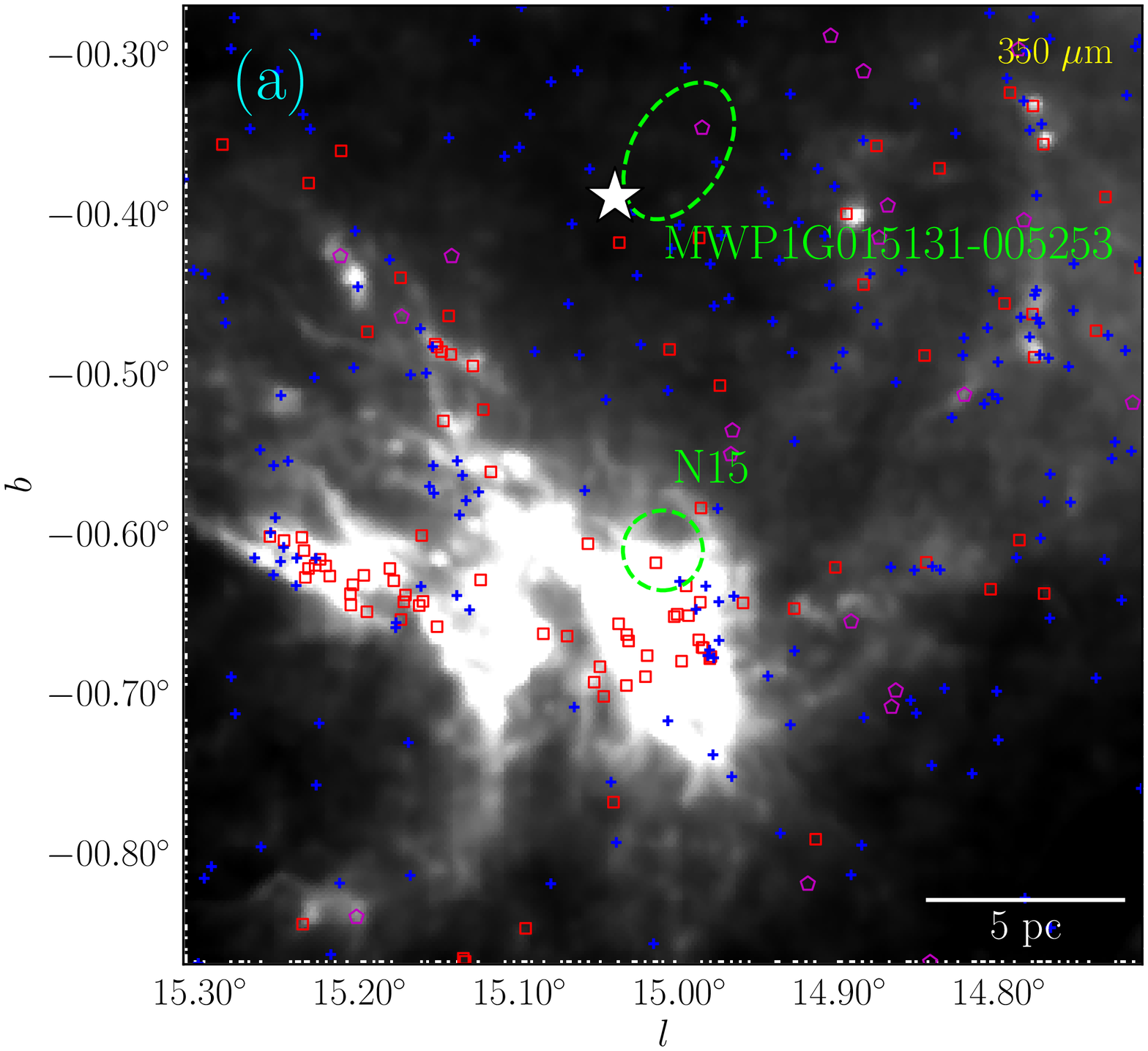}\\
\plotone{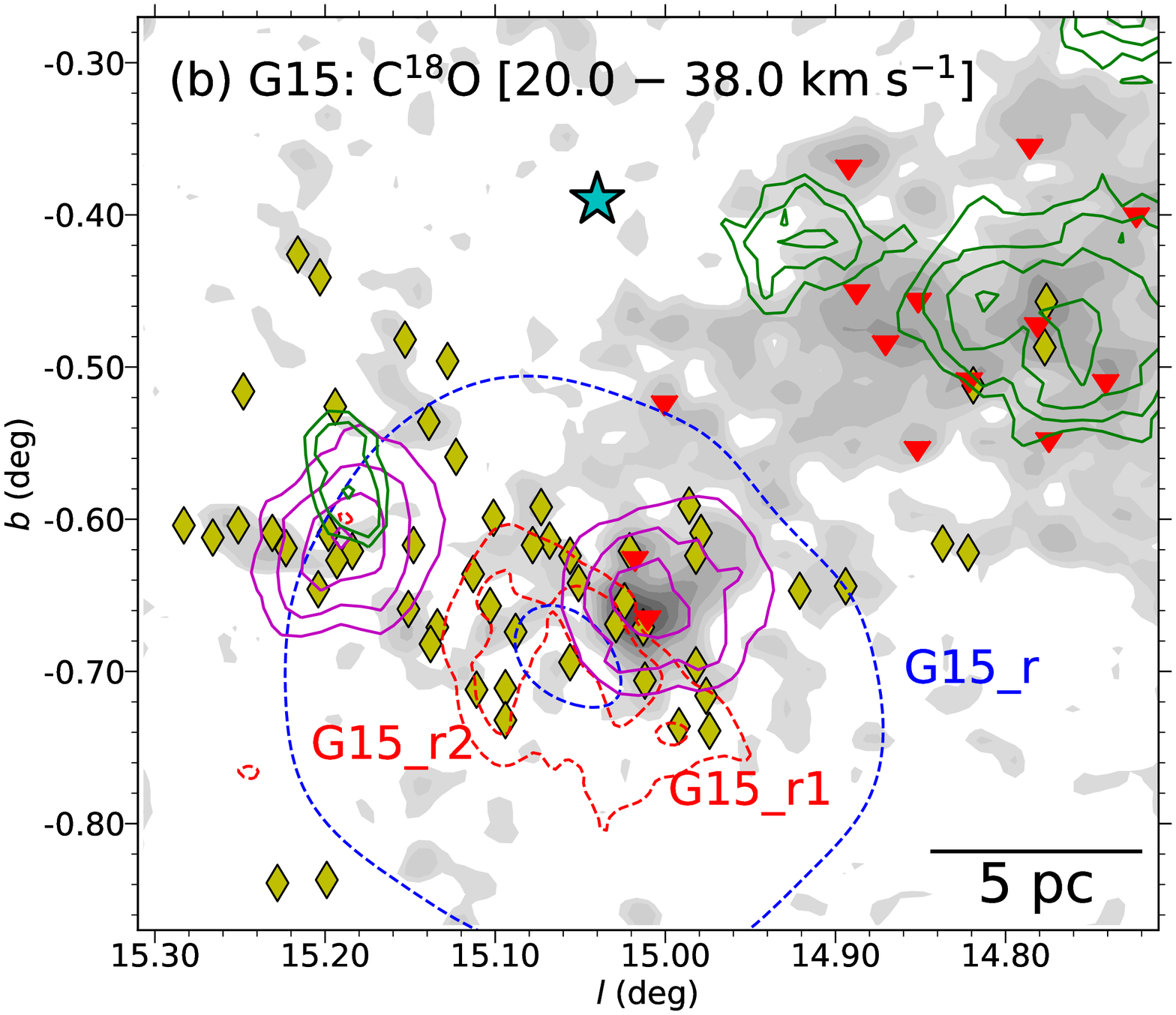}
\caption{\scriptsize (a) Class I (red squares), flat-spectrum (magenta pentagons), and Class II (blue crosses) YSOs overplotted on
 the {\sl Herschel} 350 $\mu$m image of the G15 region. The remaining symbols are similar to those in Figure~\ref{fig1}. (b) 20 nearest-neighbor
 surface density contours of Class I (and flat-spectrum) and Class II YSOs in magenta and green, respectively, are overlaid on the
 velocity-integrated C$^{18}$O map (solid gray contours). The contours are drawn at 0.45, 0.6, 0.9, and 1.4 YSO pc$^{-2}$ for Class I
 and 0.45, 0.5, 0.6, 0.8 YSO pc$^{-2}$ for Class II sources. Contour at 100 and 3,000 mJy beam$^{-1}$ of the VGPS 1.4 GHz emission
 are also shown in dashed blue to indicate the extent of the ionized gas. The NVSS 1.4 GHz contour at 300 and 3,000 mJy beam$^{-1}$ are also
 shown in dashed red which resolves two peaks, and are also labeled. The corresponding ionized regions are labeled. Positions of
 the cold condensations and dust clumps \citep{urquhart18} are shown as red triangles and yellow diamonds, respectively. The position of
 the W--R star is marked by a cyan star.}
\label{fig12}
\end{figure}

The distribution of YSO clusters in any star-forming region can help us characterize areas of active star formation. To examine
 the clustering behavior of YSOs toward our selected star-forming regions, we performed a nearest-neighbor surface density analysis of the
 identified YSOs, assuming that they are all located at a single distance. In surface density analysis, any value can be adopted
 for nearest-neighbor to examine clustering behavior. Generally, a large value of nearest-neighbor is sensitive to the large-scale
 distribution of YSOs, while a small value helps to identify small-scale density variations. This method can be thought of as a smoothing
 process to identify areas with large numbers of YSOs. We performed the nearest-neighbor surface density analysis for 20 YSOs
 as \citet{schmeja08} reported that 20 nearest-neighbor analysis can be used to identify clusters of 10--1500 YSOs.
 Figure~\ref{fig12}a shows the locations of the YSOs in the  G15 region overplotted on the {\sl Herschel} 350 $\mu$m image. The
 surface density contours of Class I (including flat-spectrum sources) and Class II YSOs in the G15 region, overlaid on the integrated
 C$^{18}$O intensity map over the velocity range 20--38 km  s$^{-1}$, are shown in Figure~\ref{fig12}b. Because of the large optical depth,
 C$^{18}$O emission detects the densest parts of the molecular clouds compared to $^{12}$CO and $^{13}$CO. Again, we perform a dendrogram
 analysis on this integrated C$^{18}$O map to identify cold molecular condensations. Cold condensations identified in the integrated
 C$^{18}$O/$^{13}$CO maps and the dust clumps \citep{urquhart18} are marked in Figure~\ref{fig12}b.
 
\begin{figure*} 
\epsscale{1.0}
\plottwo{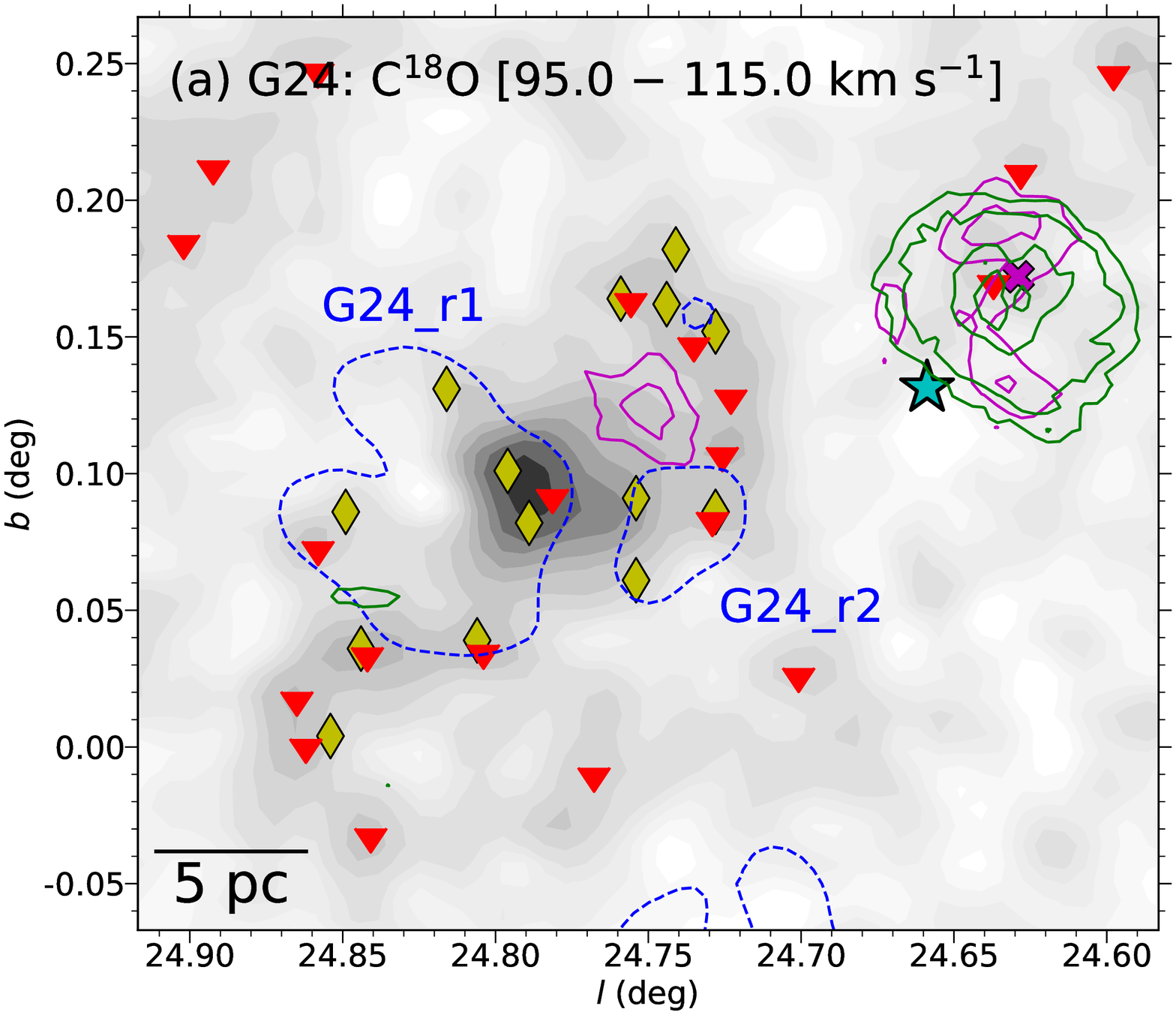}{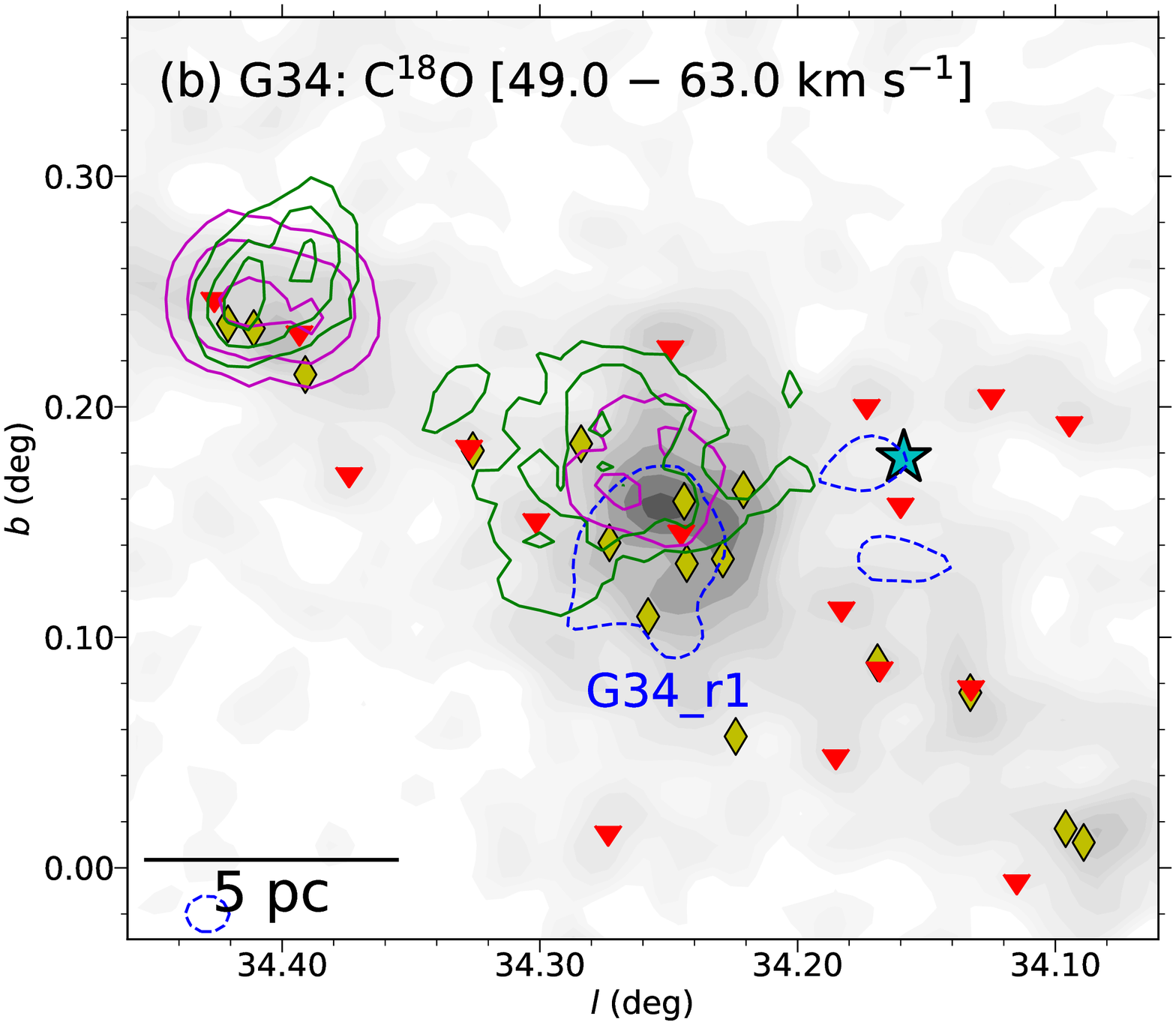}\\
\plottwo{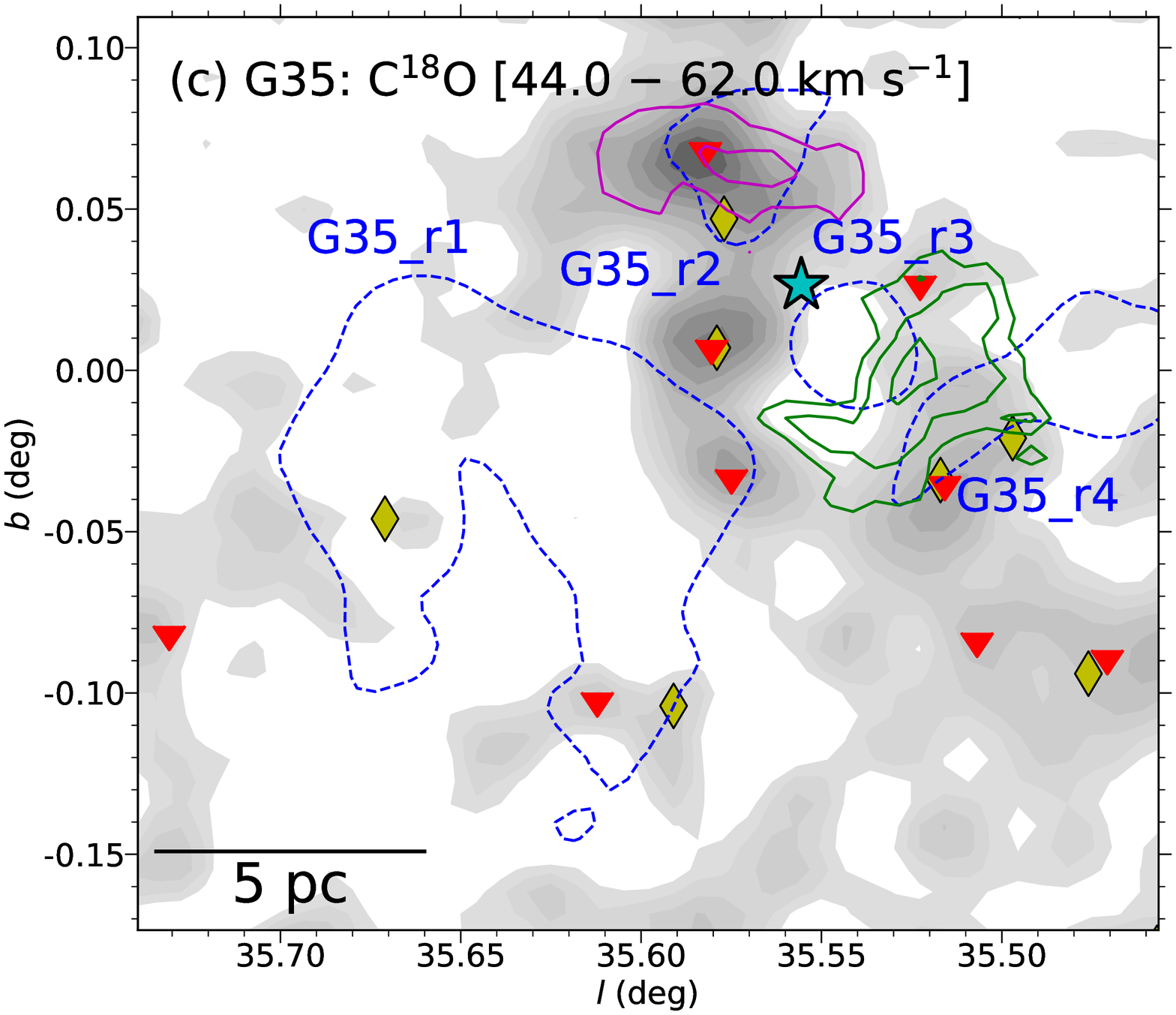}{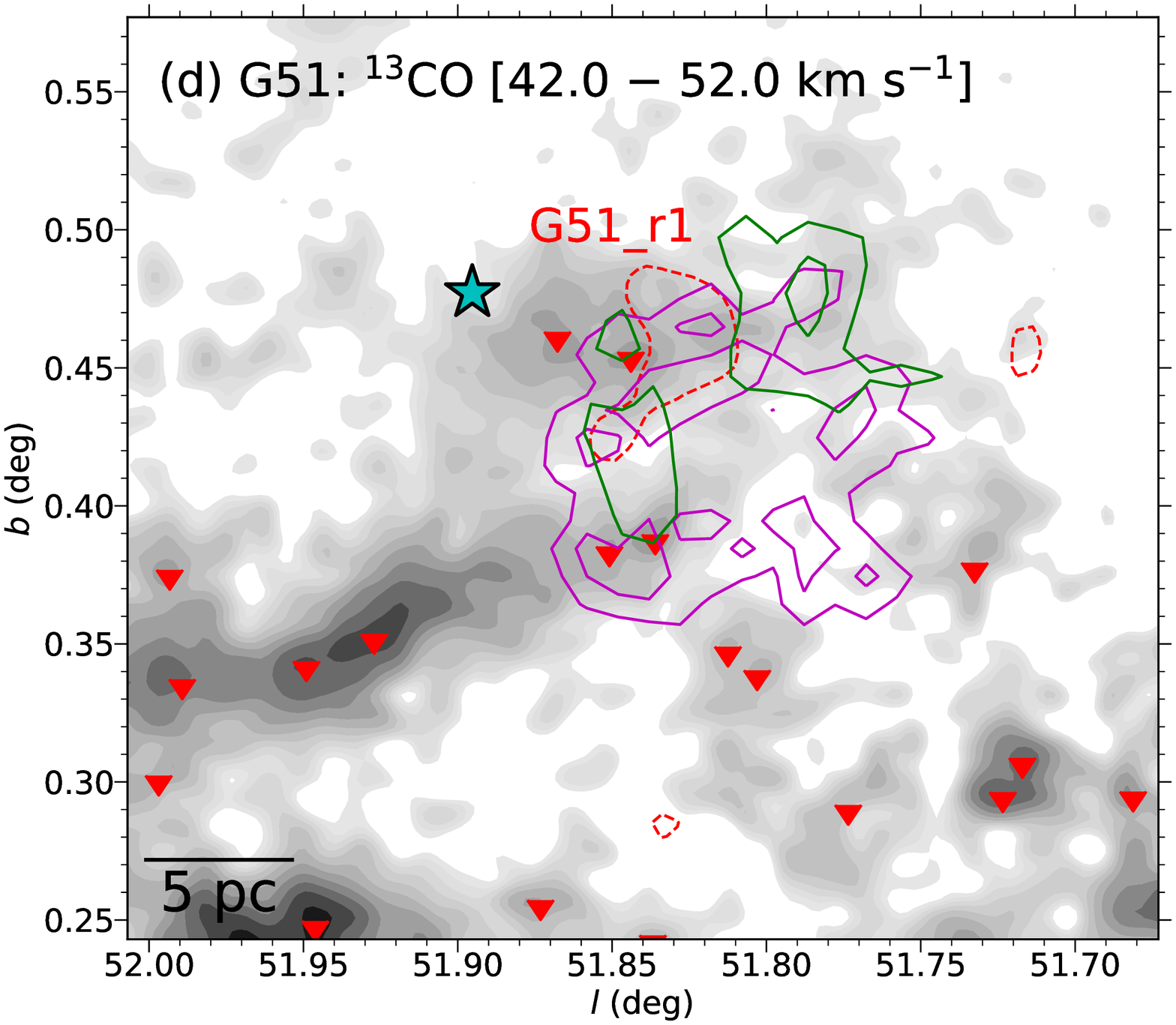}
\caption{\scriptsize The 20 nearest-neighbor surface density contours of Class I (and flat-spectrum) and Class II YSOs in magenta and green,
 respectively, overlaid on the velocity-integrated C$^{18}$O maps (solid gray contours). As C$^{18}$O data were unavailable
 for the G51 region, the contours for this region are overlaid on the velocity-integrated $^{13}$CO map. (a) G24 region:
 Class I contours are drawn at 0.20, and 0.25 YSO pc$^{-2}$, Class II contours at 0.25, 0.30, 0.45, and 0.60 YSO pc$^{-2}$,
 and a single contour of ionized gas in dashed blue at 250 mJy beam$^{-1}$ shows the extent of the ionized gas. (b) G34 region:
 Class I contours are drawn at 1.3, 2, and 4 YSO pc$^{-2}$, Class II contours at 0.65, 0.8, and 1.0 YSO pc$^{-2}$, and 180 mJy
 beam$^{-1}$ for the ionized gas. (c) G35 region: 0.4, and 0.5 YSO pc$^{-2}$ for Class I, and 0.27, 0.3, and 0.35 YSO pc$^{-2}$
 for Class II, and 170 mJy beam$^{-1}$ for the ionized gas. (d) G51 region: 0.035 and 0.04 YSO pc$^{-2}$ for Class I and 0.03
 and 0.035 YSO pc$^{-2}$ for Class II, and 90 mJy beam$^{-1}$ for the ionized gas. Ionized regions in all panels are labeled.
 Positions of cold condensations and dust clumps \citep{urquhart18} are marked as red triangles and yellow diamonds, respectively.
 Note that no dust clumps are identified in the selected area for G51. Positions of the W--R stars are marked by cyan stars.}
\label{fig13}
\end{figure*}

The 20 nearest-neighbor surface density contours for other regions, overlaid on velocity-integrated C$^{18}$O maps, are shown in Figure~\ref{fig13}.
 As C$^{18}$O data were not available for the G51 region, the 20 nearest-neighbor YSO surface density contours for this region are overplotted
 on an integrated $^{13}$CO map. The 5$\sigma$ VGPS contours of ionized gas in all regions are also shown to mark the boundaries of the ionized gas.

\subsection{Molecular shells around W--R stars}
\label{sec:CO_NH2}
{\sl Herschel} column density maps are often contaminated by foreground and background emission. Such contamination becomes
 particularly unavoidable toward the inner Galactic plane. Although this might not affect the large structures, it could overwhelm
 faint structures in the {\sl Herschel} column density maps. The FUGIN survey (see Section~\ref{GRS}) provides simultaneously observed
 spectral data of the $^{12}$CO ($J$=1--0), $^{13}$CO ($J$=1--0), and C$^{18}$O ($J$=1--0) lines. The $^{12}$CO and $^{13}$CO
 data were used to construct column density maps of the selected regions. This method to construct column density maps is expected to
 be more robust, because it generates maps for any given velocity range and thus eliminates foreground and background contributions.
 Excitation temperatures ($T_\mathrm{ex}$) were estimated from the $^{12}$CO data assuming the transition as optically thick, using
\begin{equation}
	T_\mathrm{ex} = 5.5 / \mathrm{ln}~\bigg(1 + \frac{5.5}{T_\mathrm{mb}(^{12}\mathrm{CO~peak})+0.82}\bigg).
\end{equation}

The corresponding $T_\mathrm{ex}$ was used to estimate the optical depth of the $^{13}$CO emission ($\tau_{13} (v)$) at each
  pixel and velocity ($v$) from the $^{13}$CO brightness temperature ($T_\mathrm{mb}(v)$),
\begin{equation}
	\tau_{13} (v) = -\mathrm{ln}~\bigg[1 - \frac{T_\mathrm{mb}(v)}{5.3} \bigg( \frac{1}{\mathrm{exp}(\frac{5.3}{T_\mathrm{ex}}) - 1} - 0.16\bigg)^{-1} \bigg]. 
\end{equation}

\begin{figure*}
\epsscale{1.0}
\plottwo{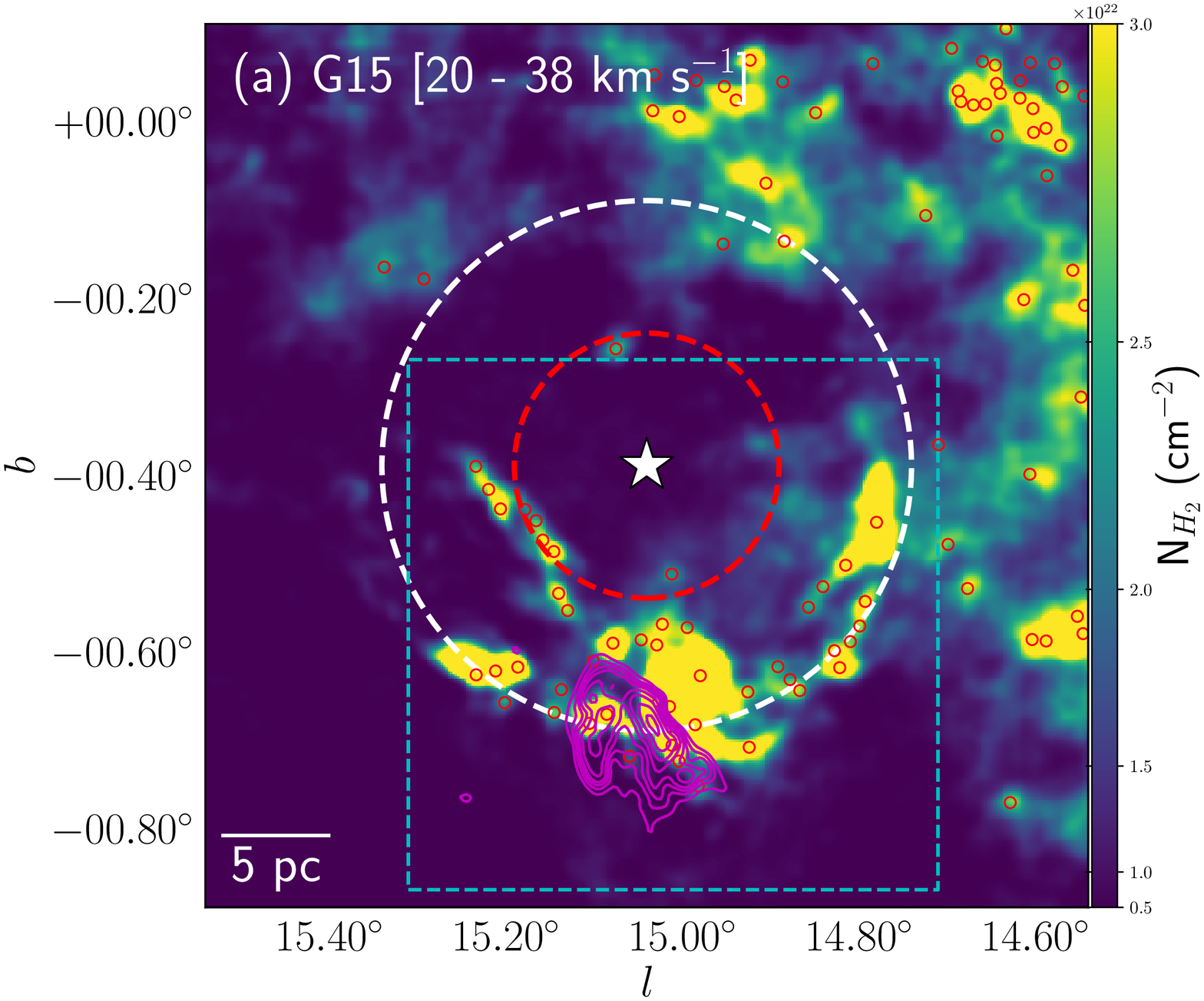}{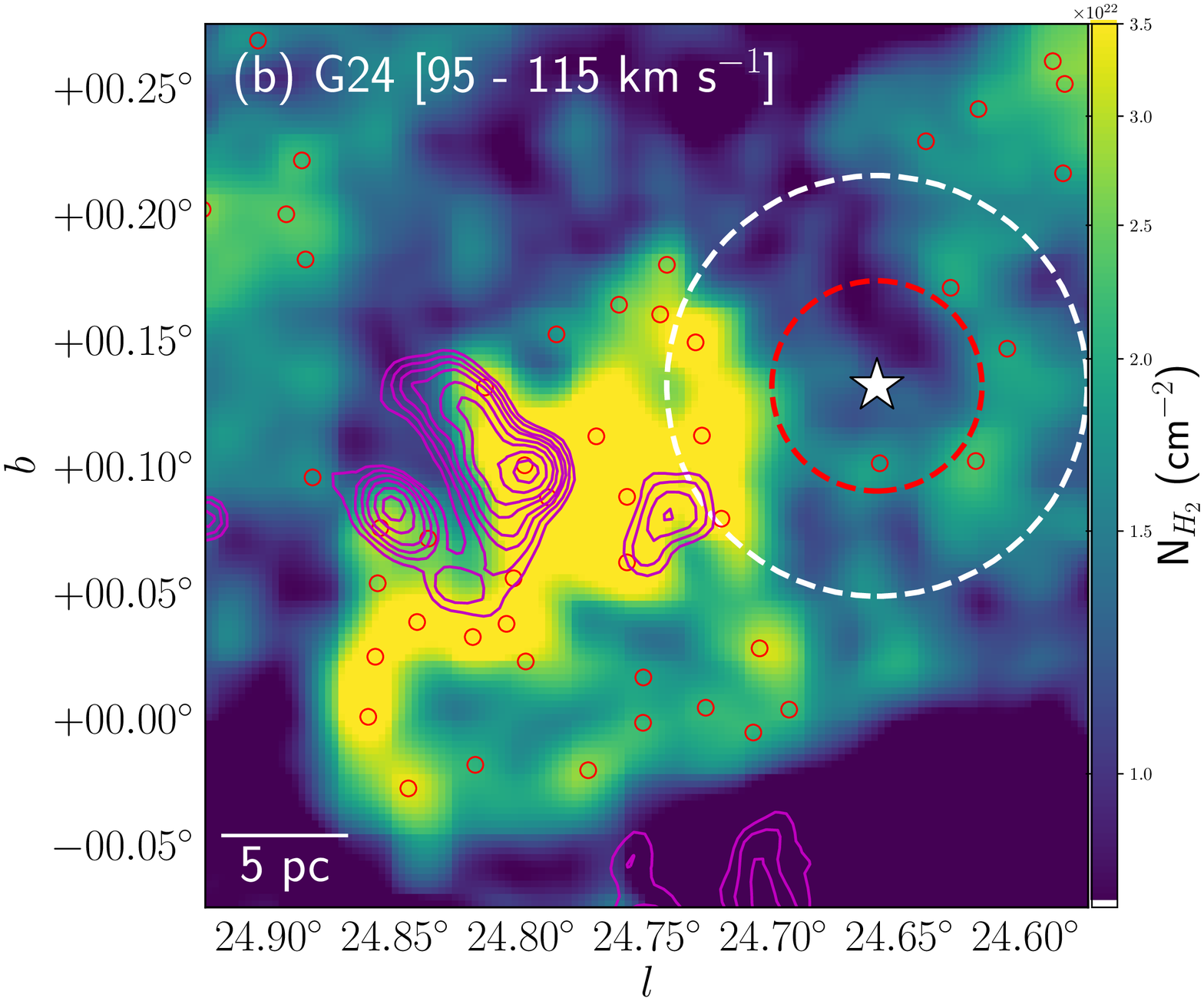} \\
\plottwo{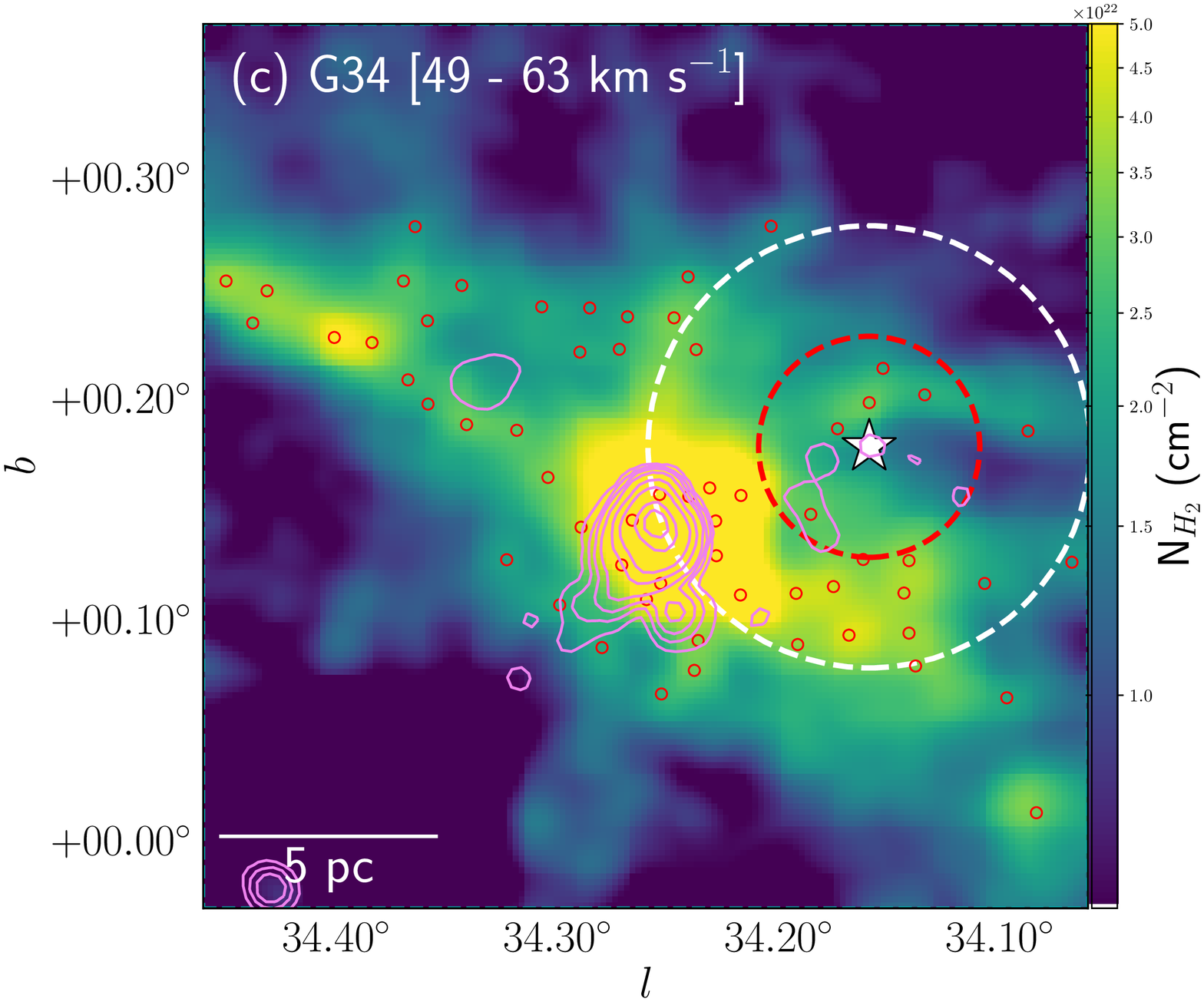}{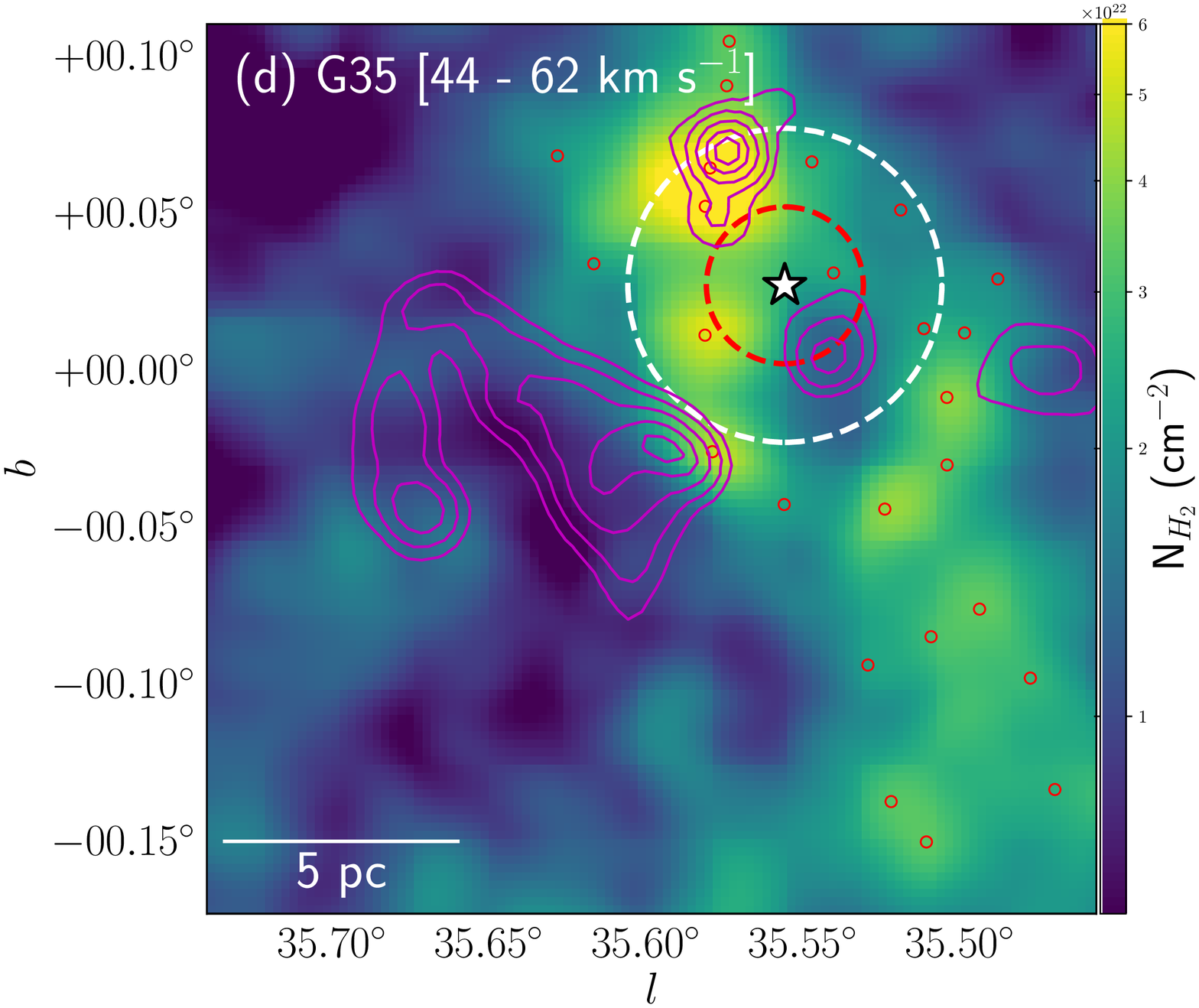} \\
\epsscale{0.5}
\plotone{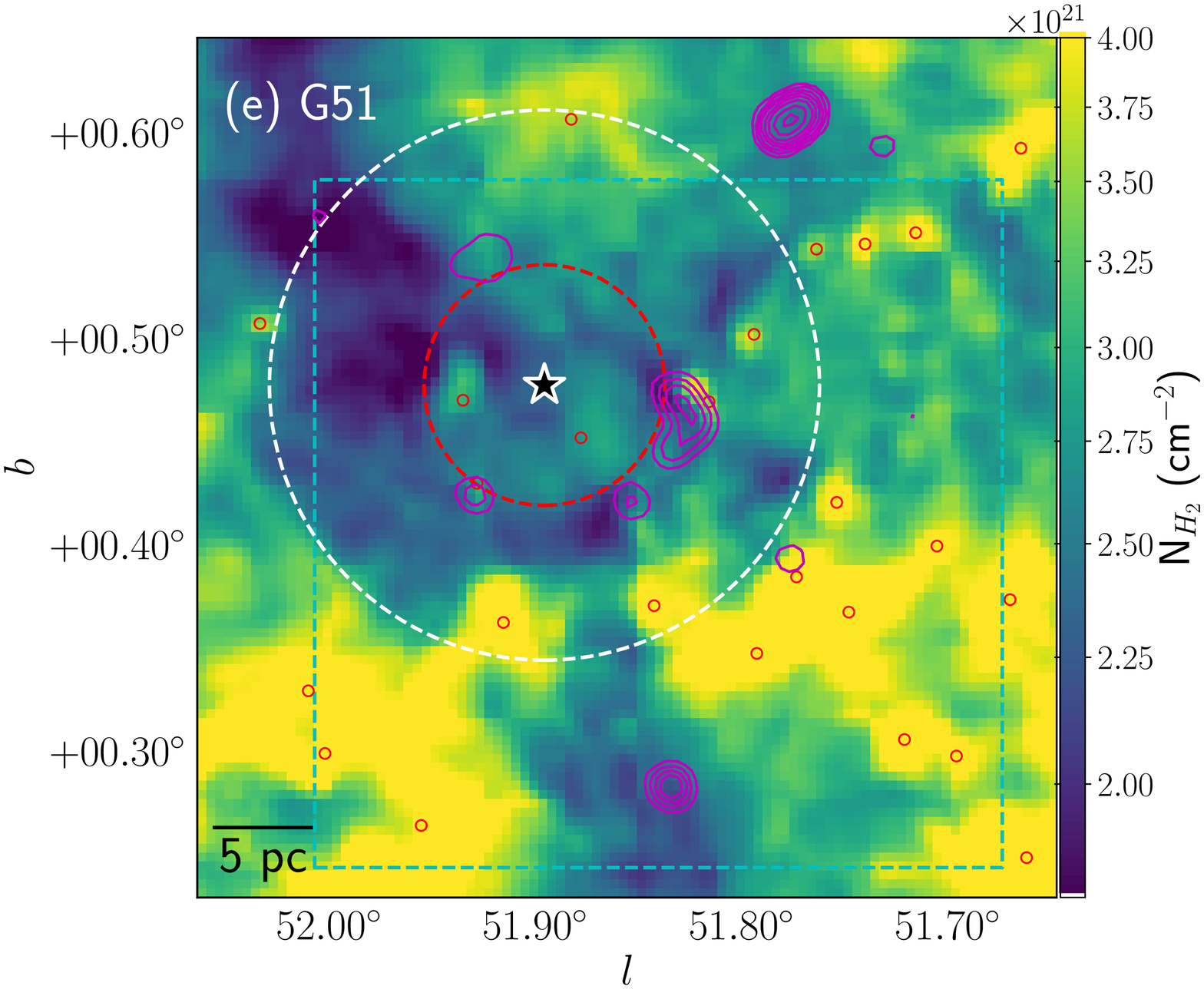}
\caption{\scriptsize Column density maps constructed using $^{12}$CO and $^{13}$CO data of all regions. Since $^{12}$CO data
 were not available for the G51 region, we show the {\sl Herschel} column density map again for a larger area. The cyan dashed boxes
 for the G15 and G51 regions show the areas of interest. For remaining regions, maps are shown for the same areas explored in this paper.
 The magenta contours in each panel depict the distribution of the ionized gas. Possible boundaries of cavities centered on the
 W--R stars are	marked as red dashed circles. Outer boundaries of shells (radii twice the radii of corresponding inner shells)
 are also marked as white dashed circles. The annular areas in red and white circles are meant to help us estimate the column
 densities of the shells. Clumps identified in these column density maps are also marked by small red open circles.}
\label{fig14}
\end{figure*}

The $^{13}$CO column density for each pixel was estimated by integrating the emission over the identified velocity range, using
\begin{equation}
	N(^{13}\mathrm{CO}) = 2.4\times10^{14}\times\sum\frac{T_\mathrm{ex}\tau_{13} (v) \Delta v}{1 - \mathrm{exp}\big(-\frac{5.3}{T_\mathrm{ex}}\big)}.
\end{equation}
Finally, $N(^{13}\mathrm{CO})$ was converted to $N(\mathrm{H_2})$ using a conversion factor of 7.7$\times$10$^5$ \citep{kohno18}. A more
 detailed description of the calculation method can be found in \citet{kohno18} and \citet{torii19}. The final column density maps for four
 regions are presented in Figure~\ref{fig14}. As FUGIN data do not cover the G51 region, we used the {\sl Herschel} column density map to
 identify the molecular shell. It is found that a few clumps seen in the dust column density maps (figures are not shown in this paper)
 are missing in these gas column density maps. Those clumps might be part of foreground and background clouds. In addition, gas column
 density maps detected detailed structures that were missing in the dust column density maps, possibly because the $^{13}$CO column density
 maps are less contaminated from the foreground and background emission.

Low-density cavities and circular shells around W--R stars are apparent in almost all regions except for G35 (see Figure~\ref{fig14}).
 Note that these shells and cavities are identified around W--R stars using molecular line data, and are different from the MIR bubbles
 seen in the {\sl Spitzer}-IRAC images.
 Possible inner boundaries of the shells are marked by red circles based on visual identification of the cavities. Although the shell is not
 very apparent in the G35 region, the immediate vicinity of the W--R star is deficient in molecular gas (see Figure~\ref{fig14}d). The outer
 boundaries of the shells are also marked by white circles of radii equal to twice the radii of the inner boundaries. Note that these outer
 boundaries do not have any physical significance, but just help us in estimating the average column densities of the shells. The cavities
 within the shells are devoid of molecular gas and have comparatively lower column densities than the annular areas marked within the red
 and white circles.

\subsection{Pressure of W--R stars}
\label{sec:pressure}
We identify the primary energetic component of the W--R stars that acts as the major agent to disperse the surrounding molecular gas.
 The expansion of the molecular gas with velocities of about 2--5 km s$^{-1}$ is detected in the \pv diagrams of all regions (except for G34;
 see Section~\ref{sec:dynamics}) which is likely caused by the energetic inputs from the central W--R stars. In general, massive stars exert
 multiple pressure components on their surrounding molecular gas (e.g. pressure owing to radiation, stellar winds, and ionized gas).

\begin{figure} 
\epsscale{1.2}
\plotone{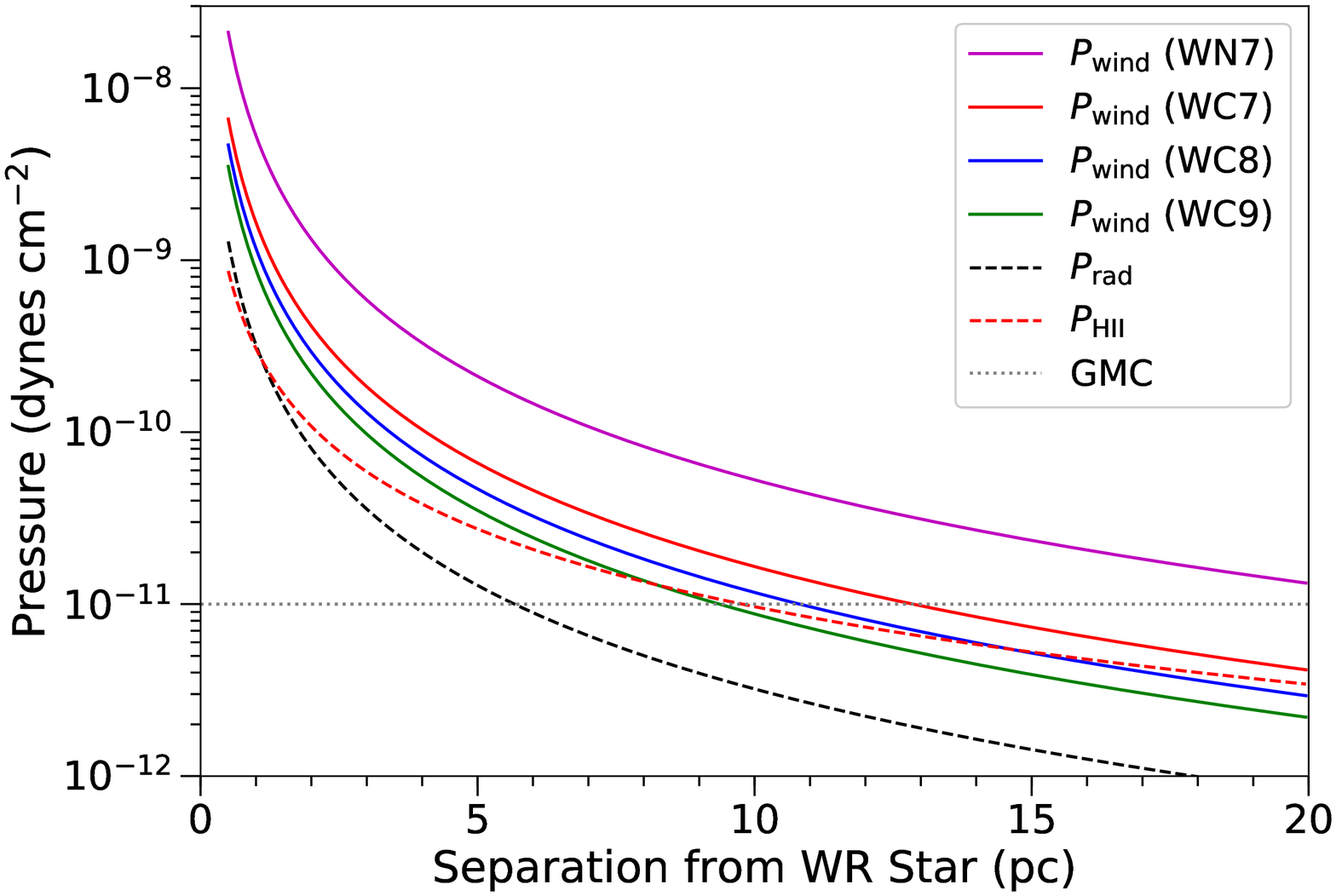}
\caption{\scriptsize Wind pressure exerted by W--R stars of different spectral types (red, green, blue, and magenta curves) at
 different physical distances. The pressure owing to radiation is shown by a single dashed black curve, because all WC7, WC8,
 WC9, and WN7 stars have a similar bolometric luminosity of about 3$\times$10$^5~L_{\odot}$. The pressure experienced by a
 typical cool giant molecular cloud is shown by the gray dotted line. In the absence of ionized peaks associated with W--R
 stars, we have shown a possible pressure component of ionized gas (red dashed line) developed by a typical O-type progenitor
 ($m_\ast \sim 20 M_\odot$). For all W--R stars, the pressure owing to winds is an order of magnitude higher than the radiation
 pressure. }
\label{fig15}
\end{figure}

To get an idea about the different pressure components and the extent of the influence of the W--R stars, we computed the pressure
 owing to stellar winds, $P_\mathrm{wind}$ ($= \dot{M_\mathrm{w}} V_\mathrm{w} /4\pi D_\mathrm{s}^2$, where $D_\mathrm{s}$ is the distance
 from the W--R star) and the pressure owing to radiation, $P_\mathrm{rad}$ ($= L_\mathrm{bol} /4\pi c D_s ^2$) for a range of distances from
 the W--R star. \citet{nugis00} studied the wind characteristics of W--R stars based on their
 stellar parameters. The stellar wind velocities for our spectral types were obtained from \citet{nugis00}. Typical mass-loss rates
 for these stars were also computed using their equation (24). Armed with those typical values of mass-loss rates ($\dot{M_\mathrm{w}}$)
 and stellar wind velocities ($V_\mathrm{w}$), we estimate $P_\mathrm{wind}$ for a separation range ($D_\mathrm{s}$) of 1--20 pc from the
 powering W--R star (see Figure~\ref{fig15}). To calculate $P_\mathrm{rad}$, the bolometric luminosities ($L_\mathrm{bol}$) of these W--R
 stars are also obtained from \citet{nugis00}. The $L_\mathrm{bol}$ values are similar ($\sim$2--5$\times$10$^5~L_{\odot}$) for all W--R spectral types in our
 sample. Hence, $P_\mathrm{rad}$ is estimated for a single $L_\mathrm{bol}$ value of 3$\times$10$^5~L_{\odot}$, for the same range of
 $D_\mathrm{s}$. Figure~\ref{fig15} shows that $P_\mathrm{wind}$ is typically an order of magnitude higher than $P_\mathrm{rad}$.
 \citet{dyson80} reported that a cool giant molecular cloud with a typical temperature of $\sim$20 K and a particle density of 10$^3$--10$^4$
 cm$^{-3}$ exhibits an outward pressure (i.e. $P_\mathrm{MC}$) of about 10$^{-12}$--10$^{-11}$ dyne cm$^{-2}$. 
 None of the W--R stars in our sample are associated with ionized gas peaks, which makes it difficult to estimate the pressure owing to
 the ionized gas from W--R stars. \citet{dewangan16} also found for the G27 region that the pressure from ionized gas is two orders
 lower compared to $P_\mathrm{wind}$ and $P_\mathrm{rad}$. We nevertheless estimated a characteristic value of pressure owing to 
 ionized gas formed by an O-type progenitor of a W--R star ($m_\ast \sim 20 M_\odot$). The ionized gas pressure can be formulated as, 
 $P_\mathrm{HII} = \mu_{\mathrm{II}} m_H c_{\mathrm{II}}^2 \left(S_{Lyc}/4\pi\beta_2 D_\mathrm{s}^3\right)^{1/2}$, where the mean molecular
 weight in an \hii region, $\mu_{\mathrm{II}}$=0.678 \citep{bisbas09}, the sound speed in an \hii region, $c_{\mathrm{II}}$ = 11 km s$^{-1}$,
 and the recombination coefficient, $\beta_2$ = 2.6$\times$10$^{-13}$ cm$^3$ s$^{-1}$. Possible pressure component caused by \hii region developed
 by a typical O-type progenitor of a W--R star \citep[$S_{Lyc}\sim$10$^{48.3}$ photon s$^{-1}$;][]{panagia73} is also plotted in Figure~\ref{fig15}.
 It can be seen that the pressure due to ionized gas is typically lower compared to $P_\mathrm{wind}$. Accordingly, in Figure~\ref{fig15}
 one can see that these W--R stars can drive the surrounding molecular gas out to a distance of about 10 pc primarily because of their stellar winds.

\begin{deluxetable*}{ccccccccccccc}
\tablewidth{0pt}
\tabletypesize{\scriptsize} 
\tablecaption{Parameters of the ionized gas and shells in all five regions.\label{table3}}
\tablehead{
\colhead{Region} & Cavity    & \multicolumn{3}{c}{$N(\mathrm{H_2})$ (10$^{21}$cm$^{-2}$)} & Ionized   & Int. Flux   & log ($S_\mathrm{Lyc}$) & $R_\mathrm{t}$ & $t_{\mathrm{dyn}} \mathrm(Myr)$ & \multicolumn{2}{c}{$P$ on shell$^a$} & \\ 
		 & size (pc) &  Cavity     &  Shell   & Inc. factor$^b$ & Region id & density (Jy)& (photon s$^{-1}$)& (pc)  &(5000 cm$^{-3}$) & $P_\mathrm{wind}$&$P_\mathrm{HII}$ &}
\startdata
G15              &  6.0  &  9.6 & 18.0  & 88\% &  G15\_r      &   571.700   &     50.435       &  4.71 &      0.965      &   14.7       &27.5   \\
	         &  6.0  &  9.6 & 18.0  & 88\% &  G15\_r1$^c$ &   130.240   &     49.793       &  2.48 &      0.451      &   14.7       &14.6   \\
                 &       &      &       &      &  G15\_r2$^c$ &   146.680   &     49.844       &  1.85 &      0.252      &   14.7       &14.4   \\
G24              &  5.0  & 13.9 & 18.2  & 31\% &  G24\_r2     &     5.738   &     49.188       &  3.35 &      1.116      &    5.1       & 9.4   \\
G34              &  3.0  & 22.2 & 30.0  & 35\% &  G34\_r1     &    10.565   &     48.866       &  0.65 &      0.070      &   18.3       &19.0   \\
G35              &  1.3  & 28.0 & 31.8  & 14\% &  G35\_r1     &     1.492   &     48.251       &  1.01 &      0.231      &   96.5       &14.3   \\
                 &       &      &       &      &  G35\_r2     &     1.797   &     48.332       &  1.36 &      0.373      &   96.5       &35.6   \\
G51              &  6.0  & 2.4  & 2.8   & 17\% &  G51\_r1$^d$ &     0.088   &     47.403       &   --  &       --        &    2.4       & --    
\enddata
\tablenotetext{$a$}{In units of 10$^{-11}$ dynes cm$^{-2}$.}
\tablenotetext{$b$}{Increasing factor of column density has $\sim$10\% uncertainty.}
\tablenotetext{$c$}{Estimated from 1.4 GHz NVSS map.}
\tablenotetext{$d$}{This ionized region is located at a distance of $\sim$10 kpc \citep{bania12}.}
\end{deluxetable*}

Although no ionized-gas peaks are exactly associated with any W--R stars in our sample, several \hii regions are present in the selected
 regions. In fact, these \hii regions are spatially distributed around the molecular shells identified in Section~\ref{sec:CO_NH2} (see also
 Figure~\ref{fig14}). Note that expansion of the ionized gas is believed to be an efficient triggering mechanism. Thus, it is important
 to explore the pressure exerted by these particular \hii regions ($P_\mathrm{HII}$) on the surrounding gas. To estimate $P_\mathrm{HII}$, we obtained
 the Lyman continuum flux ($S_\mathrm{Lyc}$ in photons s$^{-1}$) for selected ionized regions located near the W--R stars (see labels in
 Figures~\ref{fig12}b and Figure~\ref{fig13}) using the radio continuum flux, following the equation given by \citet{moran83}:
\begin{equation}
S_\mathrm{Lyc} = 8 \times 10^{43}\left(\frac{S_\nu}{\mathrm{mJy}}\right)\left(\frac{T_e}{10^4 \mathrm{K}}\right)^{-0.45}\left(\frac{D}{\mathrm{kpc}}\right)^2 \left(\frac{\nu}{\mathrm{GHz}}\right)^{0.1},
\label{lyman_flux}
\end{equation}
where $S_\nu$ is the total flux density, $T_e$ is the electron temperature, $D$ is the distance to the source, and $\nu$ is the frequency of
 the observations. The assumption was made that the region is homogeneous and spherically symmetric. The radio continuum fluxes and extents
 of the ionized regions are estimated from the VGPS 1.4 GHz maps using the {\sc jmfit} task in {\sc aips}\footnote{The NRAO Astronomical
 Image Processing System http://www.aips.nrao.edu/index.shtml}. The pressure owing to ionized gas ($P_\mathrm{HII}$) is formulated as outlined
 above. The estimated radio continuum flux densities, extent of the ionized gas in pc, and $S_\mathrm{Lyc}$ are listed in
 Table~\ref{table3}. As ionized regions are distributed around the molecular shells, they may exert an opposite pressure on the shells
 compared to the $P_\mathrm{wind}$ from the central W--R stars. Thus, we also estimated the $P_\mathrm{wind}$ from the W--R stars and 
 $P_\mathrm{HII}$ from a few surrounding \hii regions on the molecular shells, and corresponding values are listed in Table~\ref{table3}.
 However, the \hii region toward the G51 region (G51\_r1) is located at a distance of $\sim$10 kpc \citep{bania12},
 and it is therefore excluded from this analysis.

\section{Discussion}
\label{sec:SF}
The primary aim of this study is to assess the influence of W--R stars on their parent molecular clouds. The presence of nebulous or
 bubble-like features around massive W--R stars is typically interpreted as evidence of an interaction between W--R stars and their surrounding
 molecular clouds \citep{lamers99, dewangan16}. A few recent studies already reported a positive impact of W--R stars on their parent molecular
 clouds for the next generation of star formation \citep[see][and references therein]{liu12, cichowolski15, dewangan16}. Owing to their strong
 energetic (wind and radiation) impact, the W--R stars in our sample have dispersed their nearby molecular gas and created cavities that are
 primarily represented by a trough between the two emission peaks in the $^{13}$CO spectra (Section~\ref{sec:association}).

\begin{figure} 
\epsscale{1.2}
\plotone{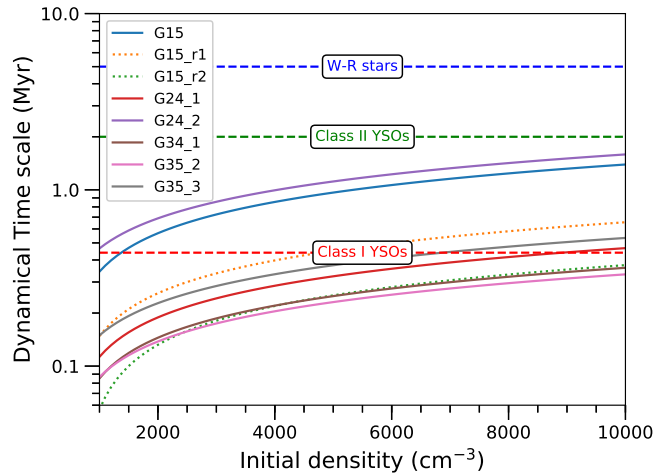}
\caption{\scriptsize Variation of the dynamical ages of the ionized gas seen in all selected regions for a range of initial
 ambient densities. The regions shown in the legend are labeled in Figures~\ref{fig12}b  and~\ref{fig13}. The mean lifetimes
 of Class I and Class II YSOs, and the typical age of a W--R star are also marked.}
\label{fig16}
\end{figure}

Ionized gas (i.e. \hii regions) is seen in all regions (see Figures~\ref{fig12}b and \ref{fig13}). In addition, in the surface density
 analysis, clusters of YSOs were also spatially found near the \hii regions. It is therefore important to examine whether these \hii regions
 are sufficiently old to trigger the formation of the surrounding YSOs. We used the parameters listed in Table~\ref{table3} and calculated
 their dynamical ages following the procedure of \citet{baug15}. The estimated dynamical age of an \hii region may substantially vary depending
 on the initial value of the ambient density. Hence, we calculated the dynamical ages for a range of initial ambient densities starting from
 1000 to 10,000 cm$^{-3}$ \citep[i.e. classical to ultra-compact \hii regions;][]{kurtz02}. The estimated dynamical ages of the \hii regions
 located near the YSO clusters are shown in Figure~\ref{fig16}. The average lifetimes of Class I and Class II YSOs \citep[i.e. $\sim$0.44 Myr
 and $\sim$2 Myr, respectively;][]{evans09} and the typical age of W--R stars \citep[$\sim$5 Myr;][]{lamers99} are also marked in Figure~\ref{fig16}.

 Given any initial ambient density, the \hii regions in G34 and G35 are not capable of initiating the formation of YSOs seen in these regions.
 The ionized gas in the G15 and G24 (G24\_r2) are characterized by a dynamical timescale that is sufficiently long to trigger the formation of
 Class I YSOs in its vicinity. A sequence of Class I YSOs and C$^{18}$O condensations can indeed be noted around the ionized gas in the G24
 region (see Figure~\ref{fig13}a). Such a sequence is a typical signature of triggered star formation by the expansion of the ionized gas.
 However, no such sequence is seen in any other regions.
 Note that although VGPS maps typically trace extended radio continuum emission compared to NVSS maps, the VGPS is generally unable to
 resolve the detailed distribution of the ionized gas because of its lower resolution. Figure~\ref{fig12}b shows that the NVSS has resolved
 two peaks of ionized gas (G15\_r1 and G15\_r2). The dynamical ages of these two ionized regions are not sufficient to initiate the
 formation of Class II YSOs, and only one is marginally old enough to initiate the formation of Class I YSOs.
 
As can be seen in Figure~\ref{fig14}, the ionized gas is typically distributed around the molecular shells developed by the W--R stars.
 Clusters of YSOs and dense condensations are also seen a few pc from the W--R stars (see Figures~\ref{fig12} and \ref{fig13}), but none are
 generally coincident with the positions of the W-R stars. This observational picture might be explained considering that all W--R stars in
 our sample have developed expanding molecular shells and created cavities within these shells. Thus, these cavities have lower gas densities,
 perhaps too low to form any YSOs or cold dust/molecular condensations. In fact, no recent star formation is found within a few pc of our sample
 W--R stars (see Figures~\ref{fig12}b and \ref{fig13}), while clumps and YSO clusters are generally projected at separations of 2--8 pc from
 the W--R stars. The ionized regions, as well as all identified YSOs, are comparatively younger than the W--R stars, and they are commonly
 distributed around the molecular shells. 

It is also important to note that with typical expansion velocities of 2--5 km s$^{-1}$, W--R
 stars can create shells of 3--8 pc in size within 2 Myr and accumulate matter around the shells, as seen in all our selected regions. In
 fact, these shells are denser compared to the inner cavities, and the column densities in these molecular shells are enhanced by a
 minimum of 14$\pm$10\% (for the G35 region) to as much as 88$\pm$10\% (for the G15 region) compared to the column densities in the
 cavities (see Table~\ref{table3}). In addition, the surrounding \hii regions have dynamical timescales $\lesssim$2 Myr, even for an initial
 ambient density of 10,000 cm$^{-3}$, which is much younger than the ages of the W--R stars. Higher values of $P_\mathrm{wind}$ exerted on
 the shells than $P_\mathrm{HII}$ from the ionized gas (see Table~\ref{table3}) indicate a possibility of further expansion of the shells.
 All these results indicate that these W--R stars may indeed be responsible for halting recent star formation in their surrounding cavities.
 An enhanced column density and the presence of active star formation around expanding shells also indicate that molecular gas was collected
 around these shells which subsequently collapsed to form stars (i.e. the `collect and collapse' scenario). The presence of ionized gas
 around the shells may trigger further star formation, as seen in the G24 region. Such star formation around W--R shells was already
 noted by \citet{liu12}.
 
M17 is regarded one of the most active Galactic star-forming regions, with a star-formation rate of $\ge 0.004~M_\odot$ yr$^{-1}$
 \citep[see][and references therein]{povich16}, which is four times higher than that of the Orion Nebula Cluster. The reason for such a
 high star-formation rate is still open to debate. \citet{jiang02} derived an upper age limit of 3 Myr for M17 which is about 2 Myr younger
 than the W--R star (typical age of $\sim$5 Myr). The expanding shell with $v_\mathrm{exp}$ of 4 km s$^{-1}$ driven by the W--R star may
 reach $\sim$8 pc even in 2 Myr, i.e. the projected separation of the M17 cluster from the W--R star. It is thus possible that the
 presence of an external driving source, like a W--R star, has caused such a high star-formation rate in M17. Thus, the influence
 of the W--R star on the M17 region in the context of rapid star formation should be investigated in greater detail.

Overall, the presence of a W--R star significantly affects the dynamics of the parent cloud. The expanding molecular shell driven
 by a strong stellar wind may create gas-deficient cavities and, thus, further star formation is quenched in those cavities. However,
 the expanding molecular shell may still help to accumulate molecular gas in the outer layer of the shell, which may collapse and result
 in further star formation. The impacts of W--R stars might be found in several other Galactic star-forming regions. It is possible that
 several such regions are still unexplored because of poorly known spectral types and distances of W--R stars. Statistics on the influence of
 W--R stars on Galactic star formation can be improved if good estimates of the spectral types and better distance calibrations are
 available for Galactic W--R stars.
\section{Conclusions}
\label{sec:conclusions}

The primary goal of this study was to explore the impact of W--R stars on their parent molecular clouds. The main results of
 this study are the following.

1. Strong stellar winds from the W--R stars have created gas-deficient cavities in the host molecular clouds. A signature of these
 cavities in the parent molecular cloud is found in the form of a trough between the two emission peaks in the $^{13}$CO spectrum
 along W--R stars. All the W--R stars in our sample are located at the velocities of the troughs.

2. In all but one region, we identified expanding molecular shells from the \pv analysis of the $^{13}$CO data. These molecular shells
 are expanding with velocities of 2--5 km s$^{-1}$, and W--R stars are identified as the primary driving sources of this expansion.
 Although a W--R star is associated with the molecular cloud toward the G34 region, no signature of expanding molecular gas is
 noted. This is possibly because the red-shifted part of the cloud is significant and sufficiently intense to drown out any
 ring- or U-like structure in the \pv diagrams.

3. The dynamical ages of none of the \hii regions are sufficiently old to trigger the formation of Class II YSOs seen in
 our selected regions. However, ionized gas in the G24 region is found to be capable of triggering the formation of Class I YSOs.
 In fact, a sequence of a cluster of YSOs and molecular condensations is seen around the ionized gas toward the G24 region,
 indicating recent star formation, possibly triggered by expansion of the ionized gas.

4. Estimation of pressure components reveals that the pressure owing to stellar winds dominates over the radiation pressure of
 W--R stars by an order of magnitude, and the wind pressure ($P_\mathrm{wind}$) is capable of driving the
 surrounding molecular gas up to a distance of about 10 pc.

5. The M17 region (located toward the G15 region) is a Galactic star-forming region with a high star-formation rate. The
 expanding shell with $v_\mathrm{exp}$ of 4 km s$^{-1}$ driven by the W--R star may reach M17 in about 2 Myr. Thus, energetic
 input from a W--R star within 10 pc could possibly explain the high star-formation rate of the M17 region. The influence of
 this W--R star on the M17 cluster should be investigated in greater detail.

6. Gas deficient cavities of 2--6 pc in size are identified in molecular column density maps. An absence of recent
 star formation in all these cavities indicates that stellar winds from W--R stars have dispersed the molecular material
 from their immediate vicinity and developed these cavities. Star formation is quenched in the cavities.

7. The column densities in cavities generally have lower values and are enhanced by a minimum of 14$\pm$10\% (in the G35
 region) to as much as 88$\pm$10\% (in the G15 region) compared to the surrounding
 shells. Molecular gas is collected around expanding shells and enhanced column densities. Clusters of YSOs (i.e. Class I--II),
 massive condensations of 10$^2$--10$^3 M_\odot$ and ionized gas are typically distributed around these shells. Star formation
 around the molecular shells might be affected by the W--R stars. Accumulation of molecular gas due to the expansion of the
 shells could have a positive impact on the recent star formation around the shells.

Overall, it is evident that the stellar winds from W--R stars have a large impact on the dynamics of their surrounding
 clouds. They are capable of developing gas-deficient cavities surrounded by expanding molecular shells. Although star
 formation is quenched in those cavities because of the impact of the W--R stars, they may also initiate further star
 formation in the outer layers of the molecular shells. Several other Galactic regions associated with W--R stars might
 be explored in greater detail to understand the influence of W--R stars on their parent molecular clouds. Such studies,
 however, require precise distance estimates to Galactic W--R stars.
\acknowledgments We thank the anonymous referee for the critical comments that have helped us improve the scientific
 content and the presentation of the paper. TB acknowledges funding from the National Natural Science Foundation of China
 (NSFC) through grant 11633005 and support from the China Postdoctoral Science Foundation through grant 2018M631241. This
 work was supported by the National Key Research and Development Program of China through grant 2017YFA0402702. We also
 acknowledge research support from the NSFC through grants U1631102 and 11373010. This publication made use of archival
 data obtained with the {\sl Spitzer} Space Telescope (operated by the Jet Propulsion Laboratory, California Institute of
 Technology under a contract with NASA). This publication makes use of molecular line data from the Boston University--FCRAO
 Galactic Ring Survey (GRS). The GRS is a joint project of Boston University and Five College Radio Astronomy Observatory,
 funded by the U.S. National Science Foundation (NSF) under grants AST-9800334, AST-0098562, and AST-0100793. 
 The National Radio Astronomy Observatory is a facility of the National Science Foundation operated under cooperative
 agreement by Associated Universities, Inc. LKD is
 supported by the Department of Space, Government of India. This research made use of {\sc astrodendro}, a Python package
 to compute dendrograms of astronomical data (http://www.dendrograms.org/). This research made use of Astropy, a
 community-developed core Python package for astronomy (Astropy Collaboration 2018). This publication makes use of data
 from FUGIN, FOREST Unbiased Galactic plane Imaging survey with the Nobeyama 45 m telescope, a legacy project of the Nobeyama
 45 m radio telescope. The FUGIN data were retrieved from the JVO portal (http://jvo.nao.ac.jp/portal/) operated by ADC/NAOJ.
 This work has made use of data from the European Space Agency's (ESA) mission {\sl  Gaia} (\url{https://www.cosmos.esa.int/gaia}),
 processed by the {\sl Gaia} Data Processing and Analysis Consortium (DPAC, \url{https://www.cosmos.esa.int/web/gaia/dpac/consortium}).
 Funding for the DPAC has been provided by national institutions, in particular the institutions participating in the {\sl Gaia}
 Multilateral Agreement. 
\appendix
\section{A. Cloud structures toward field regions}
\label{appendix}
\begin{figure*} 
\epsscale{1.0}
\plottwo{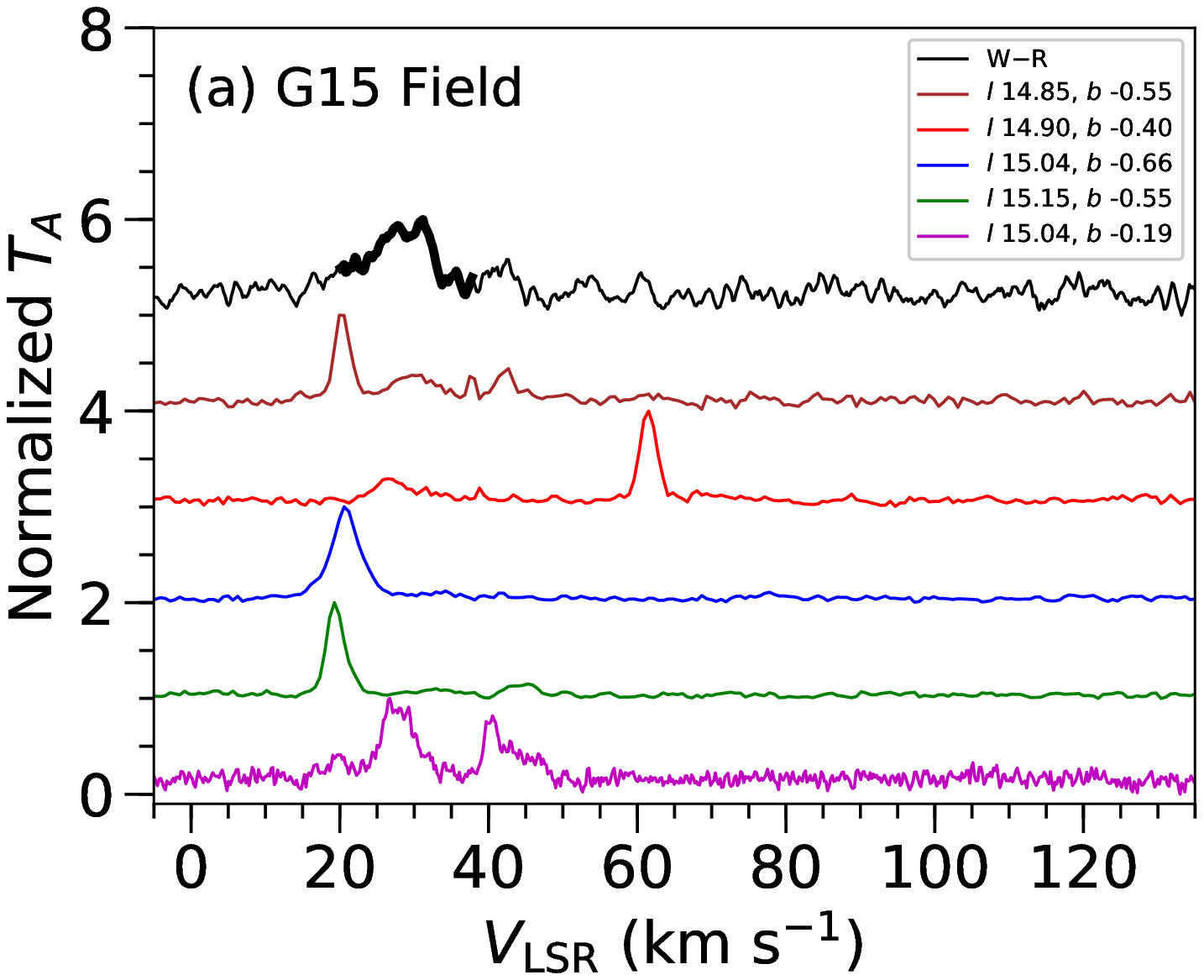}{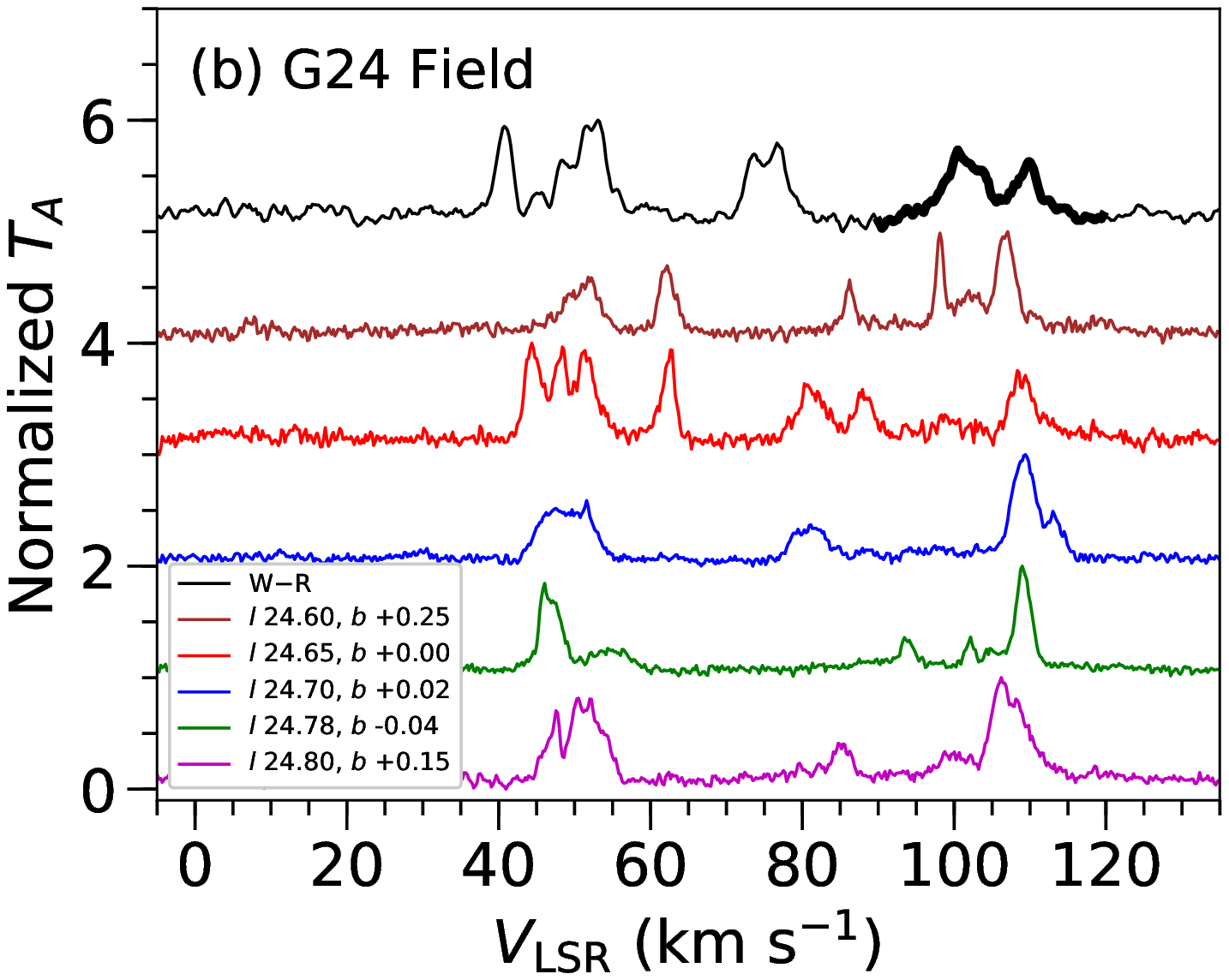} \\
\plottwo{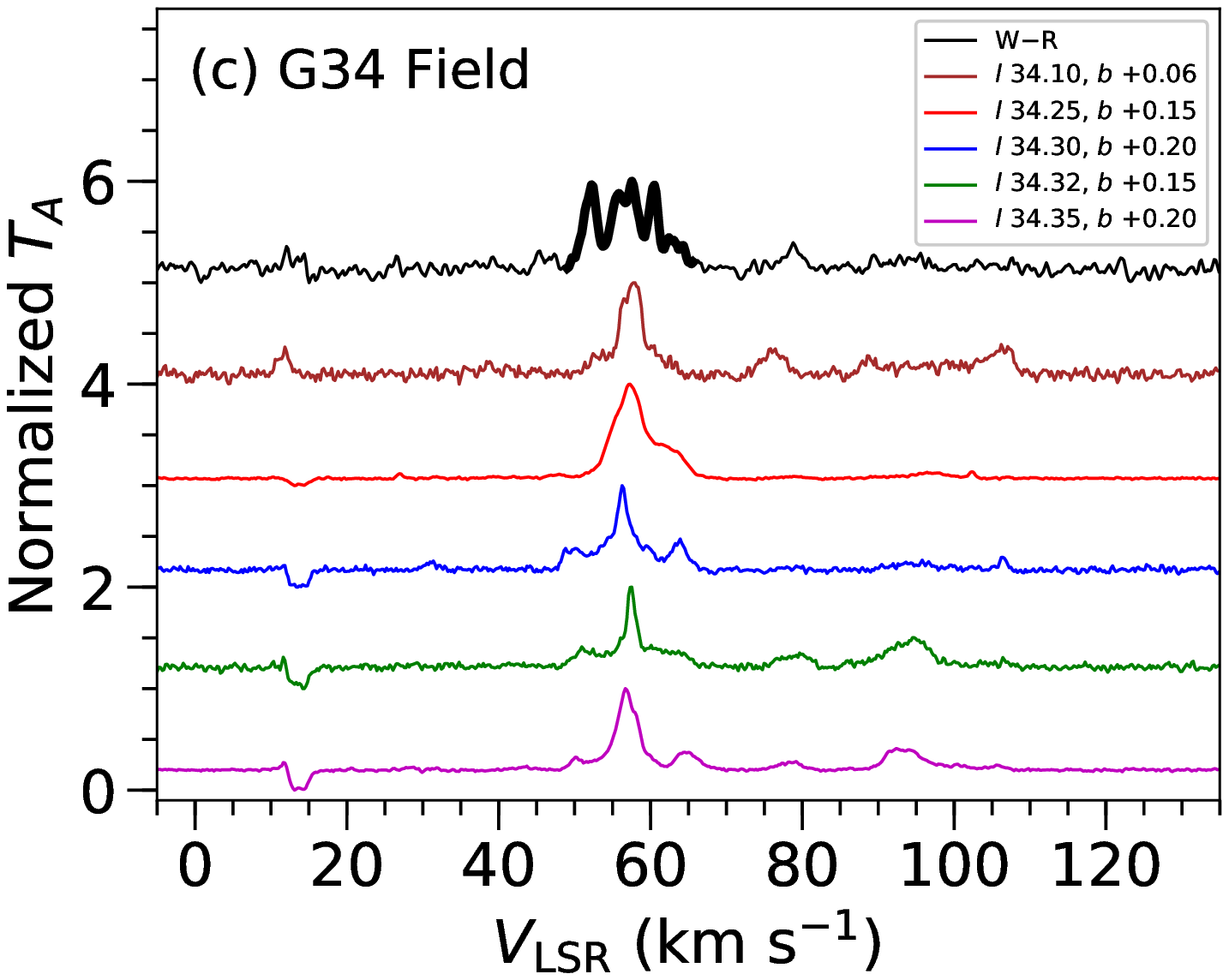}{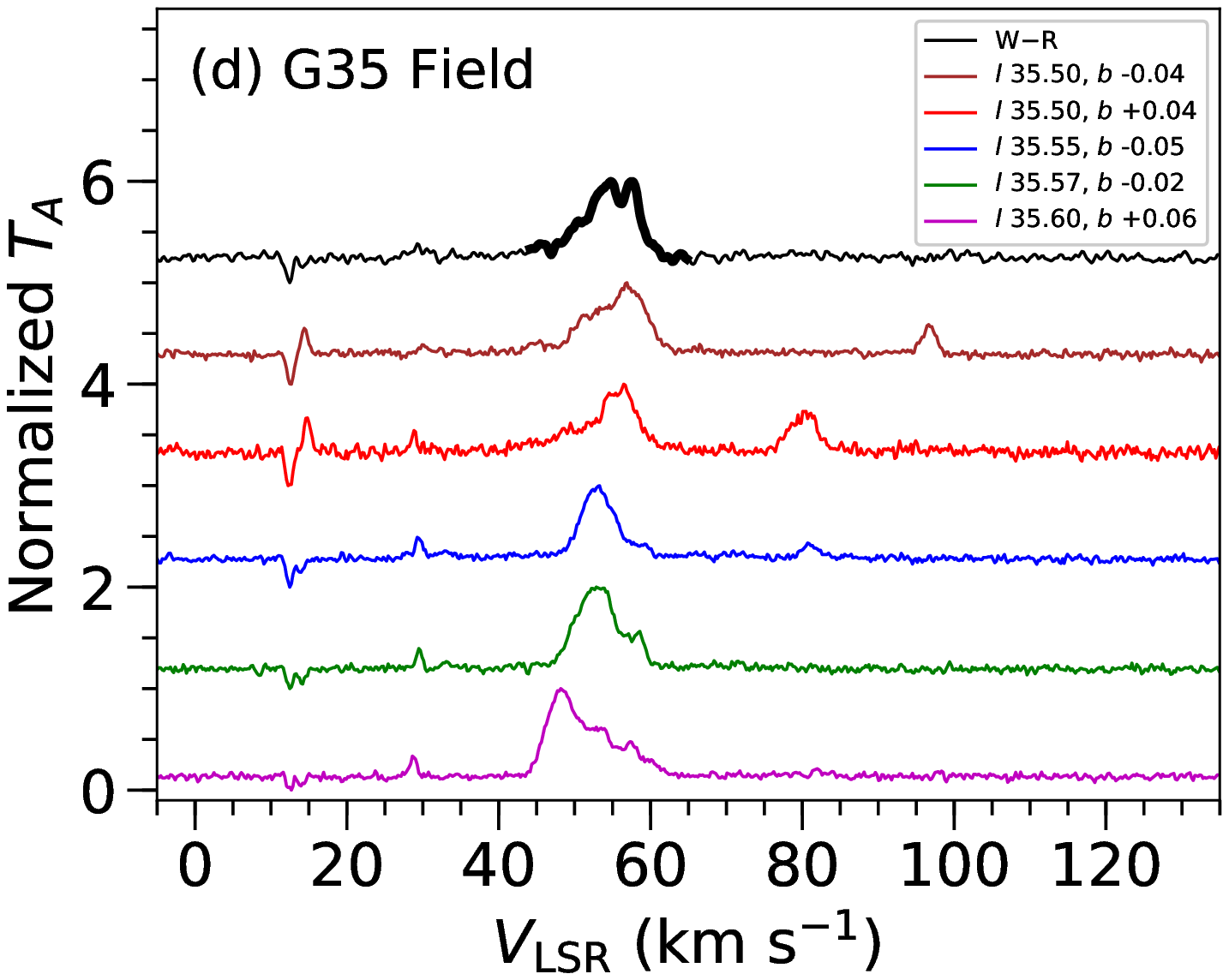} \\
\epsscale{0.5}
\plotone{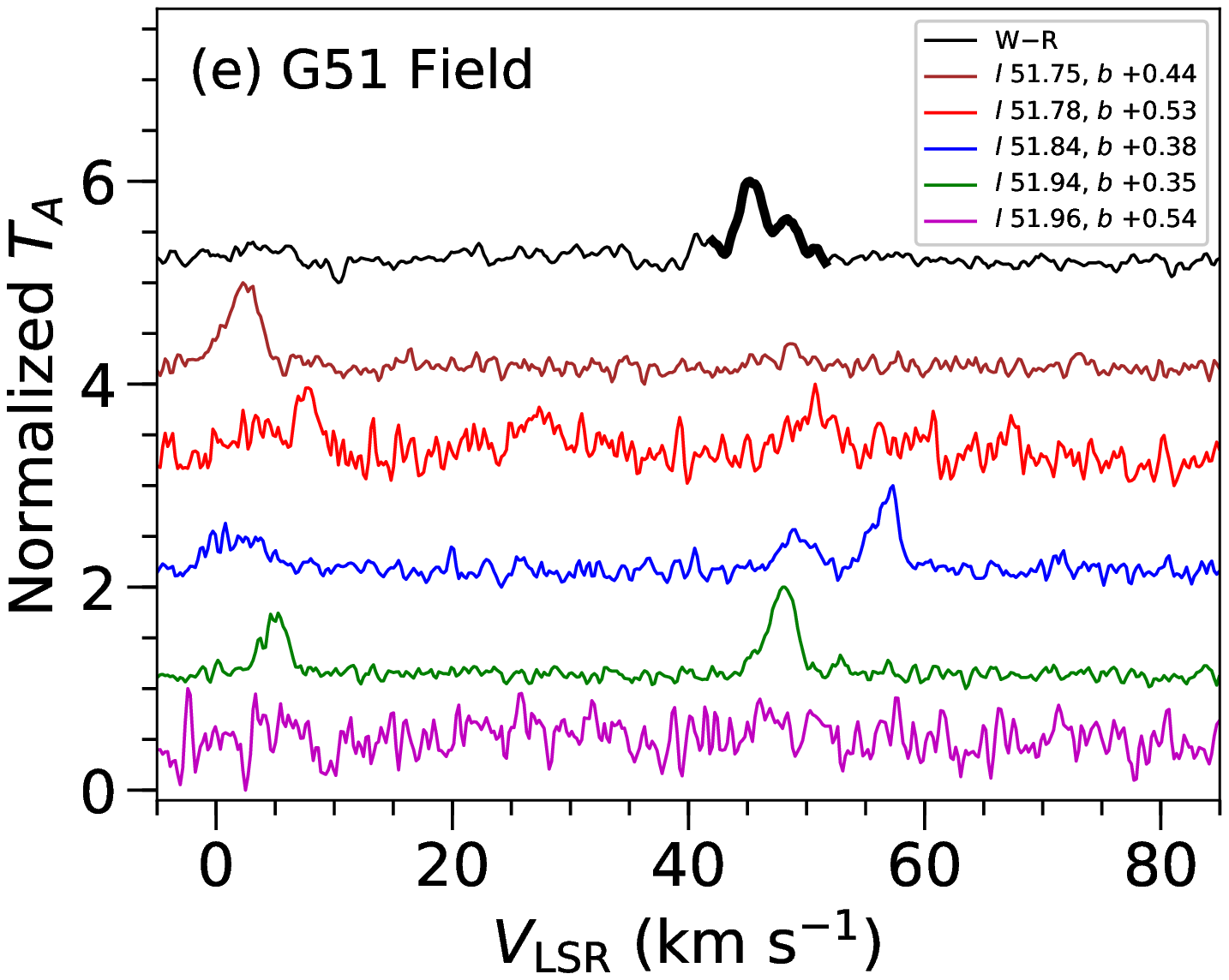}
\caption{\scriptsize \scriptsize $^{13}$CO spectra of five randomly chosen lines of sight for each region:
 (a) G15, (b) G24, (c) G34, (d) G35, and (e) G51. The spectra were constructed following the same procedure
 as that applied to the W--R stars, and plotted for the full velocity range of GRS from $-$5--135 km s$^{-1}$.
 The parts of the clouds explored in this paper are marked by bold lines. A
 few velocity regimes (particularly in the G24 region) show double or multiple peaks for which \pv diagrams
 are examined.}
\label{Apdx_fig1}
\end{figure*}

It is not necessary that the double-peaked structures seen in $^{13}$CO spectra with a trough between the peaks could only develop
 owing to the presence of W--R stars. In fact, such structures simply signify the presence of two intense parts of the molecular cloud
 along the sight lines. Similar structures could appear for several reasons. For example, molecular spectra in regions affected by
 active cloud--cloud collisions also exhibit similar structures \citep{baug16}. However, the occurrence of different mechanisms results
 in distinct signatures in the \pv diagrams. Hence, as a quality control, we constructed molecular spectra along five randomly
 distributed lines of sight within the field for every region. The corresponding spectra along with the spectrum toward the W-R star
 for the whole velocity range of the GRS are presented in Figure~\ref{Apdx_fig1}. The spectra for field directions were constructed
 by summing up all the emission within a 1$'$ diameter. We considered slightly larger diameter than the GRS beam (45$''$) to accumulate
 more emission. The sky coordinates for all spectra are also indicated in the figure. The host molecular clouds studied in this paper
 are marked with bold lines in Figure~\ref{Apdx_fig1}.

\begin{figure*}
\epsscale{1.0}
\plottwo{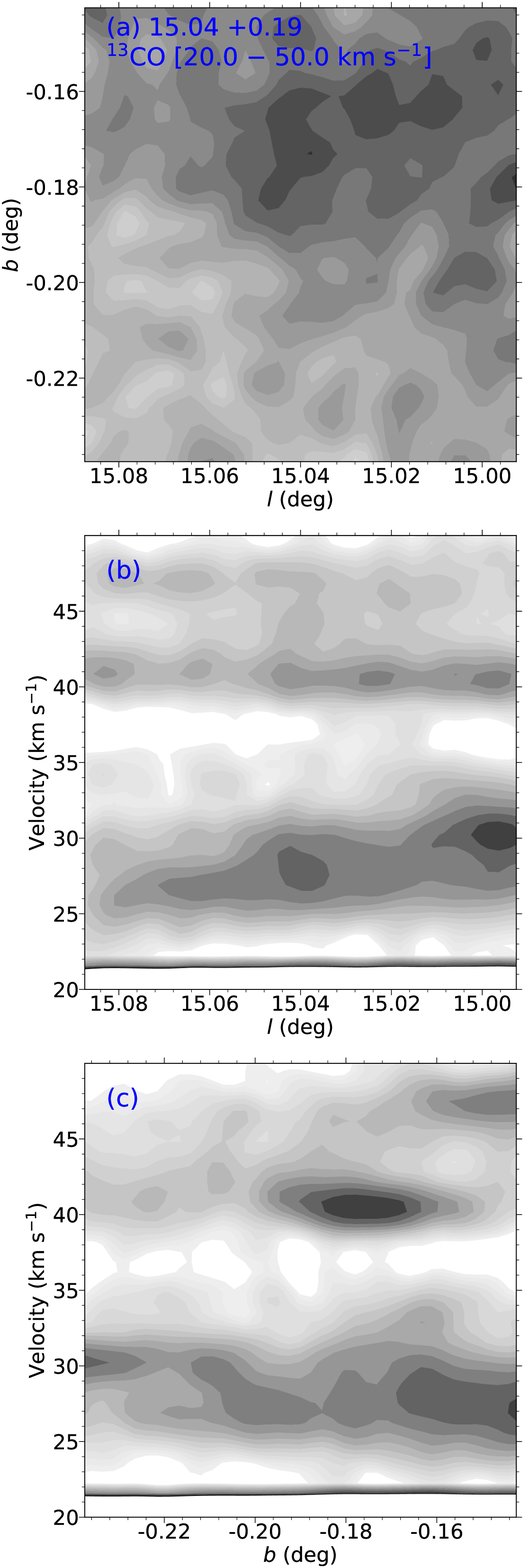}{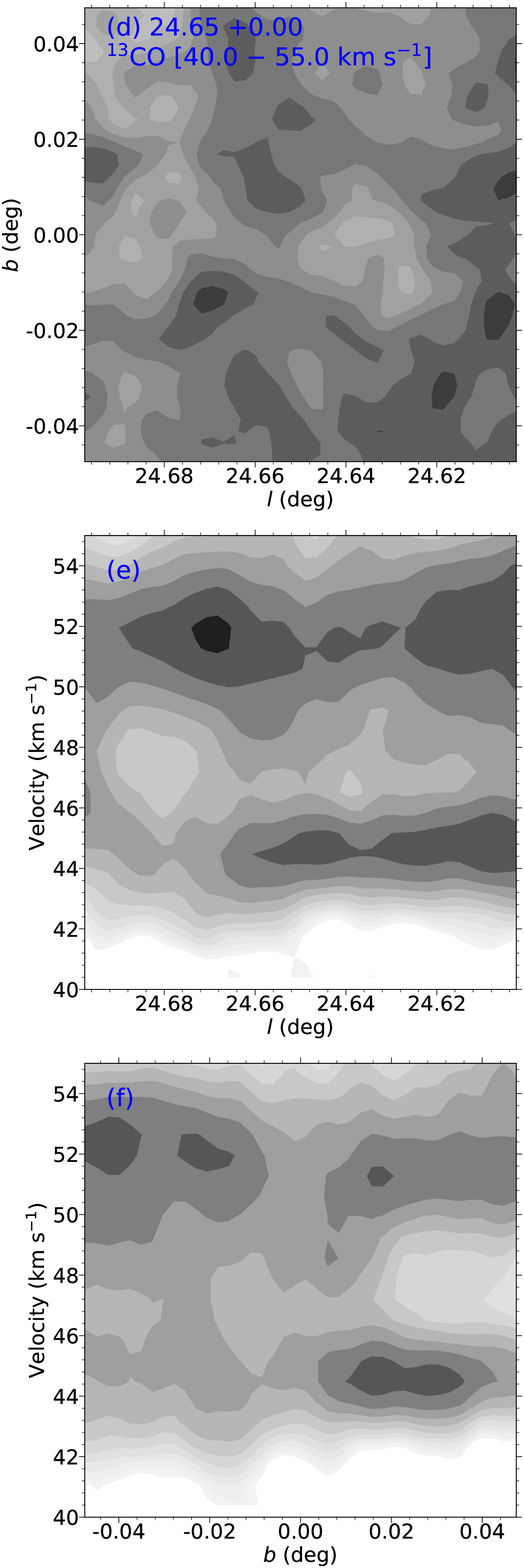}
 \caption{\scriptsize (a) and (d) Integrated intensity map of the field regions (labeled in each figure) for the respective
  velocity range (marked) where multiple peaks are seen in the $^{13}$CO spectrum (in Figure~\ref{Apdx_fig1}). (b), (c),
  (e) and (f) Corresponding \pv diagrams.}
\label{Apdx_fig2}
\end{figure*}

\begin{figure*}
\epsscale{1.0}
\plottwo{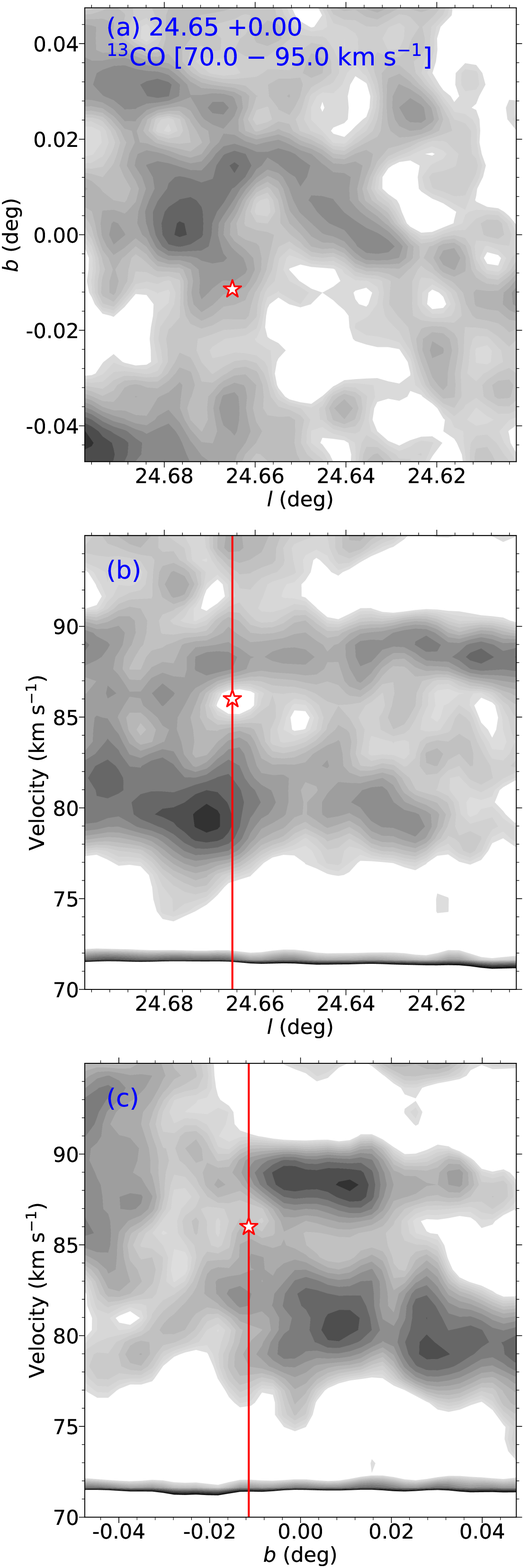}{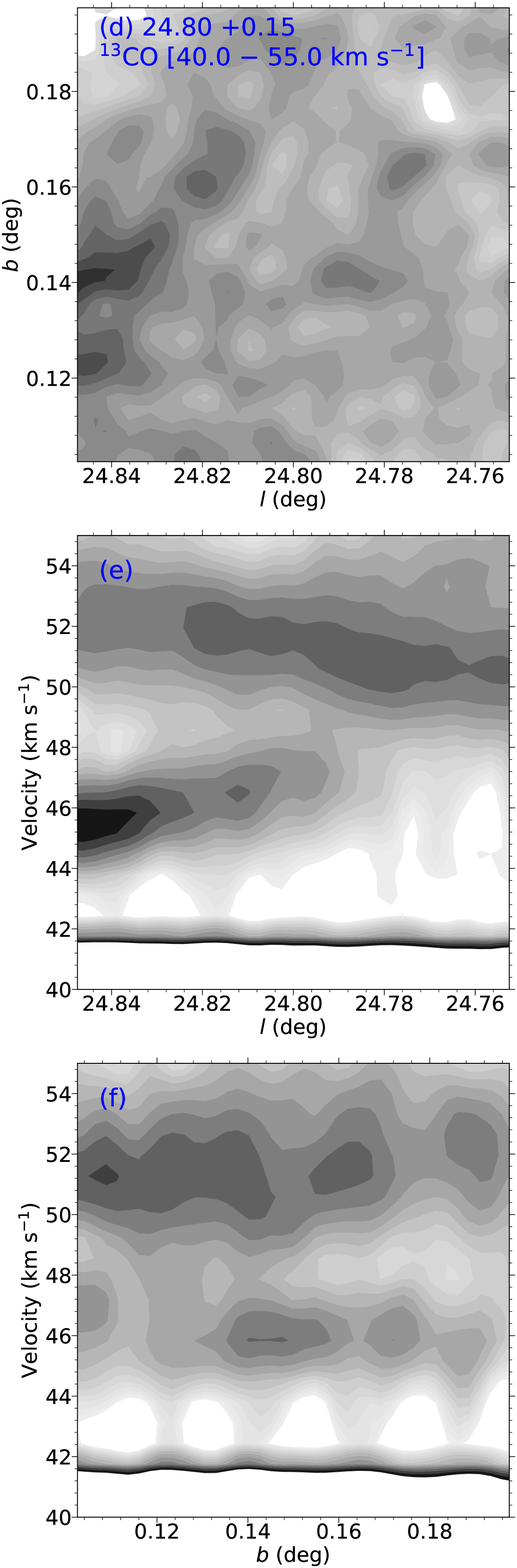}
 \caption{\scriptsize As Figure~\ref{Apdx_fig2}. In these \pv diagrams, ring-like structures are easily
 discerned. A ring-like structure can be seen in (b). The possible source (an intermediate mass YSO) of these
  ring-like structures (or expansion) is marked by a star. }
\label{Apdx_fig3}
\end{figure*}

\begin{figure*}
\epsscale{1.0}
\plottwo{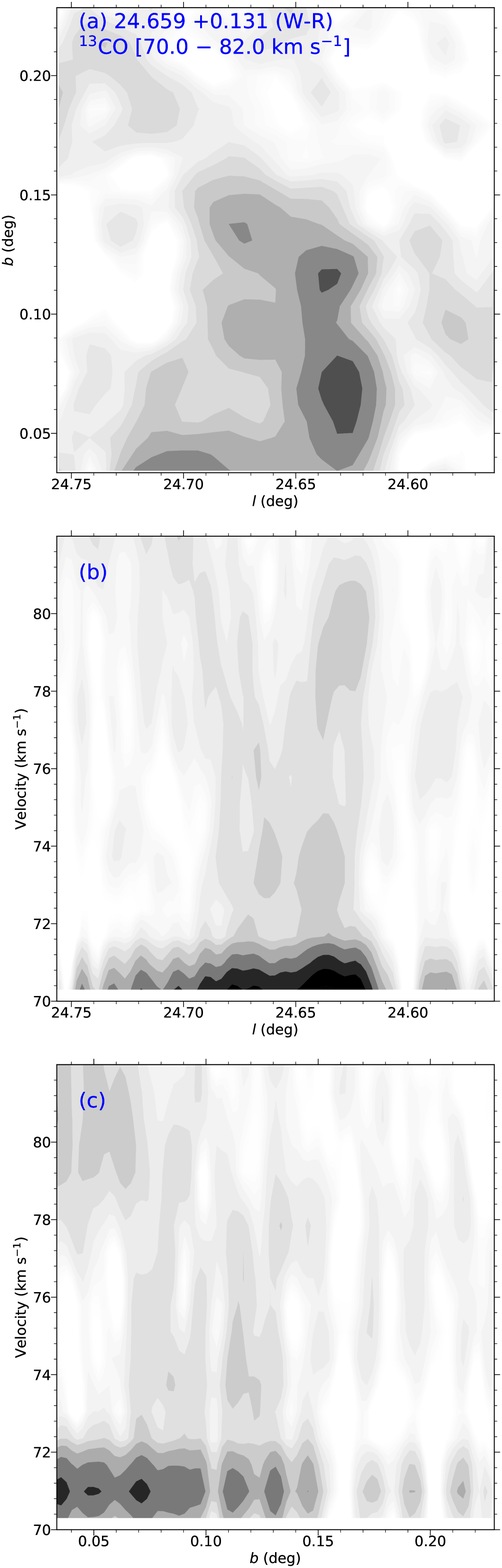}{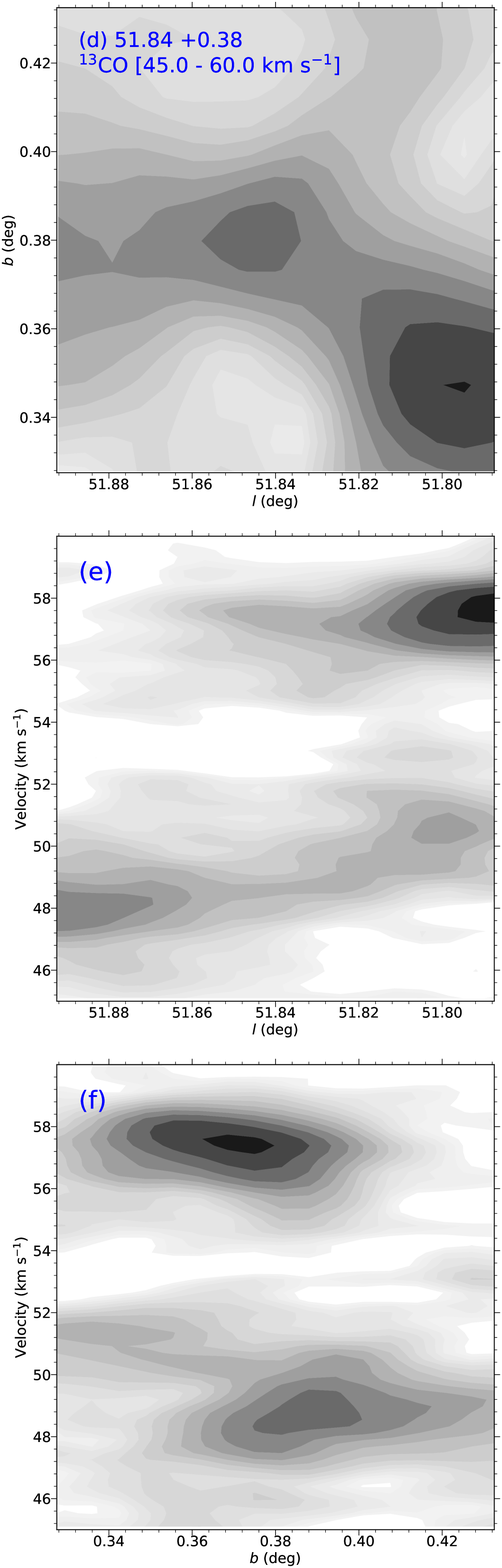}
\caption{\scriptsize As Figure~\ref{Apdx_fig2}. No ring-like structure is seen in these \pv diagrams.}
\label{Apdx_fig4}
\end{figure*}

Double-peaked structures are seen along multiple lines of sight toward the G24 region, and also toward the G15 and G51 regions.
 Another cloud with double-peaked emission features ($v_\mathrm{LSR}~\sim$ 70--82 km s$^{-1}$) is also seen in the $^{13}$CO spectrum
 along the W--R star in the G24 region. Thus, to understand the origin of these double- or triple-peaked structures, we constructed
 \pv diagrams of these particular clouds. The corresponding \pv diagrams are presented in Figures~\ref{Apdx_fig2}, \ref{Apdx_fig3},
 and \ref{Apdx_fig4}. The \pv diagrams of all but one sight line (in G24 field) indicate the presence of two distinct molecular clouds,
 but no signature of expanding molecular gas is seen along these lines of sight including the cloud at $v_\mathrm{LSR}~\sim$ 70--82
 km s$^{-1}$ toward the W--R star in the G24 region (see Figures~\ref{Apdx_fig4}d,e,f). One cloud toward the G24 field
 shows a ring-like structure in the \pv diagram (see Figure~\ref{Apdx_fig3}b) that resembles the expansion of molecular gas.
 Therefore, we explored the possible driver of the expanding gas, and found an YSO candidate reported by \citet{robitaille08}. We
 also modeled the spectral energy distribution (SED) of the source using the SED fitting tool of \citet{robitallie06} which results
 the source to be a Class I YSO with a mass of 5.0$\pm$0.5 M$_\odot$ and a surface temperature of 14,000$\pm$2200 K. This particular
 source is hot and massive enough to drive expansion of the surrounding molecular gas. All these results eventually signify that the
 ring-like structures in the \pv diagrams seen around W--R stars are signatures of expanding molecular gas driven by the strong
 energetics from W--R stars.
   
\end{document}